\documentclass[12pt]{elsarticle}                                          
\usepackage{axodraw4j}
\usepackage{hyperref}
\usepackage{graphicx}                                                         
\usepackage{a41}                                                                
\usepackage{xcolor}                     
\usepackage[rflt]{floatflt}
\usepackage{float}
\usepackage{lscape}
\usepackage{hyperref}
\hypersetup{colorlinks=true, 
linkcolor=blue, filecolor=blue,
urlcolor=mygreen}
\usepackage{breakurl} 
\usepackage{times}
\usepackage{array}
\usepackage[english]{babel}
\usepackage[latin1]{inputenc}
\usepackage[T1]{fontenc}
\usepackage{ae}
\usepackage{url}
\usepackage{amsmath, amsthm, amssymb}
\usepackage{slashed}

\usepackage{rotating}
\usepackage{graphicx}
\usepackage{comment}
\newcounter{mmacnt}
\def\restartmma{\setcounter{mmacnt}{0}}
\restartmma \catcode`|=\active
\def|#1|{\mathrm{#1}}
\catcode`|=12
\newenvironment{mma}{
\par\smallskip
\catcode`|=\active
\parskip=0pt\parindent=0pt 
\small
\def\In##1\\{%
\def\linebreak{\hfill\break\null\qquad}%
\refstepcounter{mmacnt}
\hangindent=2.5em\hangafter=0
\leavevmode
\llap{\tiny\sffamily In[\arabic{mmacnt}]:=\kern.5em}%
\mathversion{bold}\footnotesize$
\displaystyle##1$\normalsize
\mathversion{normal}\par
 }%
\def\Print##1\\{%
\def\linebreak{\hfill\break}%
\hangindent=2.5em\hangafter=0
\leavevmode ##1\par}%
\def\Out##1\\{%
\def\linebreak{$\hfill\break\null\hfill$}%
\kern\abovedisplayskip\par
\hangindent=2.5em\hangafter=0
\leavevmode
\llap{\tiny\sffamily Out[\arabic{mmacnt}]=\kern.5em}
\footnotesize$\displaystyle##1$
\normalsize\hfill\null\par
\kern\belowdisplayskip
}%
\def\Warning##1##2\\{%
\def\linebreak{\hfill\break}%
\hangindent=2.5em\hangafter=0
\leavevmode
{\scriptsize##1 : ##2}\par}%
}{%
\par\smallskip
}

\usepackage{color}
\newenvironment{fshaded}{%
\MakeFramed {\FrameRestore}
}%
{\endMakeFramed}

\makeatletter
\def\ps@pprintTitle{%
\let\@oddhead\@empty
\let\@evenhead\@empty
\def\@oddfoot{\reset@font\hfil\thepage\hfil}
\let\@evenfoot\@oddfoot
}
\makeatother
\begin{document}
\begin{frontmatter}
\title{\Large
\textbf{Decay of CP-even Higgs
$H\rightarrow h \gamma \gamma$
in Two Higgs Doublet Model: 
one-loop analytic results,
ward identity checks}}
\author[1,2]{Khiem Hong Phan}
\ead{phanhongkhiem@duytan.edu.vn}
\author[1,2]{Dzung Tri Tran}
\author[3]{Thanh Huy Nguyen}
\address[1]{\it Institute of Fundamental
and Applied Sciences, Duy Tan University,
Ho Chi Minh City $70000$, Vietnam}
\address[2]{Faculty of Natural Sciences,
Duy Tan University, Da Nang City $50000$,
Vietnam}
\address[3]
{\it VNUHCM-University of Science,
$227$ Nguyen Van Cu, District $5$,
Ho Chi Minh City $70000$, Vietnam}
\pagestyle{myheadings}
\markright{}
\begin{abstract} 
The first analytic expressions 
for loop-induced contributions 
for the decay of CP-even Higgs
$H\rightarrow h \gamma \gamma$
with $h$ being Standard-Model-like
Higgs boson within the framework of
Two Higgs Doublet Model are 
presented in this paper.
The one-loop form factors for the decay
processes are written in terms of
the scalar one-loop 
Passarino-Veltman functions
following the notations of the
packages~{\tt LoopTools} and
{\tt Collier}. Subsequently, 
physical results for the decay 
processes can be generated numerically by
using one of the above-mentioned 
packages. The analytic expressions 
shown in this paper
are verified by several
numerical checks, for examples,
the ultraviolet and 
infrared finiteness of
one-loop amplitude.
Furthermore, the amplitude 
satisfies the Ward-Takahashi identity
due to on-shell photons in final 
states. The identity is 
also verified numerically in
this work. In phenomenological
studies, the differential decay rates
as functions of the invariant mass 
of two photons in final states of
$H\rightarrow h \gamma \gamma$ are
first studied in parameter space
of the Two Higgs Doublet Models.
\end{abstract}
\end{frontmatter}
\section{Introduction}
The nature of Higgs sector 
is still an unsolved question. 
It is well-known
that the scalar Higgs potential
in the Standard Model (SM)
is selected as a
simplest form in which a scalar doublet 
is only taken into account. 
There is no fundamental principle for
determining the structure of
the Higgs sector.
In many of beyond the Standard Models 
(BSMs), the scalar Higgs potential 
is extended by introducing new 
scalar singlet and scalar multiplet.
As a result, there exist many
the additional scalar particles 
such as neutral CP-even and 
CP-odd Higges, singly (and doubly)
charged Higgses in many of
BSMs. The future colliders, 
e.g. High-luminosity Large Hadron
Colliders (HL-LHC)~\cite{Liss:2013hbb,
CMS:2013xfa} and future
Lepton Colliders
(LC)~\cite{Baer:2013cma}
are proposed as one of the main 
targets responsible for discovering
the structure of scalar Higgs
potential in the SM and in many of
BSMs, consequently for answering
the nature of electroweak
dynamic symmetry breaking (EWSB).
In the perspectives, the
measurements for decay rates
and production cross-sections
of SM-like Higgs and of the 
above-mentioned scalar particles should be
performed as accurately as possible.
Recently, loop-induced 
decay processes 
$h\rightarrow \gamma\gamma,~Z\gamma$
have been measured at the
LHC~~\cite{CMS:2014fzn, ATLAS:2015egz,
CMS:2021kom, D0:2008swt,
CMS:2013rmy, ATLAS:2017zdf,ATLAS:2020qcv}.
Furthermore, decay processes
$h\rightarrow \ell \bar{\ell}
\gamma$ have also probed at
the LHC~\cite{CMS:2015tzs,CMS:2017dyb,
CMS:2018myz, ATLAS:2021wwb}.
Probing for charged Higgs have been
reported at LHC~\cite{Cen:2018okf,CMS:2020imj,
CMS:2021wlt, CMS:2022jqc,ATLAS:2022zuc,
ATLAS:2023bzb} and references in therein.
The measurements for decay rates and
and production cross-sections for 
heavy CP-odd, CP-even Higgses 
have been performed at the 
LHC~\cite{CMS:2019kca,CMS:2019qcx,
CMS:2019ogx,ATLAS:2019tpq,ATLAS:2020zms,
ATLAS:2020gxx,ATLAS:2020tlo,ATLAS:2022rws,
ATLAS:2023zkt}, etc.

The detailed theoretical 
calculations for one-loop 
radiative corrections
to the decay widths and the
production cross-sections of
SM-like Higgs and of the additional 
scalar particles
predicted in many of BSMs 
are crucial for matching
high-precision data at future
colliders. The concerned computations
have performed in many Higgs
Extensions of the SM (HESMs) as 
reported in~\cite{Krause:2018wmo,
Athron:2021kve,Denner:2019fcr,Kanemura:2017gbi,
Kanemura:2019slf,Kanemura:2022ldq,Aiko:2023xui,
Phan:2021xwc,VanOn:2021myp,Kachanovich:2020xyg,
Hue:2023tdz,Chiang:2012qz,Benbrik:2022bol}
and references in therein. 
One-loop corrections to the production 
processes of CP-odd Higgs ($A^0$)
in the HESMs have evaluated at
LHC~\cite{Akeroyd:1999xf,
Akeroyd:2001aka, Yin:2002sq}
and at future LC~\cite{Arhrib:2002ti,
Farris:2003pn, Sasaki:2017fvk,
Abouabid:2020eik, Bernreuther:2018ynm,
Accomando:2020vbo, Aiko:2022gmz,
Akeroyd:2023kek,Esmail:2023axd,
Biekotter:2023eil,Phan:2024zus}.
More recently, one-loop corrections
for the decay process
$H\rightarrow h h$
have considered in the 
papers~\cite{Brignole:1992zv,
He:2016sqr,
Falaki:2023tyd}.
Furthermore,
di-Higgs productions
at the LHC as well as
at the LC
have computed in 
Refs.~\cite{Moretti:2004wa,
Binoth:2006ym,
Lopez-Val:2009xtx,
Ahmed:2021crg}, etc.
In this work, 
we present the
first analytic results for
one loop-induced contributions
for the decay processes of
CP-even Higgs
$H\rightarrow h \gamma \gamma$
within the Two Higgs Doublet Model.
One-loop amplitudes
are expressed in terms of
the scalar Passarino-Veltman
(PV) functions following the general
conventions of the 
packages~{\tt LoopTools}~\cite{Hahn:1998yk}
and {\tt Collier}~\cite{Denner:2016kdg}.
Physical results are hence evaluated
numerically by using one of
the mentioned packages.
In phenomenological analysis,
the differential decay rates
as functions of the invariant mass
of two final photons are
first studied in parameter space
of the Two Higgs Doublet Models.

The paper is
arranged as follows.
In section $2$, we review briefly
the Two Higgs Doublet Models.
In the Section $3$, the detailed
evaluations for one-loop
contributions to the decay amplitudes
$H\rightarrow h\gamma \gamma$ are shown.
Phenomenological results
are examined in concrete section $4$.
Conclusions and outlook for the paper
are addressed in section $5$.
\section{Two Higgs Doublet Model} 
In this section, we review shortly the
Two Higgs Doublet Model (THDM), the model
with adding an complex Higgs doublet
possessing the hypercharge $Y = 1/2$
into the scalar sector of the SM. 
We refer the paper Ref.~\cite{Branco:2011iw} 
for reviewing theory and phenomenology 
of the THDM in further detail. Following the
renormalizable conditions and
the requirements of gauge invariance,
the scalar Higgs potential reads
the general form as follows:
\begin{eqnarray}
\label{V2HDM}
\mathcal{V}(\Phi_1,\Phi_2) &=&
m_{11}^2\Phi_1^\dagger \Phi_1
+ m_{22}^2\Phi_2^\dagger \Phi_2-
\Big[
m_{12}^2\Phi_1^\dagger \Phi_2
+{\rm H.c.}
\Big]
\nonumber
\\
&&
+ \frac{\lambda_1}{2}
(\Phi_1^\dagger \Phi_1)^2
+\frac{\lambda_2}{2}
(\Phi_2^\dagger \Phi_2)^2
+ \lambda_3(\Phi_1^\dagger \Phi_1)
(\Phi_2^\dagger \Phi_2)
 \\
&&
+ \lambda_4(\Phi_1^\dagger \Phi_2)
(\Phi_2^\dagger \Phi_1)
+ \frac{1}{2}
[
\lambda_5~(\Phi_1^\dagger \Phi_2)^2
+ ~{\rm H.c.}
].
\nonumber
\end{eqnarray}
Considering
the model being CP-conserving version, 
all parameters in the scalar
potential hence are being 
real ones in this case.
Furthermore, the scalar potential is
contemplated to be symmetric under the
$Z_2$-transformation, e.g. $\Phi_1
\leftrightarrow \Phi_1$ and
$\Phi_2\leftrightarrow -\Phi_2$, 
allowing the soft breaking terms,
$m_{12}^2\Phi_1^\dagger \Phi_2
+{\rm H.c}$.
The parameter
$m_{12}^2$ plays a key role
for soft breaking scale of
the $Z_2$-symmetry.

For the EWSB, two scalar
doublets can be
parameterized
into the form of
\begin{eqnarray}
\Phi_1 &=&
\begin{bmatrix}
\phi_1^+ \\
(v_1+\rho_1+i\eta_1)/\sqrt{2}
\end{bmatrix}
,
\\
\Phi_2 &=&
\begin{bmatrix}
\phi_2^+ \\
(v_2+\rho_2+i\eta_2)/\sqrt{2}
\end{bmatrix}.
\label{representa-htm}
\end{eqnarray}
The vacuum expectation value
is fixed at
$v=\sqrt{v_1^2+v_2^2}\sim 246.$ GeV
for agreement with the SM.
After the EWSB, the spectrum of the
THDM includes of two CP-even Higgs
bosons $h$ and $H$ in which
$h$ is identified with the SM-like
Higgs boson discovered at the LHC,
a CP-odd Higgs $A^0$, and a pair of
singly charged ones $H^\pm$.
In order to obtain the physical masses
for all scalar particles, one first
diagonalizes the mass matrices in their
flavor bases. In detail, the relations of
the masses and flavor bases for all
scalar Higgs bosons are given by
\begin{eqnarray}
\begin{pmatrix}
\rho_1\\
 \rho_2
\end{pmatrix}
&=&
\mathcal{R}(\alpha)
\begin{pmatrix}
H\\
h
\end{pmatrix},
\\
\begin{pmatrix}
\phi_1^{\pm}\\
 \phi_2^{\pm}
\end{pmatrix}
&=&
\mathcal{R}(\beta)
\begin{pmatrix}
G^{\pm}\\
H^{\pm}
\end{pmatrix},
\\
\begin{pmatrix}
\eta_1\\
 \eta_2
\end{pmatrix}
&=&
\mathcal{R}(\beta)
\begin{pmatrix}
G^{0}\\
A^0
\end{pmatrix}.
\end{eqnarray}
Where $\alpha$ is the mixing
angle between two neutral Higgses
and $\beta$ is corresponding to
the mixing angle between
charged scalar particles with
charged Goldstone bosons $G^{\pm}$
(and CP-odd Higgs
with neutral Goldstone boson
$G^0$ as well). The mixing
angle $\beta$ is given by $t_{\beta}
\equiv \tan \beta= v_2/v_1$. While
$\alpha$ is considered as free
parameter which will be
constrained by experimental
data. All rotation matrices
are taken the form of
\begin{eqnarray}
\mathcal{R}(\theta)
 =
\begin{pmatrix}
c_{\theta}
&
-s_{\theta}
\\
s_{\theta}
&
c_{\theta}
\end{pmatrix}
\end{eqnarray}
where $\theta$ stand
for $\alpha$ and $\beta$
in this case.

The masses of Higgs bosons
are written in terms of the
pare parameters as follows:
\begin{eqnarray}
M_{A^0}^{2}  &=&
M^{2}-\lambda_{5}v^{2},
\\
M_{H^{\pm}}^{2}
&=&
M^{2}-
\frac{1}{2}(\lambda_{4}
+\lambda_{5})v^{2},
\\
M_{h}^{2} &=& M_{11}^{2}
s_{\beta-\alpha}^{2}
+ M_{22}^{2}c_{\beta-\alpha}^{2}
+M_{12}^{2}s_{2(\beta-\alpha)},
\\
M_{H}^{2} &=&
M_{11}^{2}c_{\beta-\alpha}^{2}
+M_{22}^{2}s_{\beta-\alpha}^{2}
-M_{12}^{2}s_{2(\beta-\alpha)}
\end{eqnarray}
where the parameter $M^2$
is used as
$M^{2}=m_{12}^{2}/
(s_{\beta}c_{\beta})$.
The matrix elements $M_{ij}$
for $i,j =1,2$ are shown
explicitly as
\begin{eqnarray}
M_{11}^{2}&=&
(\lambda_{1}c_{\beta}^{4}
+\lambda_{2}s_{\beta}^{4})v^{2}
+\frac{v^{2}}{2}\;
\lambda_{345}
\; s_{2\beta}^{2}, \\
M_{22}^{2}&=&M^{2}
+ \frac{v^{2}}{4}
\Big[
\lambda_{1}+\lambda_{2}
-2\lambda_{345}
\Big]
s_{2\beta}^{2}, \\
M_{12}^{2} &=&M_{21}^{2}
=
-\frac{v^{2}}{2}
\Big[
\lambda_{1}c_{\beta}^{2}
-\lambda_{2}s_{\beta}^{2}
-
\lambda_{345}\; c_{2\beta}
\Big]
s_{2\beta}.
\end{eqnarray}
Here, we have used
$\lambda_{345} = \lambda_{3} 
+\lambda_{4}+\lambda_{5}$.

All couplings
relating to the calculations
for one-loop contributions to
the decay amplitudes
$H\rightarrow \gamma \gamma$, 
$H\rightarrow h \gamma \gamma$
in the THDM are shown in Tables
~\ref{THDM-coupling1},~\ref{THDM-coupling2}.
Here $A_{\mu}$ is the photon field,
$p^\pm$ is the incoming momentum of $H^\pm$,
$s_W (c_W)$ is sine (and cosine)
of the Weinberg's angle, respectively.
Deriving all couplings in
Tables~\ref{THDM-coupling1},
\ref{THDM-coupling2} for THDM are
shown in further detail in the
appendix $D$.
\begin{table}[H]
\centering
{\begin{tabular}{|l|l|l|}
\hline \hline
\textbf{Vertices}
&\textbf{Notations}
& \textbf{Couplings}\\
\hline \hline
&&\\
$hW_{\mu}W_{\nu}$
& 
$
g_{hWW}
\cdot
g_{\mu\nu}
$
&
$
i
\left(
\dfrac{2 M_W^2 }{v}
\; s_{\beta-\alpha}
\right)
\cdot
g_{\mu\nu}$
\\
\hline
&&
\\
$HW_{\mu}W_{\nu}$
& $g_{HWW}
\cdot
g_{\mu\nu}
$
&
$
i
\left(
\dfrac{2 M_W^2 }{v}
\; c_{\beta-\alpha}
\right)
\cdot
g_{\mu\nu}$ \\
\hline
&& \\
$hZ_{\mu}Z_{\nu}$
& $g_{hZZ}
\cdot
g_{\mu\nu}
$
&
$i
\left(
\dfrac{2M_Z^2 }{v}
\; s_{\beta-\alpha}
\right)
\cdot
g_{\mu\nu}$ \\
\hline
&&
\\
$HZ_{\mu}Z_{\nu}$
& 
$g_{HZZ}
\cdot
g_{\mu\nu}
$
&
$
i
\left(
\dfrac{2M_Z^2 }{v}
\; c_{\beta-\alpha}
\right)
\cdot
g_{\mu\nu}$ \\
\hline
&&
\\
$hH^{\pm}H^{\mp}$
& $
g_{hH^{\pm}H^{\mp}}
$
&
$
i
\Bigg[
\dfrac{
c_{\alpha+\beta}
(4M^2-3M_h^2 - 2M_{H^\pm}^2)
}
{2vs_{2\beta}}
$
\\
&&
$
\hspace{2cm}
+
\dfrac{
(2M_{H^\pm}^2-M_h^2)
c_{(\alpha-3\beta)}
}
{2vs_{2\beta}}
\Bigg]
$\\
\hline
&&
\\
$Z_{\mu}H^{\pm}
(p^{+})H^{\mp}(p^{-})$
& 
$g_{ZH^{\pm}H^{\mp}}
\cdot
(p^+- p^-)_{\mu}
$
&
$
i
\left(
\dfrac{M_Z}{v}
\; c_{2W}
\right)
\cdot
(p^+- p^-)_{\mu}$
\\
\hline
&&
\\
$A_{\mu} H^{\pm}(p^{+})
H^{\mp}(p^{-})$
& 
$g_{A H^{\pm}H^{\mp}}
\cdot
(p^+- p^-)_{\mu}
$
&
$
i
\left(
\dfrac{M_Z
}{v}\; s_{2W}
\right)
\cdot
(p^+- p^-)_{\mu}
$
\\
\hline\hline
\end{tabular}}
\caption{
\label{THDM-coupling1}
All couplings giving one-loop
contribution to
the decay amplitudes
$H\rightarrow \gamma \gamma,
~h \gamma \gamma$
in the THDM. Here $A_{\mu}$
is the photon field,
$p^\pm$ is the incoming
momentum of $H^\pm$,
$s_W (c_W)$ is sine
(and cosine)
of the Weinberg's
angle, respectively.
}
\end{table}
\begin{table}[H]
\centering
{\begin{tabular}{|l|l|l|}
\hline \hline
\textbf{Vertices}
&\textbf{Notations}
& \textbf{Couplings}
\\
\hline
\hline
&&
\\
$H H^{\pm}H^{\mp}$
&
$g_{H H^{\pm}H^{\mp}}$
&
$
i
\Bigg[
\dfrac{
s_{\alpha+\beta}
(4M^2-3M_H^2-2M_{H^\pm}^2)
}
{2v\; s_{2\beta}}
$
\\
&
&
$
\hspace{2cm}
+
\dfrac{
(2M_{H^\pm}^2 - M_H^2)
s_{\alpha-3\beta}}
{2v\; s_{2\beta}}
\Bigg]
$
\\
\hline
&&
\\
$hHH$
&
$g_{hHH}$
&
$
i
\dfrac{
\Big[
s_{2\alpha}(3M^2-M_h^2-2M_H^2)
+
M^2s_{2\beta}
\Big]
s_{\alpha-\beta}
}{v\; s_{2\beta}}
$
\\
\hline
&&
\\
$Hhh$
&
$g_{Hhh}$
&
$
i
\dfrac{
\Big[
s_{2\alpha}(3M^2-M_H^2-2m_h^2)
-M^2s_{2\beta}
\Big]
c_{\alpha-\beta}
}{2vs_{2\beta}}
$
\\
\hline
&&
\\
$
H(p_H) H^{\pm}
(p_{H^{\pm}} )
W^{\mp}_{\mu}$
&
$g_{H H^{\pm} W}
\cdot
(p_H-p_{H^{\pm}})_{\mu}
$
&
$
\pm i
\left(
\dfrac{M_W}{v}
s_{\beta - \alpha}
\right)
\cdot
(p_H-p_{H^{\pm}})_{\mu}
$
\\
\hline
&&
\\
$h(p_h) H^{\pm}(p_{H^{\pm}})
W^{\mp}_{\mu}$
& $g_{h H^{\pm} W}
\cdot
(p_h-p_{H^{\pm}})_{\mu}
$
& $ \mp i
\left(
\dfrac{M_W}{v}
c_{\alpha-\beta}
\right)
\cdot
(p_h-p_{H^{\pm}})_{\mu} $ \\
\hline
&&
\\
$H^{\pm} H^{\mp} A_{\mu}A_{\nu}$
& $g_{H^{\pm} H^{\mp}AA}
\cdot
g_{\mu\nu}
$
&
$i
\left(
\dfrac{4 M_W^2 s_W^2}{v^2}
\right)
\cdot
g_{\mu\nu}$ \\
\hline
&&
\\
$H H^{\pm} W^{\mp}_{\mu}
A_{\nu}$
& $g_{H H^{\pm} W A}
\cdot
g_{\mu\nu}$
&
$
i
\left(
\dfrac{2 M_W^2 s_W
}{v^2}
s_{\alpha-\beta}
\right)
\cdot
g_{\mu\nu}
$ \\
\hline
&&
\\
$
h H^{\pm} W^{\mp}_{\mu}A_{\nu}$
& $g_{h H^{\pm} W A}
\cdot
g_{\mu\nu}
$
& $
i
\left(
\dfrac{2 M_W^2
s_W}{v^2}
c_{\alpha-\beta}
\right)
\cdot
g_{\mu\nu}
$ \\
\hline
&&
\\
$hH H^{\pm}H^{\mp}$
& $g_{HhH^{\pm}H^{\mp}}$
&
$i\;
\lambda_{HhH^{\pm}H^{\mp}}$
[in Eq.~(\ref{gHhSS})]
\\
\hline\hline
\end{tabular}}
\caption{
\label{THDM-coupling2}
All couplings giving one-loop
contribution to
the decay amplitudes
$H\rightarrow \gamma \gamma,
~h \gamma \gamma$
in the THDM. Here $A_{\mu}$ is
the photon field,
$p^\pm$ is the incoming
momentum of $H^\pm$,
$s_W (c_W)$ is sine
(and cosine)
of the Weinberg's
angle, respectively.
}
\end{table}

The Yukawa Lagrangian is then
expressed in terms of the mass
eigenstates is as
in~\cite{Phan:2024zus,Phan:2024vfy}
\begin{eqnarray}
\label{YukawaTHDM}
{\mathcal{L}}_{\rm Yukawa} =
-\sum_{f=u,d,\ell}
\left(g_{hff}\cdot
\bar{f} f h +
g_{Hff}\cdot  \bar{f} f H
- i g_{A^0ff}\cdot
\bar{f} \gamma_5 f A^0 \right)
+ \cdots,
\end{eqnarray}

The Yukawa couplings for
four different types of the THDM
are then given in
Table~\ref{YukawaTHDM},
see Refs.~\cite{Phan:2024zus, 
Phan:2024vfy}
for further detail.
\begin{table}[H]
\centering
\begin{tabular}{|l|l|l|l|}
\hline\hline
Type
&
$g_{huu}$
&$g_{hdd}$
&$g_{h\ell\ell}$
\\
\hline
\hline
&&&
\\
I
&
$\dfrac{m_u}{\sqrt{2}v}
\dfrac{c_\alpha}{s_\beta}$
&
$\dfrac{m_d}{\sqrt{2}v}
\dfrac{c_\alpha}{s_\beta}$
&
$\dfrac{m_{\ell}}{\sqrt{2}v}
\dfrac{c_\alpha}{s_\beta}$
\\
\hline
&&&
\\
II
&
$\dfrac{m_u}{\sqrt{2}v}
\dfrac{c_\alpha}{s_\beta}$
&
$-\dfrac{m_d}{\sqrt{2}v}
\dfrac{s_\alpha}{c_\beta}$
&
$-\dfrac{m_{\ell}}{\sqrt{2}v}
\dfrac{s_\alpha}{c_\beta}
$
\\
\hline
&&&
\\
X
&
$\dfrac{m_u}{\sqrt{2}v}
\dfrac{c_\alpha}{s_\beta}$
&
$\dfrac{m_d}{\sqrt{2}v}
\dfrac{c_\alpha}{s_\beta}$
&
$-\dfrac{m_{\ell}}{\sqrt{2}v}
\dfrac{s_\alpha}{c_\beta}$
\\
\hline
&&&
\\
Y
&
$\dfrac{m_u}{\sqrt{2}v}
\dfrac{c_\alpha}{s_\beta}$
&
$-\dfrac{m_d}{\sqrt{2}v}
\dfrac{s_\alpha}{c_\beta}$
&
$\dfrac{m_{\ell}}{\sqrt{2}v}
\dfrac{c_\alpha}{s_\beta}$
\\
\hline
\hline
\end{tabular}
\caption{\label{YukawaTHDM}
The Yukawa
couplings in THDMs with
type I, II, X, and Y
respectively. In this Table,
$f \equiv u$ for up quarks, 
$f \equiv d$ for up quarks
and $f \equiv \ell$ for leptons.
While the Yukawa couplings 
of CP-even $H$ to fermion pair 
($g_{Hff}$) are obtained
by replacing 
$c_\alpha \rightarrow s_\alpha$
and vice versa in $g_{hff}$.
}
\end{table}

The physical parameter space
$\mathcal{P}_{\rm THDM}$ for THDM
is set as follows:
\begin{eqnarray}
	\mathcal{P}_{\rm THDM} =
	\{M_h^2\sim 125. \textrm{GeV}
	, M_H^2, M_{A^0}^2,
	M^2_{H^{\pm}}, m_{12}^2,
	t_{\beta}, s_{\beta-\alpha} \}.
\end{eqnarray}
We note that we are interested in
the alignment limit of the SM. 
Therefore, we are going to
take $s_{\beta-\alpha} \rightarrow 1$ in this
work. Moreover, the role of CP-odd Higgs
$A^0$ is not related to these processes.
All physical results generated in the next
sections are independent of $M_{A^0}$.
The parameter $m_{12}^2$ plays a role of
the soft breaking scale of the
$Z_2$-symmetry. In the numerical results,
we will fix this value appropriately.
As a result, the decay widths for 
the processes are
scanned over two parameters like
$M^2_{H^{\pm}},~t_{\beta}$ for a example.
\section{Loop-induced contributions         
for $H \rightarrow h \gamma \gamma$ in the 
THDM}
\begin{figure}[H]
\centering
\includegraphics[scale=0.5]
{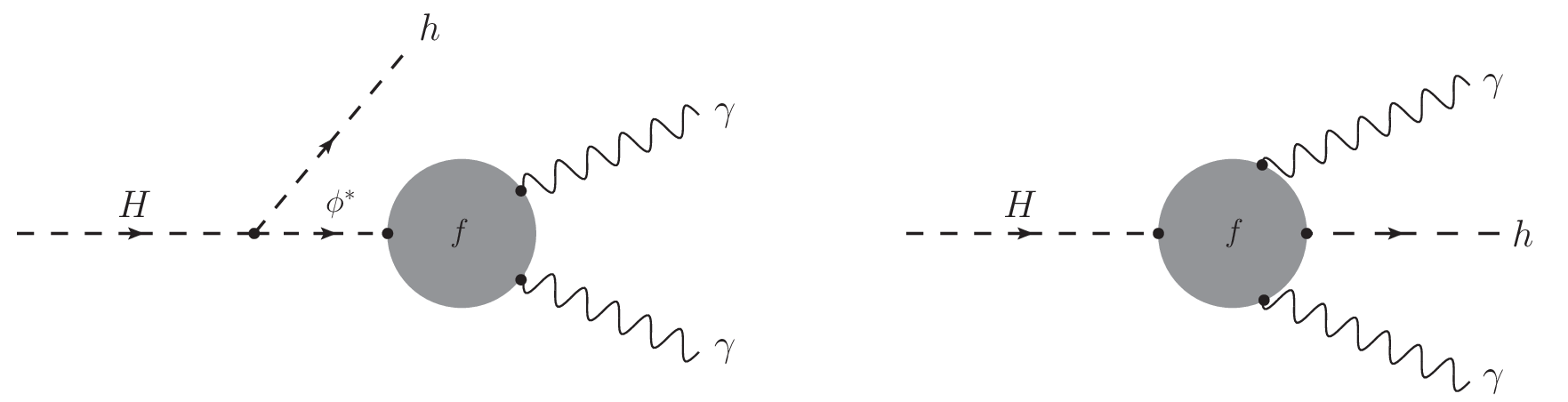}
\caption{\label{Gf} In the group $G_1$, 
one-loop Feynman diagrams include 
triangles having poles
$\phi^* \equiv h^*,~H^*$ (left) and boxes
(right) with exchanging fermions $f$ in loop
contributing to the processes.}
\end{figure}

\begin{figure}[H]
\centering
\includegraphics[scale=0.445]
{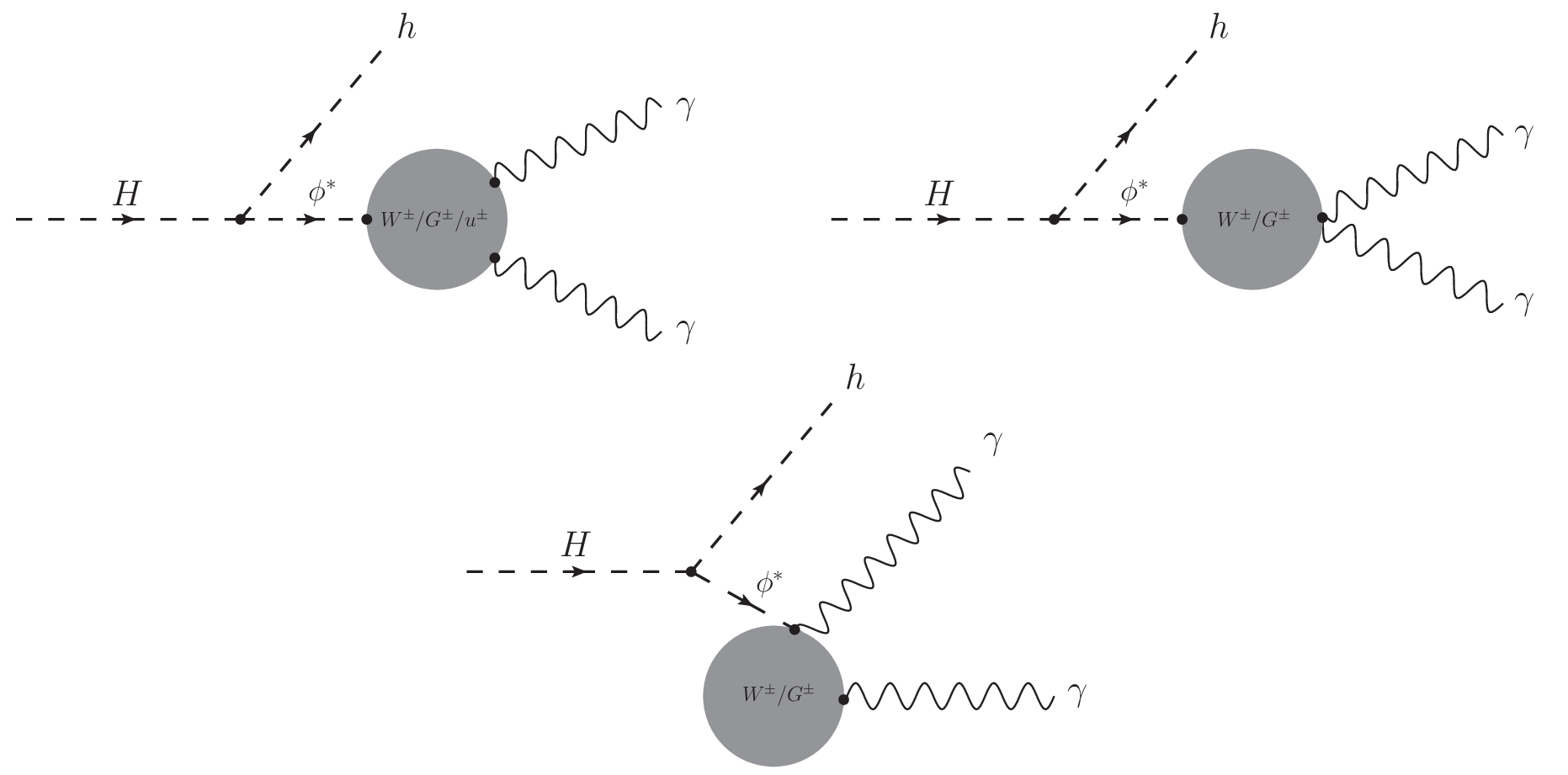}
\caption{\label{Gwtrig} 
In the
group $G_2~(a)$, we consider all
one-loop triangle Feynman diagrams
with the poles of
$\phi^* \equiv h^*,~H^*$
contributing to the processes. 
Within the HF gauge, all 
$W^\pm$ bosons, Goldstone
bosons $G^\pm$ and
Ghost particles $u^\pm$ 
are propagated in the loop.
}
\end{figure}

\begin{figure}[H]
\centering
\includegraphics[scale=0.445]
{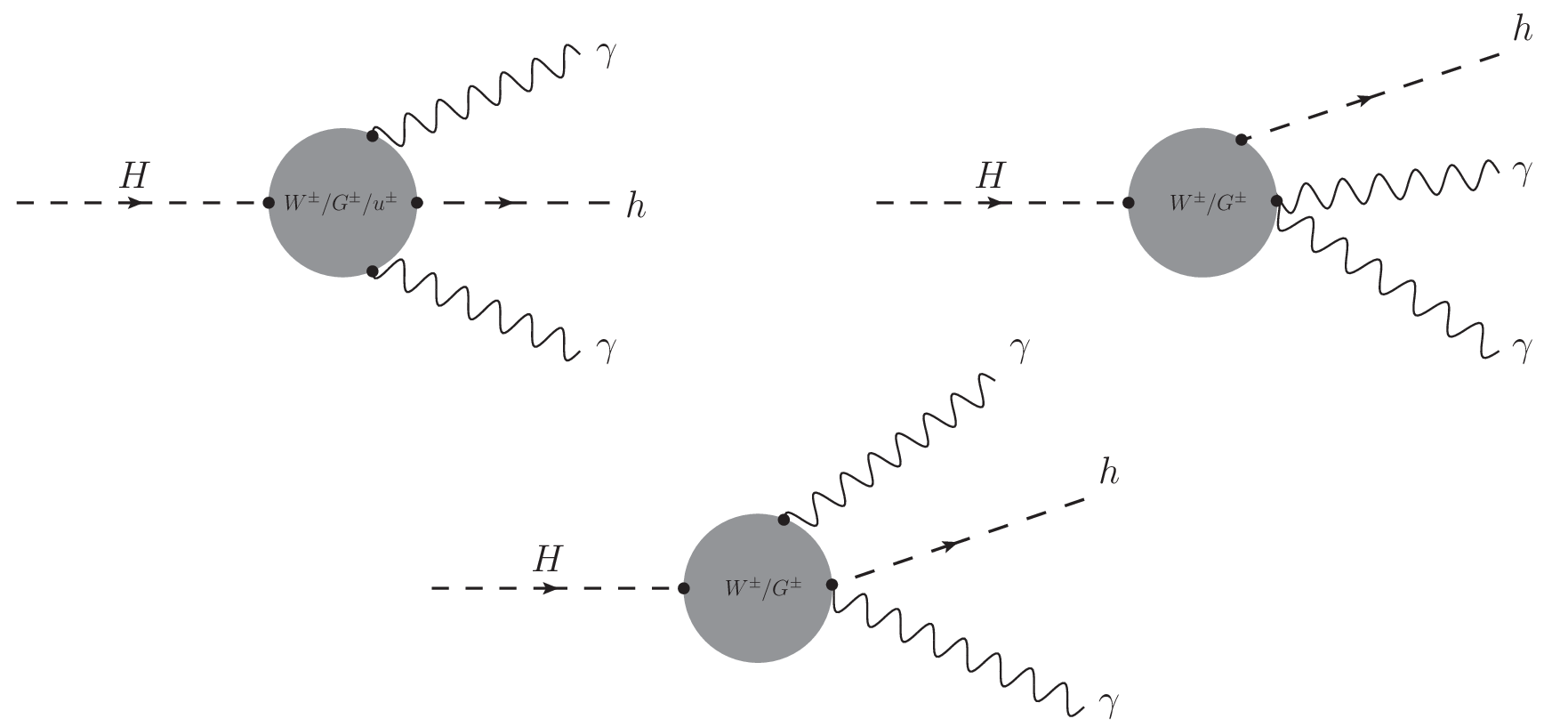}
\centering
\includegraphics[scale=0.465]
{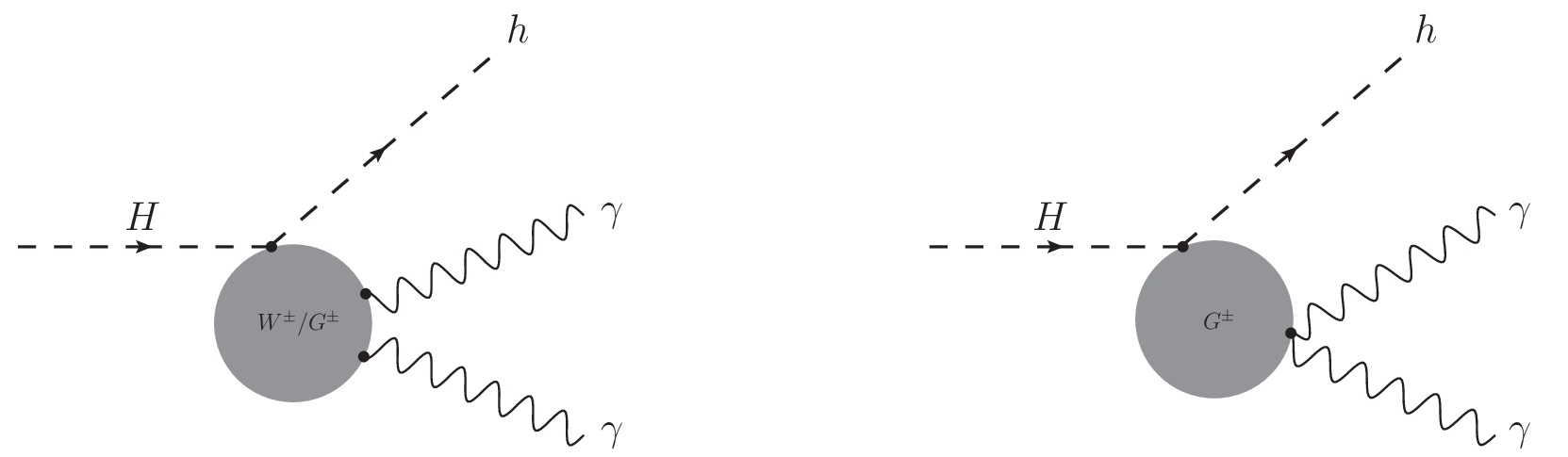}
\centering
\includegraphics[scale=0.455]
{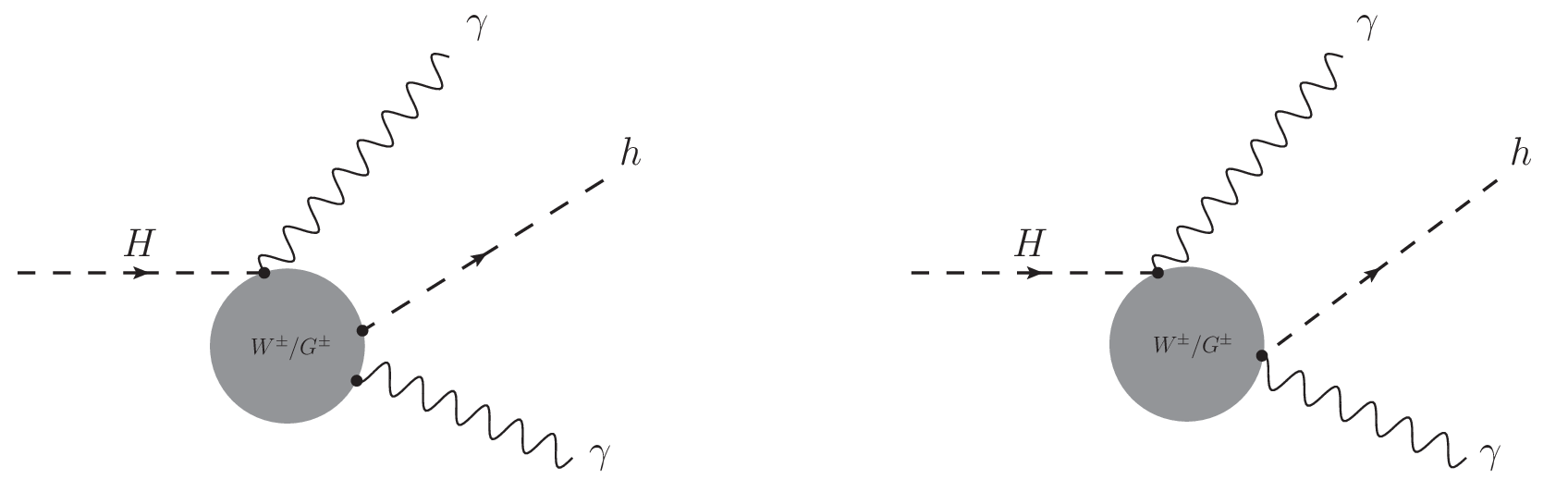}
\caption{\label{Gwbox} In the
group $G_2~(b)$, we consider all
one-loop box diagrams with
exchanging $W^\pm$ boson,
Goldstone boson $G^\pm$ and
ghost particles $u^\pm$ in loop
contributing to the
processes.}
\end{figure}

\begin{figure}[H]
\centering
\includegraphics[scale=0.495]
{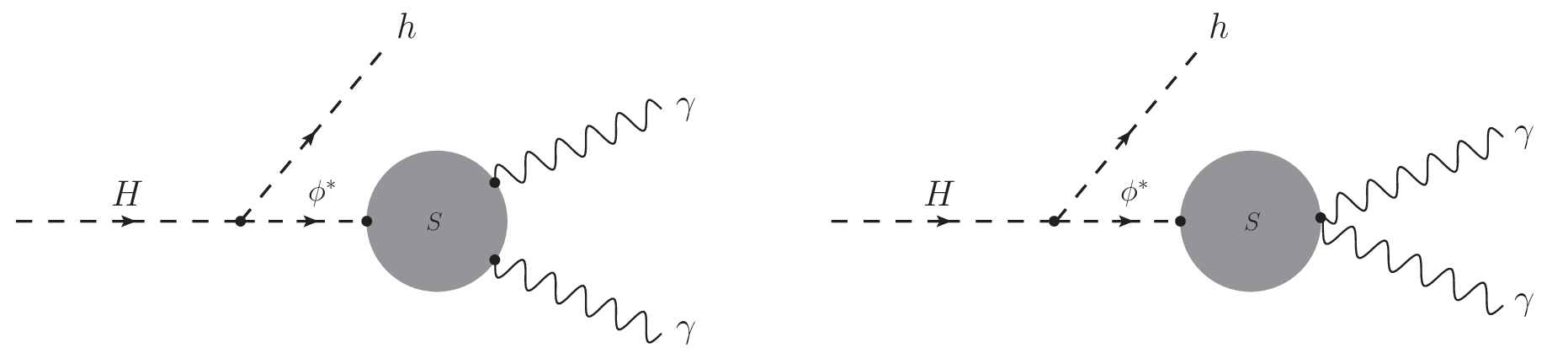}
\centering
\includegraphics[scale=0.5]
{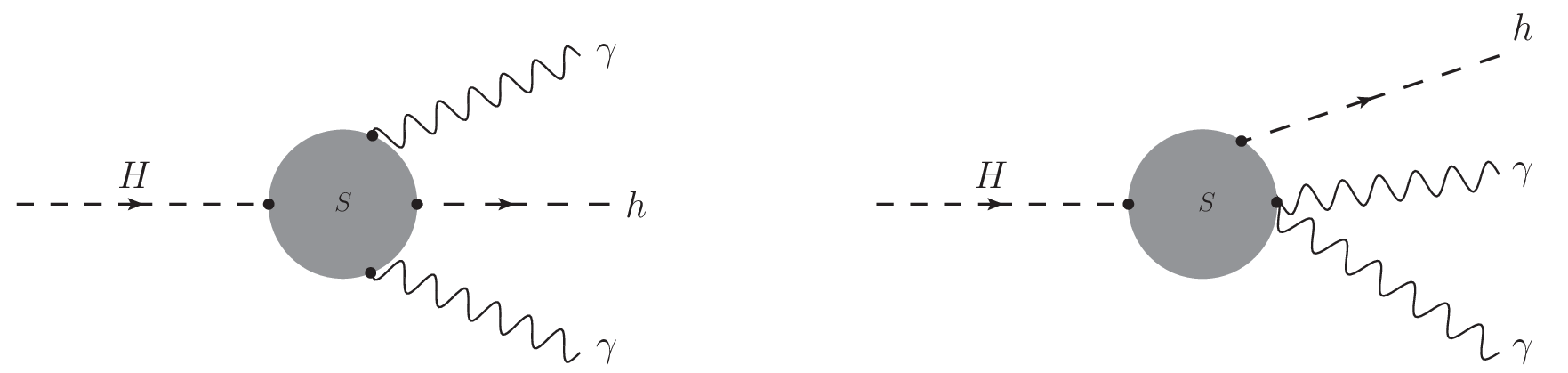}
\centering
\includegraphics[scale=0.5]
{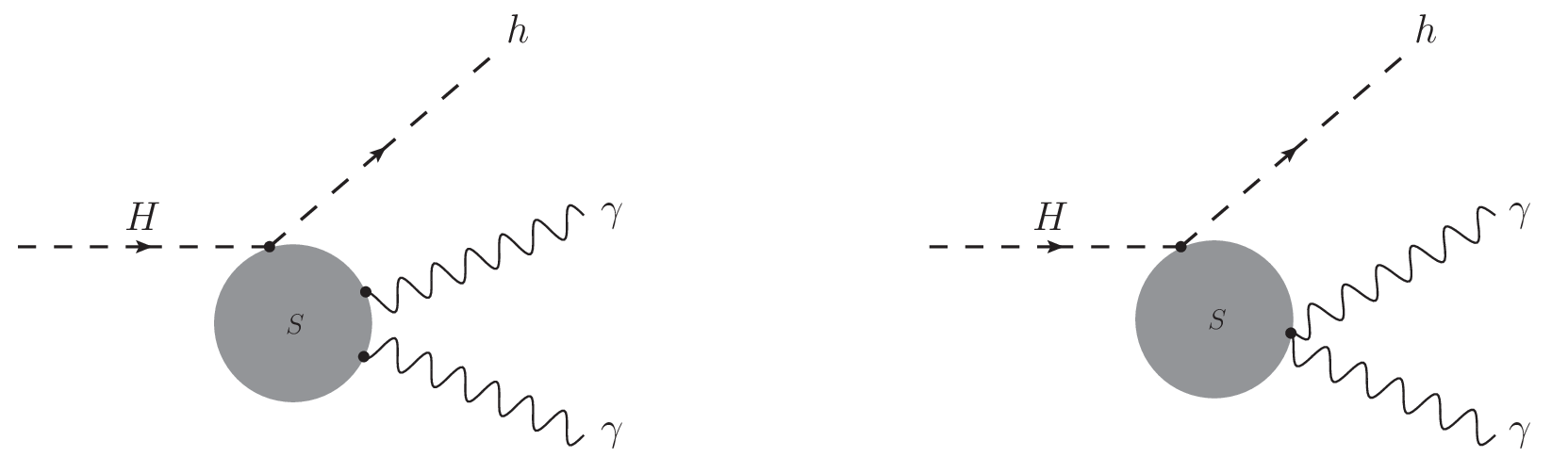}
\caption{\label{Gs}
In the group
$G_3$, one-loop triangle
Feynman diagrams with the
poles $\phi^* \equiv
h^*,~H^*$ (two first diagrams)
and one-loop box diagrams 
(four last diagrams)
exchanging charged scalar boson
$S \equiv H^\pm$ in the loop.}
\end{figure}

\begin{figure}[H]
\centering
\includegraphics[scale=0.5]
{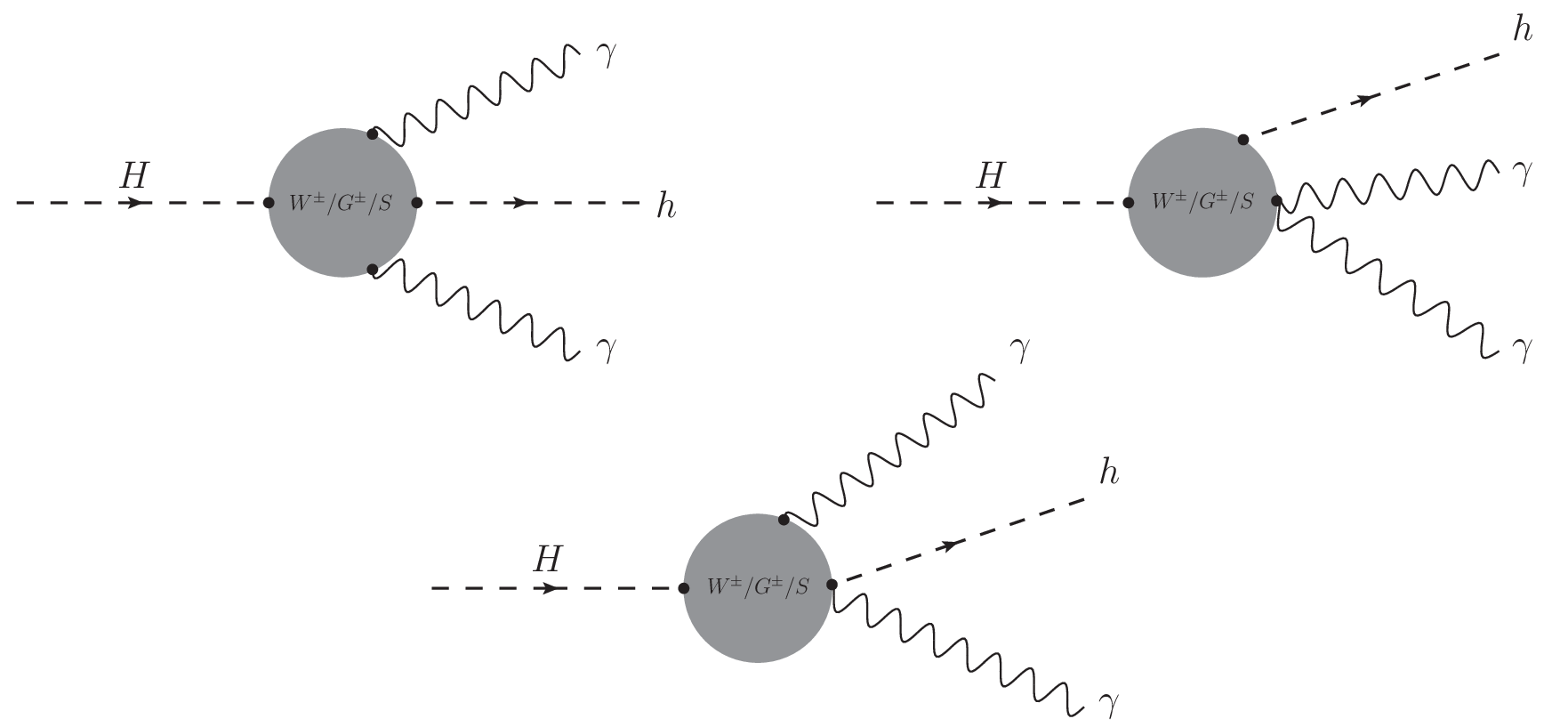}
\centering
\includegraphics[scale=0.5]
{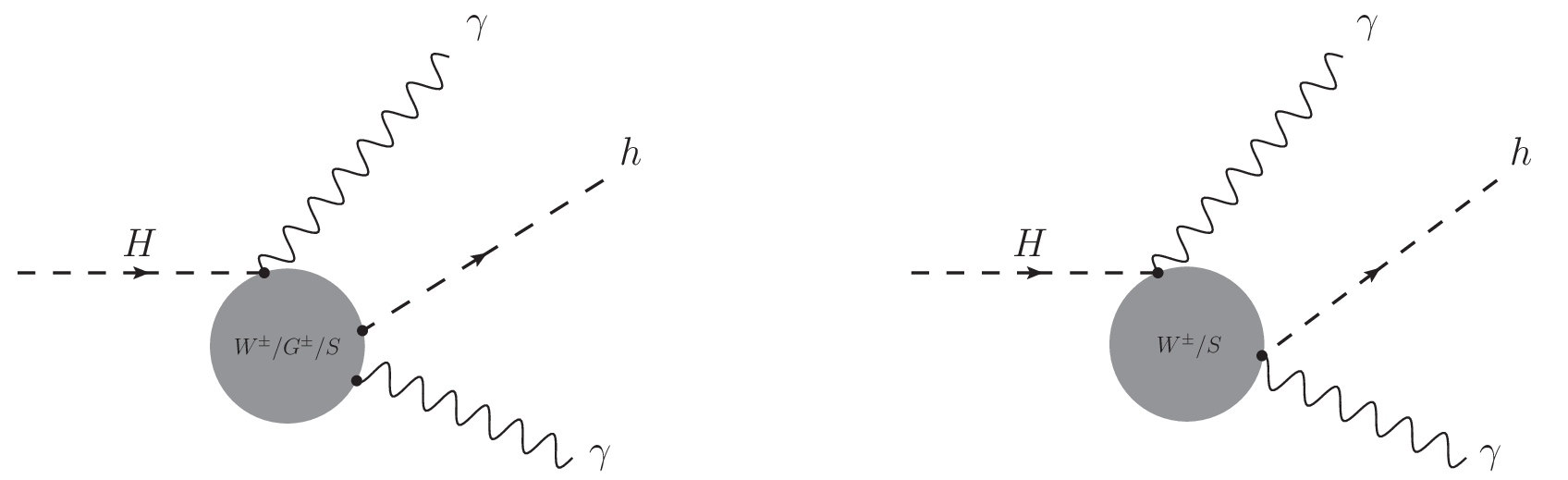}
\caption{\label{Gws}
In the group $G_4$,
one-loop box Feynman diagrams
exchanging $W^\pm$ bosons, 
Goldstone bosons $G^\pm$ 
and charged scalar 
bosons $S \equiv H^\pm$
in the loop.}
\end{figure}

First, we emphasize that general 
expressions for
$\phi \rightarrow \gamma \gamma$ with
$\phi \equiv h,~H$ in the THDM are presented in
appendix~\ref{AppendixB}. In this section, 
we focus on the detail evaluations 
for loop-induced contributing 
to decay processes of CP-even Higgs 
$H \rightarrow h \gamma \gamma$ in the THDM. 
The calculations are handled 
in the 't Hooft--Feynman gauge (HF). Within 
the HF gauge, the decay process 
$H \rightarrow h \gamma \gamma$
includes the following groups of
Feynman diagrams. In the group
$G_1$ (seen Fig.~\ref{Gf}), 
one includes one-loop triangle
Feynman diagrams with the poles of
$\phi^* \rightarrow \gamma \gamma$
for 
$\phi^*\equiv h^*,~H^*$ (left)
and one-loop box diagrams (right). 
In both topologies, all fermions $f$ 
are exchanged in the loop.
In the group $G_2$, we
consider all one-loop triangle
Feynman diagrams including
the poles of
$\phi^*$
(shown in Fig.~\ref{Gwtrig})
and all one-loop
box diagrams (presented 
in Fig.~\ref{Gwbox}) in which
all $W^\pm$ bosons, Goldstone
bosons $G^\pm$ and
Ghost particles $u^\pm$ 
are propagated in the loop.
We next consider group $G_3$
in which all
charged scalar Higgses
propagating in the loop
(depicted as in Fig.~\ref{Gs})
are taken into account in the
processes. As above groups, 
we also have one-loop triangle 
Feynman diagrams with the poles
$\phi^*$ and one-loop box diagrams
in this group.
We finally include group $G_4$
(seen Fig.~\ref{Gws})
in which one-loop box diagrams
with mixing of $W^\pm$ boson,
Goldstone boson $G^\pm$ with
charged scalar boson $S \equiv H^\pm$
exchanging in loop are contributed in
the channels. Different from all
the above-mentioned groups, in the group
$G_4$, we have no contributions of 
one-loop triangle diagrams. 
In general, one-loop amplitude
for $H(p)
\rightarrow
h(q_1)
\gamma_\mu (q_2)
\gamma_\nu (q_3)$ is
expressed in terms of
Lorentz structure as follows:
\begin{eqnarray}
\label{ampgen}
 \mathcal{A}_{H \rightarrow
 h \gamma \gamma}
 &=&
 \Big[
 F_{00} g^{\mu\nu}
 + F_{11} q_1^{\mu}q_1^{\nu}
 + F_{12} q_1^{\mu}q_2^{\nu}
 + F_{13} q_3^{\mu}q_1^{\nu}
 + F_{23} q_3^{\mu}q_2^{\nu}
\Big]\varepsilon^*_{\mu}(q_2)
\varepsilon^*_{\nu}(q_3).
\end{eqnarray}
In the amplitude, 
$\varepsilon^*_{\mu}(q_2)$,
$\varepsilon^*_{\nu}(q_3)$
are corresponding to 
the polarization vectors of 
external photons
and the scalar coefficients 
$F_{00}$ as well as $F_{ij}$ for 
$i<~j=1,\cdots, 3$ 
are one-loop form factors. 
The factors are expressed as 
functions of the following
kinematic invariant variables:
\begin{eqnarray}
\label{kine}
q_{12} &=& (q_1+q_2)^2
= M_{h}^2 + 2 q_1\cdot q_2,
\\
q_{13} &=& (q_1+q_3)^2
= M_{h}^2 + 2 q_1\cdot q_3,
\\
q_{23} &=& (q_2+q_3)^2
= 2 q_2\cdot q_3.
\end{eqnarray}
The above-mentioned 
kinematic invariants 
obey the following relation
\begin{eqnarray}
\label{kinRelation}
 q_{12} + q_{13}
 + q_{23}
 = M_{H}^2 + M_{h}^2.
\end{eqnarray}
With on-shell photons in final states,
the amplitude follows 
the Ward-Takahashi identity. 
In other words, the amplitude in 
Eq.~\ref{ampgen} will be vanished 
when we replace 
$\varepsilon^*_{\mu}(q_2) 
\rightarrow q_{2, \mu}$ or
$ \varepsilon^*_{\nu}(q_3)
\rightarrow q_{3,\nu}$. 
As a result, we derive
the following relations as:
\begin{eqnarray}
\label{ffrelation1}
F_{13}
&=&
-\frac{q_{12}
- M_{h}^2}
{q_{23} } F_{11},
\\
\label{ffrelation2}
F_{00}
&=&
-\frac{q_{12}
- M_{h}^2 }{2}
F_{12}
-\frac{q_{23} }{2} F_{23},
\\
\label{ffrelation3}
F_{12}
&=&
-\frac{q_{13}
- M_{h}^2 }{q_{23} } F_{11},
\\
\label{ffrelation4}
F_{00}
&=&
-\frac{q_{13}
- M_{h}^2 }{2} F_{13}
-\frac{q_{23} }{2} F_{23}.
\end{eqnarray}
Using the above relations,
the amplitude can presented
via two independent one-loop
factors, e.g. taking $F_{11}$ and
$F_{23}$ as example.
In detail, the amplitude can
be expressed as follows:
\begin{eqnarray}
\label{ampred}
\mathcal{A}_{H \rightarrow
h \gamma \gamma}
&=&
\Bigg\{
\Bigg[
\frac{(q_{12} - M_{h}^2)
(q_{13} - M_{h}^2)}{2 q_{23}}
\cdot
g^{\mu\nu}
+
q_1^{\mu}q_1^{\nu}
+
\frac{(M_{h}^2-q_{13})}{q_{23}}
\cdot
q_1^{\mu}q_2^{\nu}
\\
&&\hspace{0cm}
+ \frac{(M_{h}^2-q_{12})}{q_{23}}
\cdot
q_3^{\mu}q_1^{\nu}
\Bigg] \cdot F_{11}
+
\Big[
q_3^{\mu}q_2^{\nu}
-
\frac{q_{23}}{2}
\cdot
 g^{\mu\nu}
\Big]
\cdot F_{23}
\Bigg\}
\,
\varepsilon^*_{\mu}(q_2)
\varepsilon^*_{\nu}(q_3).
\nonumber
\end{eqnarray}
In this paper, we will collect
two independent
one-loop form factors $F_{11}$,
$F_{23}$
in Eq.~\ref{ampred}.
They are expressed in terms of the
scalar Passarino-Veltman (PV) one-loop
integrals in following the notations
of the package~{\tt LoopTools} and
the library
{\tt Collier}~\cite{Denner:2016kdg}.
Consequently, the decay rates and
its distributions can be evaluated
numerically
by using one of the above-mentioned
packages. We emphasize that all
phenomenological results shown
in the following sections are generated
by using the package~{\tt Collier}.

The one-loop form factors are 
separated into
two contribitions which are calculated
from one-loop triangle and box diagrams.
In detail, the factors are computed
as follows:
\begin{eqnarray}
 F_{ab} & = & F_{ab}^{\textrm{Trig}}
 +  F_{ab}^{\textrm{Box}}
\end{eqnarray}
for $ab = \{ 11, 23\}$. The
form factors
$F_{ab}^{\textrm{Trig}}$
are first collected in the next paragraphs.
They are calculated from
all one-loop triangle Feynman diagrams
given the above-paragraphs.
The form factors $F_{ab}^{\textrm{Trig}}$
are contributed from fermions $f$, 
$W$ bosons and charged Higgses $H^{\pm}$ 
propagating in the loop. 
It is easy to confirm that
one-loop form factors
$F_{11}^{\textrm{Trig}}$ are 
absent from one-loop triangle
Feynman diagrams. 
Another one-loop form
factors $F_{23}^{\textrm{Trig}}$
are taken the form of: 
\begin{eqnarray}
\label{FFtrig}
F_{23}^{\textrm{Trig}}
&=&
\sum \limits_{\phi=h^*, H^*}
\frac{
g_{H h \, \phi}
}{
\Big[
q_{23}
-
M_{\phi}^2
+
i \Gamma_{\phi} M_{\phi}
\Big]}
\times
\left\{
F_{23,f}^{\textrm{Trig}}
+
F_{23,W}^{\textrm{Trig}}
+
F_{23,H^\pm}^{\textrm{Trig}}
\right\}.
\end{eqnarray}
All the coefficients factors
$F_{23,f/W/H^\pm}^{\textrm{Trig}}$
are presented in terms of the scalar
PV-functions in appendix C.
By expanding PV-functions as in 
Eq.~\ref{f-PV}, we arrive
at the final results for
$F_{23}^{\textrm{Trig}}$ as follows:
\begin{eqnarray}
\label{offPHIgg}
F_{23}^{\textrm{Trig}}
&=&
\dfrac{\alpha}{4 \pi}
\sum \limits_{\phi=h^*, H^*}
\dfrac{g_{H h \, \phi}}{
\Big[
q_{23}
-
M_{\phi}^2
+
i \Gamma_{\phi} M_{\phi}
\Big]}
\times
\nonumber
\\
&&
\times
\Bigg\{
\sum \limits_f
(
N^C_f
Q_f^2
)
\cdot
g_{\phi ff}
\cdot
(4\lambda_f)
\Big[
1
+
\big(
1
-
\lambda_f
\big)
f ( \lambda_f )
\Big]
\nonumber
\\
&&
+
\dfrac{
g_{\phi \, WW}
}{M_W^2}
\cdot
\Big[
2 \rho_\phi
+
3 \lambda_W
+
\lambda_W
\big(
8
-
2 \rho_\phi
-
3 \lambda_W
\big)
f ( \lambda_W )
\Big]
\nonumber
\\
&&
-
\left(
\dfrac{
4\;g_{\phi H^\pm H^\mp}
}
{
q_{23}
}
\right)
\cdot
\Big[
1
-
\lambda_{H^\pm}
f ( \lambda_{H^\pm} )
\Big]
\Bigg\}.
\end{eqnarray}
Here $N^C_f$ ($Q_f$) is number of color
(charged) for fermion $f$, respectively.
In all above equations, we have used
the variables like $\rho_\phi
= M_{\phi}^2 / q_{23}$
and $\lambda_i = 4 m_i^2/q_{23}$
with $m_i$ being
$m_f, M_W, M_{H^\pm}$ in this case.
The formulas for $f$-function given
in above is shown explicitly 
in appendix A.

We next collect one-loop
form factors $F_{ab}^{\textrm{Box}}$
which are attributed from
all one-loop box diagrams appear in
the above groups.
The factors can be also decomposed 
as follows:
\begin{eqnarray}
\label{FFbox}
F_{ab}^{\textrm{Box}}
&=&-
\dfrac{e^2}{2\pi^2}
\sum \limits_f
(N^C_f\; m_f^2\;
Q_f^2)
\cdot
g_{h ff}
\cdot
g_{H ff}
\times
F_{ab,f}^{\textrm{Box}}
\nonumber\\
&&
+
\left(
\dfrac
{e^2}
{8 \pi^2}
\right)
\cdot
g_{hWW}
\cdot
g_{HWW}
\times
F_{ab,W}^{\textrm{Box}}
\nonumber\\
&&
-
\left(
\dfrac{e^2}{8\pi^2 M_W^2}
\right)
\cdot
g_{h H^\pm W^\mp}
\cdot
g_{H H^\pm W^\mp}
\times
F_{ab,W H^\pm}^{\textrm{Box}}
\nonumber\\
&&
+
\left(
\dfrac{e^2}
{4 \pi^2}
\right)
\times
F_{ab,H^{\pm}}^{\textrm{Box}}
\end{eqnarray}
for $ab = \{ 11, 23\}$.
All one-loop form factors
attributing in Eq.~\ref{FFbox},
$F_{ab,f/W/WH^{\pm}/H^{\pm}
}^{\textrm{Box}}$
are presented in detail in the
appendix C.

After collecting the one-loop
form factors, we perform the
numerical checks for the 
computations. We verify that
one-loop form factors are ultraviolet
(UV) and infrared (IR) finiteness.
Furthermore, due to the on-shell
photons in final states, the amplitude
for the decay channels also follows
the Ward identity. The
identity is also confirmed
numerically in this work.
One finds that the results are good
stability. The numerical test
are presented in the next section.
Having all correctness one-loop
form factors, the decay rates
are then evaluated as follows:
\begin{eqnarray}
\Gamma_{H \rightarrow
h \gamma \gamma}
&=&
\dfrac{1}{256 \pi^3 M_H^3}
\int \limits_{q_{23}^\text{min}}^{q_{23}^\text{max}}
d q_{23}
\int \limits_{q_{13}^\text{min}}^{q_{13}^\text{max}}
d q_{13}
\,
\sum \limits_\text{pol.}
\big|
\mathcal{A}_{H \rightarrow
h \gamma \gamma}
\big|^2
\end{eqnarray}
where total amplitude in
squared is given by 
\begin{eqnarray}
\sum \limits_\text{pol.}
\big|
\mathcal{A}_{H \rightarrow
h \gamma \gamma}
\big|^2
&=&
\Bigg[
\dfrac{
\big(
M_h^2-q_{13}
\big)
\big(
M_H^2-q_{13}
\big)
}{2q_{23}^2}
\Big[
2
q_{23}
\cdot
q_{13}
+
(M_h^2-q_{13})
(M_H^2-q_{13})
\Big]
\\
&&
+
\big(
q_{13}^2
+
M_h^4
\big)
\Bigg]
\cdot 
\big|
F_{11}
\big|^2
-
q_{23}
\;
M_h^2
\;
\cdot 
\mathcal{R}e
\Big[
F_{11}
\times
F_{23}^*
\Big]
+
\dfrac{q_{23}^2
}{2}
\cdot 
\big|
F_{23}
\big|^2.
\nonumber
\end{eqnarray}
The integration regions 
are expressed as follows:
\begin{eqnarray}
q_{23}^\text{min}
&=& 0,
\\
q_{23}^\text{max}
&=& (M_H - M_h)^2,
\\
q_{13}^\text{max(min)}
&=&
\dfrac{1}{2}
\Bigg\{
M_H^2
+
M_h^2
-
q_{23}
\pm
\sqrt{
\big(
M_H^2
+
M_h^2
-
q_{23}
\big)^2
-
4 M_H^2
M_h^2
}
\Bigg\}.
\end{eqnarray}
\section{Phenomenological results}
We turn our attention to
discuss the phenomenological
results for the CP-even Higgs
decay process $H\rightarrow
h \gamma \gamma$ in the THDM.
As usual procedure, 
we first summary
the current constraints for
the parameter space of the THDM.
Taking into account 
the theoretical conditions 
and experimental constraints
in the THDM, we find the current
regions for the parameter
space for the phenomenological 
analyses. We first discuss on the 
theoretical conditions. 
The subjects rely on the perturbative 
regime and tree-level unitarity 
of gauge theories
as well as vacuum stability conditions
for the scalar Higgs potential.
These theoretical constraints
have studied in the following
Refs.~\cite{Nie:1998yn,
Kanemura:1999xf, Akeroyd:2000wc,
Ginzburg:2005dt, Kanemura:2015ska}
and references therein.
In the aspects of 
the experimental constraints, 
we take into consideration 
the electroweak precision
tests (EWPT) for the THDM
which the topics have implicated
at  the LEP~\cite{Bian:2016awe, Xie:2018yiv}.
Furthermore, the direct and indirect
searches for the masses of
scalar particles for the THDM have
performed at the LEP, the
Tevaron and the LHC as summarized
in the Ref.~\cite{Kanemura:2011sj}.
In additional, implicating for
one-loop induced decays
of $h\rightarrow \gamma\gamma$
and $h\rightarrow Z\gamma$ in 
the THDM have performed in
Refs.~\cite{Chiang:2012qz,
Benbrik:2022bol} and references
therein. Combining all the above
constraints, we can take logically
the parameter space for our analysis 
in the THDM as follows. We select
$126$ GeV $\leq M_H \leq 1000$ GeV,
$60$ GeV $\leq M_{A^0} \leq 1000$ GeV
and $80$ GeV $\leq M_{H^{\pm}}
\leq 1000$ GeV in the type I
and type X of the THDM.
For the Type-II and
Y, we can scan consistently
the physical parameters
as follows: $500$ GeV
$\leq M_H \leq 1000$ GeV,
$500$ GeV $\leq M_{A^0} \leq 1000$ GeV
and $580$ GeV $\leq M_{H^{\pm} }
\leq 1500$ GeV. In both types, one takes
$2 \leq t_{\beta} \leq 20$,
$0.95\leq s_{\beta-\alpha} \leq 1$ for
the alignment limit of the SM 
and can be taken the $Z_2$-breaking
parameter as
$m_{12}^2 =M_H^2 s_\beta c_\beta$.
Last but not least,
from flavor experimental data,
the further limitations
on $t_{\beta}$, $M_{H^{\pm}}$
have also performed for the THDM
with the softly broken $Z_2$-symmetry
in Ref.~\cite{Haller:2018nnx}.
Following the results in
Ref.~\cite{Haller:2018nnx},
we find that the small
values of $t_{\beta}$ are
favoured for matching
the flavor experimental data.
To complete our discussions,
we are also interested in
considering the small values of
$t_{\beta}$ (scan it
reasonably over the
region of $2 \leq t_{\beta} \leq 10$
in some cases, for examples). 
In our analysis,
the decay widths
of SM-like Higgs is taken from
experimental value
(as indicated in below).

Secondly, other input parameters
in this work are used as follows.
For gauge sector, we take
$M_Z = 91.1876$ GeV, $\Gamma_Z  = 2.4952$ GeV,
$M_W = 80.379$ GeV, $\Gamma_W  = 2.085$ GeV.
For SM-like Higgs boson, we use
$M_{h^0} =125.1$ GeV,
$\Gamma_{h^0} =4.07\cdot 10^{-3}$ GeV.
The masses of all fermions are
applied as $m_e =0.00052$ GeV,
$m_{\mu}=0.10566$ GeV and
$m_{\tau} = 1.77686$ GeV.
In the quark sector, their masses
are taken as $m_u= 0.00216$ GeV,
$m_d= 0.0048$ GeV,
$m_c=1.27$ GeV, $m_s = 0.93$ GeV,
$m_t= 173.0$ GeV, and $m_b= 4.18$ GeV.
In the $G_{\mu}$-scheme,
the Fermi constant is considered
as an input parameter, $G_{\mu}
=1.16638\cdot 10^{-5}$ GeV$^{-2}$
is taken.
The electroweak constant is then
obtained subsequently:
\begin{eqnarray}
\alpha = \sqrt{2}/\pi G_{\mu}
M_W^2(1-M_W^2/M_Z^2)
= 1/132.184.
\end{eqnarray}
Lastly, in order to avoid 
the numerical instabilities
when $q_{23} \rightarrow 0$, 
we apply
the further cut as follows
$q_{23}^{\textrm{min}} = 4\left(
E_{\gamma}^{\textrm{cut}} \right)^2
= 100 \; \textrm{GeV}^2$ 
(corresponding to 
$E_{\gamma}^{\textrm{cut}}=5$ GeV
in this case).
For all numerical results presented in
the following
subsection, we take
$s_{\beta -\alpha}=0.98$ for
the alignment THDM limit.
\subsection{Numerical checks:
the $UV$, $IR$-finiteness,
the ward identity}
We first present the numerical
checks for all one-loop form factors.
The ultraviolet(UV) and infrared 
(IR) finiteness checks for the 
one-loop form factors are 
performed in this subsection.
In order to
regularize the infrared divergences, 
the parameter $\lambda$ is introduced as 
a fictitious mass for virtual photon
(the photon exchanging in 
the loop)~\cite{Hahn:1998yk,
Denner:2016kdg,Denner:2005nn}.
We check that there are not 
virtual photon internal lines 
in all one-loop Feynman diagrams 
contributing in the processes
under consideration. As a result,
one-loop amplitude is independent 
of the parameter $\lambda$. 
Without loss the generality 
of the test, one can take 
$\lambda^2=1$ in this case. 
In the numerical tests 
presented in below, we vary the 
ultraviolet parameter 
$C_{UV} = 1/\varepsilon 
-\gamma_{E} + \log(4\pi)$
where $\gamma_{E}$ is Euler's constant
and the renormalization scale parameter 
$\mu^2$~\cite{Hahn:1998yk,
Denner:2016kdg}. 
In Tables~\ref{F23UVIRMUP},
~\ref{F23UVIRMUM}, 
we show the numerical checks for
the factors $F_{23}$ and 
$F_{11}$.
In these Tables of data, we
select the input parameters
as follows:
$M_H = 500 \, \text{GeV},
M_{H^\pm} = 200 \, \text{GeV},
q_{23} = 100^2 \, \text{GeV}^2,
q_{13} = -50^2 \, \text{GeV}^2$
and $s_{\beta - \alpha} = 0.98,
t_\beta = 5$ for an example.
In Table~\ref{F23UVIRMUP}
(in Table~\ref{F23UVIRMUM}),
we show correspondingly to
the numerical results
for $F_{23}$ and $F_{11}$ 
for the case of
$M^2 = +150^2$ GeV$^2$
(for $M^2 = -150^2$ GeV$^2$),
respectively. From the data,
one finds the numerical results
for the tests are good stability
with varying $C_{UV}$ and $\mu^2$.
\begin{table}[H]
\centering
\begin{tabular}
{|l|l|l|}
\hline\hline
$\big(
C_{UV},
\mu^2
\big)$
&
$F_{23}$
&
$F_{11}$
\\
\hline \hline
$(0,1)$
&
$+ 3.983500715637517
\cdot 10^{-7}$
&
$-1.358365377521415
\cdot 10^{-8} $
\\
&
$- 2.06279418460594
\cdot 10^{-7} \, i$
&
$- 8.71538805431069
\cdot 10^{-9} \, i$
\\ \hline
$(10^2, 10^4)$
&
$+ 3.983500715637535
\cdot 10^{-7}$
&
$- 1.358365377521405
\cdot 10^{-8}$
\\
&
$- 2.06279418460594
\cdot 10^{-7} \, i$
&
$- 8.71538805431069
\cdot 10^{-9} \, i$
\\ \hline
$(10^4, 10^6)$
&
$+ 3.983500715637995
\cdot 10^{-7}$
&
$- 1.358365377521443
\cdot 10^{-8}$
\\
&
$- 2.06279418460594
\cdot 10^{-7} \, i$
&
$- 8.71538805431069
\cdot 10^{-9} \, i$
\\
\hline\hline
\end{tabular}
\caption{
\label{F23UVIRMUP}
Numerical checks for form 
factors $F_{23}$ and $F_{11}$ 
for the case of 
$M^2 = + 150^2$ GeV$^2$.}
\end{table}
\begin{table}[H]
\centering
\begin{tabular}
{|l|l|l|}
\hline\hline
$\big(
C_{UV},
\mu^2
\big)$
&
$F_{23}$
&
$F_{11}$
\\
\hline \hline
$(0,1)$
&
$+ 4.200542451680727
\cdot 10^{-7}$
&
$- 2.747533469117127
\cdot 10^{-8} $
\\
&
$- 4.593167056981461
\cdot 10^{-7} \, i$
&
$- 1.71470502697223
\cdot 10^{-8} \, i$
\\ \hline
$(10^2, 10^4)$
&
$+ 4.200542451680789
\cdot 10^{-7}$
&
$- 2.747533469117178
\cdot 10^{-8} $
\\
&
$- 4.593167056981461
\cdot 10^{-7} \, i$
&
$- 1.71470502697223
\cdot 10^{-8} \, i$
\\ \hline
$(10^4, 10^6)$
&
$+ 4.200542451680563
\cdot 10^{-7}$
&
$- 2.747533469117238
\cdot 10^{-8} $
\\
&
$- 4.593167056981461
\cdot 10^{-7} \, i$
&
$- 1.71470502697223
\cdot 10^{-8} \, i$
\\
\hline\hline
\end{tabular}
\caption{
\label{F23UVIRMUM}
Numerical checks for form 
factors $F_{23}$ and $F_{11}$ 
for the case of $M^2 = - 150^2$ GeV$^2$.}
\end{table}
Additionally, due to the on-shell
photons in final states, the one-loop 
amplitude for the decay channels 
also obeys the Ward-Takahashi identity. 
The identity is verified numerically 
in this work. This can be done as follows.
We are going to collect analytic
results for all the above-mentioned 
one-loop form factors independently
which they are expressed in terms 
of the scalar PV-functions.
We then confirm the relations 
from Eq.~\ref{ffrelation1}
to Eq.~\ref{ffrelation4} numerically.
In this subsection, we present
the numerical results
for the following relation 
given in Eq.~\ref{wardind}
as a typical example. 
The mentioned 
relation is given by
\begin{eqnarray}
\label{wardind}
F_{00} &=& F_{T}
=\dfrac{(q_{12} - M_h^2)
(q_{13} - M_h^2)}
{2 q_{23}} F_{11}
- \dfrac{q_{23}}{2} F_{23}.
\end{eqnarray}
To arrive this relation,
we have already replaced
$F_{13}$ by $F_{11}$ in
Eq.~\ref{ffrelation1}.
We verify numerically
the identity in Eq.~\ref{wardind}.
The results of this test are shown
in Tables~\ref{wardMUP},
\ref{wardMUM}. 
\begin{table}[H]
\centering
\begin{tabular}{|l|l|l|}
\hline\hline
&
$F_{11}$
&
$-$
\\
$\big(
q_{13}
,
q_{23}
,
M^2
\big)$
&
$F_{23}$
&
$-$
\\
&
$F_{T}$
&
$F_{00}$
\\
\hline \hline
&
$-1.524721445905201
\cdot 10^{-7}$
&
$-$
\\
&
$- 7.110585958979337
\cdot 10^{-8} \, i$
&
\\
&
&
\\
$\big(
+150^2
,
+250^2
,
+150^2
\big)$
&
$+3.651043549471353
\cdot 10^{-7}$
&
$-$
\\
&
$-3.437209227618915
\cdot 10^{-7} \, i$
&
\\
&
&
\\
&
$-0.01278816221085315$
&
$-0.01278816221085316$
\\
&
$+0.01009834059249554 \, i$
&
$+0.01009834059249555 \, i$
\\  \hline
&
$-8.89913100855076
\cdot 10^{-8}$
&
$-$
\\
&
$-6.375931724486408
\cdot 10^{-8} \, i$
&
\\
&
&
\\
$\big(
-150^2
,
+250^2
,
+150^2
\big)$
&
$+2.775629657896212
\cdot 10^{-7}$
&
$-$
\\
&
$-2.777884080549074
\cdot 10^{-7} \, i$
& \\
&
&
\\
&
$-0.002970210139871302$
&
$-0.002970210139871301$
\\
&
$+0.01276735098373022 \, i$
&
$+0.01276735098373023 \, i$
\\
\hline
&
$+8.14172209386663
\cdot 10^{-8}$
&
$-$
\\
&
$+3.110963119073888
\cdot 10^{-8} \, i$
&
\\
&
&
\\
$\big(
+150^2
,
-250^2
,
+150^2
\big)$
&
$+1.766382780308712
\cdot 10^{-7}$
&
$-$
\\
&
$-1.386843033642122
\cdot 10^{-7} \, i$
&
\\
&
&
\\
&
$+0.004226065602186965$
&
$+0.004226065602186968$
\\
&
$-0.004828278017271411
\, i$
&
$-0.004828278017271414
\,i$
\\
\hline
&
$+4.98006304470054
\cdot 10^{-8}$
&
$-$
\\
&
$+2.575736201977261
\cdot 10^{-8} \, i$
&
\\
&
&
\\
$\big(
-150^2
,
-250^2
,
+150^2
\big)$
&
$+1.453614520978374
\cdot 10^{-7}$
&
$-$
\\
&
$-1.309372710452731
\cdot 10^{-7} \, i$
&
\\
&
&
\\
&
$+0.00963426277087704$
&
$+0.00963426277087701$
\\
&
$-0.001458304822241892 \, i$
&
$-0.001458304822241893 \, i$
\\
\hline\hline
\end{tabular}
\caption{
\label{wardMUP}
The ward identity check that $F_{T} =F_{00}$
for the case of $M^2= +150^2$ GeV$^2$.}
\end{table}

\begin{table}[H]
\centering
\begin{tabular}{|l|l|l|}
\hline\hline
&
$F_{11}$
&
$-$
\\
$\big(
q_{13}
,
q_{23}
,
M^2
\big)$
&
$F_{23}$
&
$-$
\\
&
$F_{T}$
&
$F_{00}$
\\
\hline \hline
&
$-3.164637261836014
\cdot 10^{-7}$
&
\\
&
$-1.446908683148855
\cdot 10^{-7} \, i$
&
$-$
\\
&
&
\\
$\big(
+150^2
,
+250^2
,
-150^2
\big)$
&
$+9.06539595717883
\cdot 10^{-7}$
&
\\
&
$-7.511627252884878
\cdot 10^{-7} \, i$
&
$-$
\\
&
&
\\
&
$-0.03119082320101478$
&
$-0.03119082320101476$
\\
&
$+0.02216554224388151
\, i$
&
$+0.02216554224388154
\, i$
\\
\hline
&
$-1.932366818004465
\cdot 10^{-7}$
&
\\
&
$-1.323298998771165
\cdot 10^{-7} \, i$
&
$-$
\\
&
&
\\
$\big(
-150^2
,
+250^2
,
-150^2
\big)$
&
$+6.979798741398192
\cdot 10^{-7}$
&
\\
&
$-5.980635610637261
\cdot 10^{-7} \, i$
&
$-$
\\
&
&
\\
&
$-0.00942694241053888$
&
$-0.00942694241053889$
\\
&
$+0.0271707764493079 \, i$
&
$+0.0271707764493075 \, i$
\\
\hline
&
$+1.502250124801403
\cdot 10^{-7}$
&
\\
&
$+7.086353703138775
\cdot 10^{-8} \, i$
&
$-$
\\
&
&
\\
$\big(
+150^2
,
-250^2
,
-150^2
\big)$
&
$+3.880760921445276
\cdot 10^{-7}$
&
\\
&
$-3.197203563280603
\cdot 10^{-7} \, i$
&
$-$
\\
&
&
\\
&
$+0.00974000546640239$
&
$+0.00974000546640232$
\\
&
$-0.01111742282172064
\, i$
&
$-0.01111742282172063
\, i$
\\
\hline
&
$+1.011443268310633
\cdot 10^{-7}$
&
\\
&
$+5.619181764807639
\cdot 10^{-8} \, i$
&
$-$
\\
&
&
\\
$\big(
-150^2
,
-250^2
,
-150^2
\big)$
&
$+3.072281617154063
\cdot 10^{-7}$
&
\\
&
$-2.933174559185206
\cdot 10^{-7} \, i$
&
$-$
\\
&
&
\\
&
$+0.01994208102813598$
&
$+0.01994208102813591$
\\
&
$-0.003421005171538428 \, i$
&
$-0.003421005171538422 \, i$
\\
\hline\hline
\end{tabular}
\caption{
\label{wardMUM}
The ward identity check 
that
$F_{T} =F_{00}$ for
$M^2= -150^2$ GeV$^2$.
}
\end{table}
After the numerical checks, 
it is important to emphasize that
we only present the 
analytical expressions
for $F_{11},~F_{23}$ 
which they are used
for generating 
the decay rates in this 
paper. All physical
results shown in the 
next subsections 
are then generated
via the factors
$F_{11},~F_{23}$.
\subsection{Phenomenological
analyses}
\begin{figure}[H]
\centering
\begin{tabular}{cc}
\hspace{-3.2cm}
$\frac{1}{\Gamma_{
H\rightarrow \gamma\gamma
}
}\frac{\Gamma_{H\rightarrow
h \gamma\gamma}}{
d m_{\gamma\gamma}}$
[GeV$^{-1}$]
&
\hspace{-3.2cm}
$\frac{1}{\Gamma_{
H\rightarrow \gamma\gamma
}
}\frac{\Gamma_{H\rightarrow
h \gamma\gamma}}{
d m_{\gamma\gamma}}$
[GeV$^{-1}$]
\\
\includegraphics[width=8cm, height=8cm]
{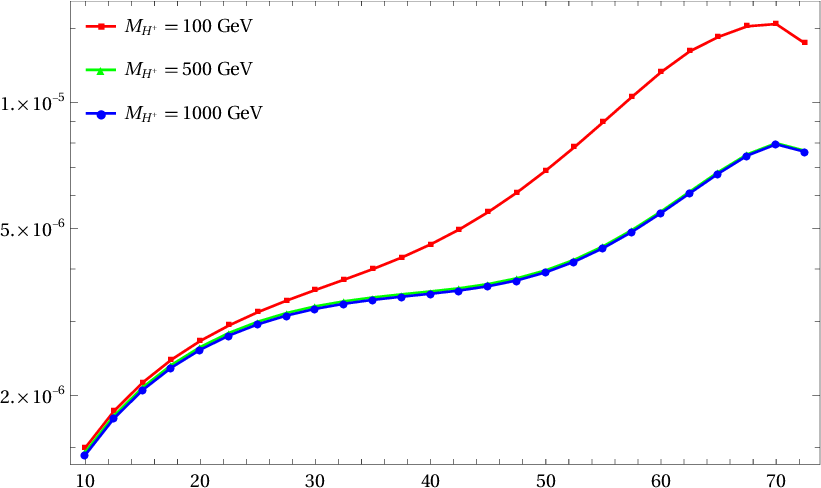}&
\includegraphics[width=8cm, height=8cm]
{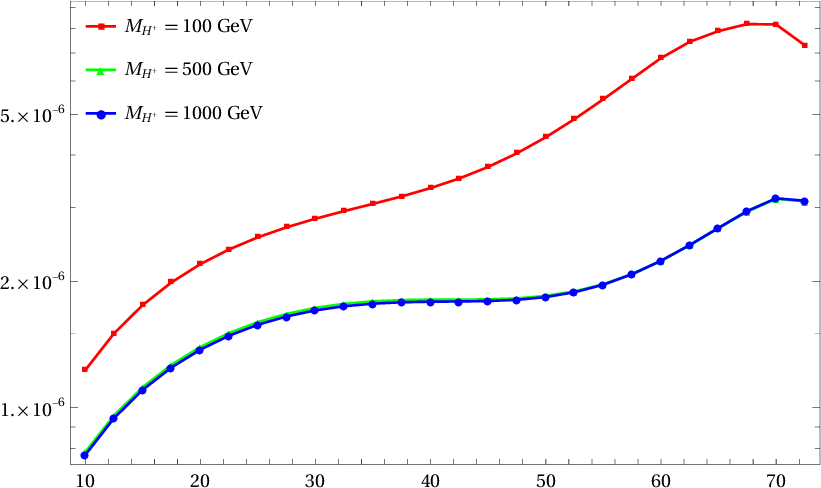}
\\
\hspace{6cm}
$m_{\gamma\gamma}$ [GeV]
&
\hspace{6cm}
$m_{\gamma\gamma}$
[GeV]
\end{tabular}
\caption{\label{dmyy200}
The values of
$\mathcal{R}_{\gamma\gamma}$
for the case of $M_H=200$ GeV
at $t_{\beta} =2$ (left) and 
at  $t_{\beta} =8$ (right).
In the plots, the rectangle
points with red line
are shown for
$\mathcal{R}_{\gamma\gamma}$
at $M_{H^{\pm}}=100$ GeV.
The triangle points with green
line are presented for
$\mathcal{R}_{\gamma\gamma}$
at $M_{H^{\pm}}= 500$ GeV.
While the circle points
with blue line are for
$\mathcal{R}_{\gamma\gamma}$
at $M_{H^{\pm}}=1000$ GeV.
}
\end{figure}

In the phenomenological studies, 
we are interested in
the differential decay rates of
$H \rightarrow h \gamma \gamma$ 
with respect to
the invariant mass of two photons
$m_{\gamma\gamma}$ and they are
is normalized by the decay width 
of $H \rightarrow \gamma \gamma$. 
The quantity is presented as 
functions of $M_{H^{\pm}}$
and $t_{\beta}$. In detail, we
are going to concern the 
$\mathcal{R}$-factor as follows:
\begin{eqnarray}
\mathcal{R}_{\gamma\gamma}
(M_H, M_{H^{\pm}}, t_{\beta})
=
\frac{1}{\Gamma_{
H\rightarrow \gamma\gamma
}
}\frac{\Gamma_{H\rightarrow
h \gamma\gamma}}{
d m_{\gamma\gamma}}
\; [\textrm{GeV}^{-1}].
\end{eqnarray}
In the above formula, 
$\Gamma_{
H\rightarrow \gamma\gamma
}$ is computed by using
the one-loop form factors given 
in Eq.~\ref{onshellPHIgg}.
In group $G_1$,
we confirm that the contributions
of top quark loop are dominant
in comparison with other fermions.
As a result, we take only top loop
into account in this work.
The Yukawa couplings between 
CP-even Higgs to top quark pair 
are the same in four types 
of the THDM.
In Fig.~\ref{dmyy200},
we investigate the factor 
$\mathcal{R}_{\gamma\gamma}$
in the case of $M_H=200$ GeV
at $t_{\beta} =2$ (left) and 
at $t_{\beta} =8$ (right). 
In the plots, the rectangle 
points connecting with
red line are shown for 
the values
of $\mathcal{R}_{\gamma\gamma}$
at $M_{H^{\pm}}=100$ GeV.
The triangle points with green
line are presented for the factor
$\mathcal{R}_{\gamma\gamma}$ 
at $M_{H^{\pm}}= 500$ GeV.
While the circle points
with blue line are for
$\mathcal{R}_{\gamma\gamma}$
at $M_{H^{\pm}}=1000$ GeV.
We find that 
the decay rates of 
$\Gamma_{H\rightarrow h \gamma\gamma}$
develop with $m_{\gamma\gamma}$ in 
this case. In general, decay rates
are proportional to $t_{\beta}^{-1}$ 
and the charged Higgs mass $M_{H^{\pm}}$.
In the high-mass regions 
of $m_{\gamma\gamma}$, the decay 
rates are more dependent on 
the charged Higgs masses $M_{H^{\pm}}$.
It is also interested to observe
that the decay rates are 
being unchanged in the 
large mass regions of the 
charged Higgs.
It shows that the contributions
of charged Higgs to the
processes become smaller in 
the concerned regions
in comparison with 
other attributions.

\begin{figure}[H]
\centering
\begin{tabular}{cc}
\hspace{-3.4cm}
$\frac{1}{\Gamma_{
H\rightarrow \gamma\gamma
}
}\frac{\Gamma_{H\rightarrow
h \gamma\gamma}}{
d m_{\gamma\gamma}}$
[GeV$^{-1}$]
&
\hspace{-3.4cm}
$\frac{1}{\Gamma_{
H\rightarrow \gamma\gamma
}
}\frac{\Gamma_{H\rightarrow
h \gamma\gamma}}{
d m_{\gamma\gamma}}$
[GeV$^{-1}$]
\\
\includegraphics[width=8cm, height=8cm]
 {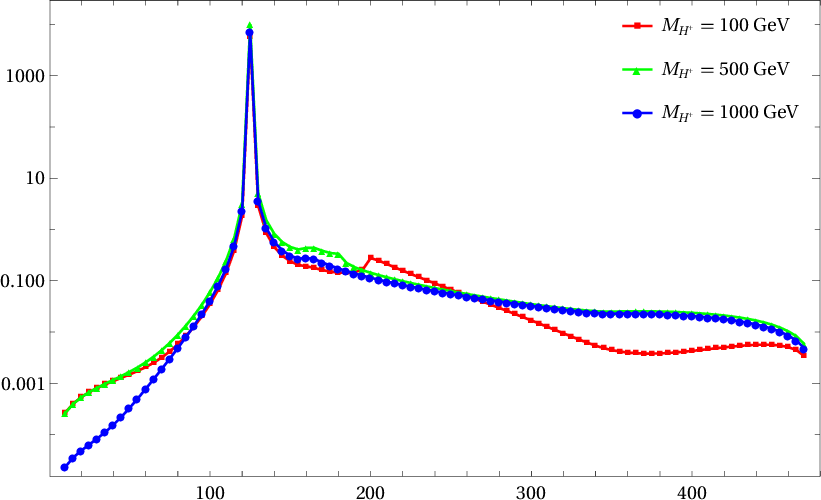}&
 \includegraphics[width=8cm, height=8cm]
 {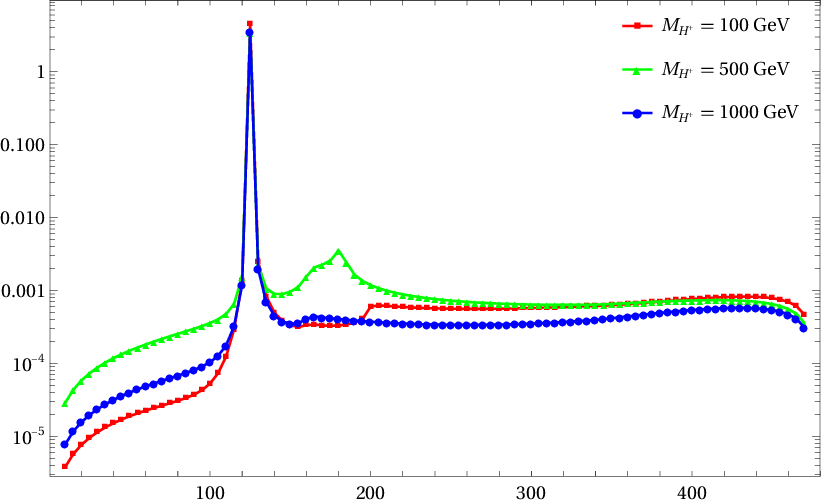}
 \\
 \hspace{6cm}
 $m_{\gamma\gamma}$ [GeV]
 &
 \hspace{6cm}
$m_{\gamma\gamma}$ [GeV]
\end{tabular}
\caption{\label{dmyy600}
The values of
$\mathcal{R}_{\gamma\gamma}$
for the case of $M_H = 600$ GeV
at $t_{\beta} =2$ (left)
and at
$t_{\beta} =8$ (right).
In the plots, the rectangle
points with red line
are shown for
$\mathcal{R}_{\gamma\gamma}$
at $M_{H^{\pm}}=100$ GeV.
The triangle points with green
line are presented for
$\mathcal{R}_{\gamma\gamma}$
at $M_{H^{\pm}}= 500$ GeV.
While the circle points
with blue line are for
$\mathcal{R}_{\gamma\gamma}$
at $M_{H^{\pm}}=1000$ GeV.
}
\end{figure}

\begin{figure}[H]
\centering
\begin{tabular}{cc}
\hspace{-3.7cm}
$\frac{1}{\Gamma_{
H\rightarrow \gamma\gamma
}
}\frac{\Gamma_{H\rightarrow
h \gamma\gamma}}{
d m_{\gamma\gamma}}$
[GeV$^{-1}$]
&
\hspace{-3.7cm}
$\frac{1}{\Gamma_{
H\rightarrow \gamma\gamma
}
}\frac{\Gamma_{H\rightarrow
h \gamma\gamma}}{
d m_{\gamma\gamma}}$
[GeV$^{-1}$]
\\
\includegraphics[width=8cm, height=8cm]
 {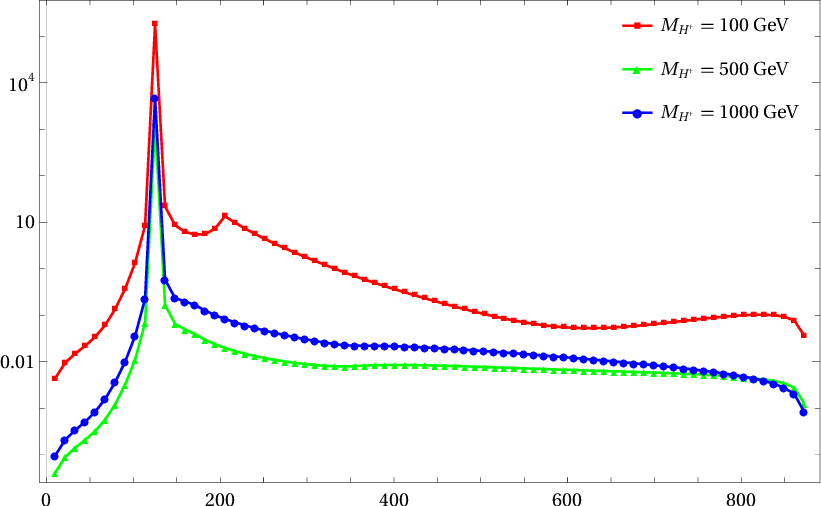}&
 \includegraphics[width=8cm, height=8cm]
 {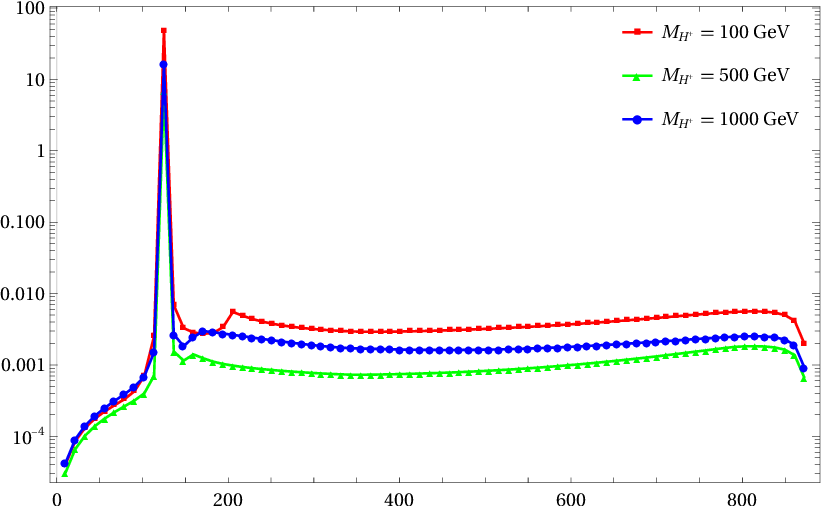}
 \\
 \hspace{6cm}
 $m_{\gamma\gamma}$ [GeV]
 &
 \hspace{6cm}
$m_{\gamma\gamma}$ [GeV]
\end{tabular}
\caption{\label{dmyy1000}
The values of
$\mathcal{R}_{\gamma\gamma}$
for the case of $M_H=1000$ GeV
at $t_{\beta} =2$ (left) and at
$t_{\beta} =8$ (right).
In the plots, the rectangle
points with red line
are shown for
$\mathcal{R}_{\gamma\gamma}$
at $M_{H^{\pm}}=100$ GeV.
The triangle points with green
line are presented for
$\mathcal{R}_{\gamma\gamma}$
at $M_{H^{\pm}}= 500$ GeV.
While the circle points
with blue line are for
$\mathcal{R}_{\gamma\gamma}$
at $M_{H^{\pm}}=1000$ GeV.
}
\end{figure}

In Fig.~\ref{dmyy600}, 
we show for the values of 
$\mathcal{R}_{\gamma\gamma}$
at $M_{H}=600$ GeV with
$t_{\beta} =2$ (left)
and at $t_{\beta} =8$ (right). 
We use the same previous notations
for the linespoints plotting styles
which are indicated for the values of
$\mathcal{R}_{\gamma\gamma}$
corresponding to the
charged Higgs mass $M_{H^{\pm}}$
in these Figures. One finds a peak
of $m_{\gamma\gamma}=M_h=125$ GeV
in both cases of $t_{\beta}$. 
In general, we find that 
the decay rates develop to the peak
and decrease rapidly beyond the peak.
In overall, the decay rates depend
on $t_{\beta}^{-1}$.
Different from the previous case,
in the higher-mass regions of $M_H$, 
the dependence of decay rates 
on $M_{H^{\pm}}$
is more complicated.
We next consider
the values of $\mathcal{R}_{\gamma\gamma}$
at $M_H=1000$ GeV as functions of 
the charged Higgs masses
and $t_{\beta}$, as shown in
Fig.~\ref{dmyy1000}. The distributions
are generated at $t_{\beta} =2$ for the 
left panel plot 
and at $t_{\beta} =8$ for the 
right panel plot.
The same notations for the linespoints 
of the data are labeled as previous plots.
In all cases, one finds
a peak of $m_{\gamma\gamma}=M_h=125$ GeV.
The decay rates develop to the peak
and decrease rapidly beyond the peak.
In general, the decay rates depend
on $t_{\beta}^{-1}$. As same as the case 
of $M_H=500$ GeV, we find that 
the dependence of decay rates 
on $M_{H^{\pm}}$
is more complicated in the
higher-mass regions of $M_H$.
Since the present article 
focuses mainly on the analytic
structure as well as the test 
for the computations. Detailed 
phenomenological analyses for
the processes under investigation 
which we study correlations
between decay rates of $H\rightarrow
h\gamma \gamma$
and $H\rightarrow \gamma \gamma$,
$h\rightarrow \gamma \gamma$, 
will be discussed in more concrete.
Furthermore, the implications for
the processes with combining
the updated experimental data at
the colliders should be examined in
further detail. These topics are far
from the scope of this paper. We may
be to address the mentioned topics in
our future publications.
\section{Conclusions} 
We have presented
the first analytical results
for one-loop induced contributions
to the important decay channels
of CP-even Higgs $H\rightarrow
\gamma \gamma$ and $H\rightarrow
h \gamma \gamma$ with $h$ being
standard model-like Higgs boson
within the framework of Two
Higgs Doublet Model. Analytical
expressions for one-loop form
factors are written in terms
of the basis PV-functions
following the standard notations
of the packages~{\tt LoopTools}
and {\tt Collier}.
Subsequently, physical results
can be evaluated numerically
by using one of the mentioned
packages. Analytical formulas
shown in this paper have verified
by the numerical checks such as
the UV and IR-finiteness of
one-loop amplitude. Furthermore,
due to on-shell photons in final
states, the corresponding one-loop
amplitude obeys the Ward-Takahashi 
identity. The identity has also
tested numerically. In phenomenological
studies, the differential decay rates
as functions of the invariant mass 
of two photons in final states
are examined in parameter space
of the Two Higgs Doublet Models.
\\

\noindent
{\bf Acknowledgment:}~
This research is funded by Vietnam
National Foundation for Science and
Technology Development (NAFOSTED) under
the grant number $103.01$-$2023.16$.
Khiem Hong Phan and Dzung Tri Tran
express their gratitude to all the
valuable support from Duy Tan
University, for the 30th anniversary
of establishment (Nov. 11, 1994 -
Nov. 11, 2024) towards "Integral,
Sustainable and Stable Development".
\section*{Appendix A: Tensor reduction 
for one-loop integrals }               
In this work, we follow
tensor reduction method developed in
Ref.~\cite{Denner:2005nn}.
The method is reviewed briefly in
this appendix. In general,
tensor one-loop integrals with
$N$-external lines
can be decomposed
into the basis integrals such as
scalar one-loop one-, two-, three-
and four- point integrals
(they are noted hereafter as
$A_0,~B_0,~C_0$ and $D_0$),
respectively. The definition for
tensor one-loop integrals
with rank $P$ are as follows:
\begin{eqnarray}
\{A; B; C; D\}^{\mu_1\mu_2\cdots \mu_P}
= (\mu^2)^{2-d/2}
\int \frac{d^dk}{(2\pi)^d}
\dfrac{k^{\mu_1}k^{\mu_2}
\cdots k^{\mu_P}}{\{D_1; D_1 D_2;D_1D_2D_3;
D_1D_2D_3D_4\}}.
\end{eqnarray}
Where the numerator is
expressed in terms of
the Lorentz structure of
the corresponding
tensor one-loop integrals
with rank $P$. While $D_j$
($j=1,\cdots, 4$) are for
the inverse Feynman propagators
in the denominators.
We show explicitly
the Feynman propagators
as follows:
\begin{eqnarray}
D_j^{-1} = \dfrac{1}
{[(k+ q_j)^2 -m_j^2 +i\rho]}.
\end{eqnarray}
In this definition,
$q_j = \sum\limits_{i=1}^j p_i$,
$p_i$ are the external momenta
and $m_j$ are internal masses
in the loops. Dimensional
regularization for one-loop integrals
are performed within the space-time
dimension $d=4-2\varepsilon$ for
$\varepsilon \rightarrow 0$.
The parameter $\mu^2$ in the
integrals play role of a
renormalization scale.
Several reduction
formulas for one-loop
one-, two-, three- and
four-point tensor
integrals up to rank
$P=3$~\cite{Denner:2005nn}
are shown explicitly,
for examples,
\begin{eqnarray}
A^{\mu}        &=& 0, \\
A^{\mu\nu}     &=& g^{\mu\nu} A_{00}, \\
A^{\mu\nu\rho} &=& 0,\\
B^{\mu}        &=& q^{\mu} B_1,\\
B^{\mu\nu}     &=& g^{\mu\nu} B_{00}
+ q^{\mu}q^{\nu} B_{11}, \\
B^{\mu\nu\rho} &=& \{g, q\}^{\mu\nu\rho} B_{001}
+ q^{\mu}q^{\nu}q^{\rho} B_{111}, \\
C^{\mu}        &=& q_1^{\mu} C_1 + q_2^{\mu} C_2
 = \sum\limits_{i=1}^2q_i^{\mu} C_i,
\\
C^{\mu\nu}    &=& g^{\mu\nu} C_{00}
 + \sum\limits_{i,j=1}^2q_i^{\mu}q_j^{\nu} C_{ij},
\\
C^{\mu\nu\rho} &=&
\sum_{i=1}^2 \{g,q_i\}^{\mu\nu\rho} C_{00i}+
\sum_{i,j,k=1}^2 q^{\mu}_i q^{\nu}_j q^{\rho}_k C_{ijk},\\
D^{\mu}        &=& q_1^{\mu} D_1 + q_2^{\mu} D_2 + q_3^{\mu}D_3
 = \sum\limits_{i=1}^3q_i^{\mu} D_i, \\
 D^{\mu\nu}    &=& g^{\mu\nu} D_{00}
 + \sum\limits_{i,j=1}^3q_i^{\mu}q_j^{\nu} D_{ij},
\\
D^{\mu\nu\rho} &=&
	\sum_{i=1}^3 \{g,q_i\}^{\mu\nu\rho} D_{00i}+
	\sum_{i,j,k=1}^3 q^{\mu}_i q^{\nu}_j q^{\rho}_k D_{ijk}.
\end{eqnarray}
In these expression, we have used the
notation
$\{g, q_i\}^{\mu\nu\rho}$ as in~\cite{Denner:2005nn}.
The tensor is defined
as follows:
$\{g, q_i\}^{\mu\nu\rho} = g^{\mu\nu} q^{\rho}_i
+ g^{\nu\rho} q^{\mu}_i + g^{\mu\rho} q^{\nu}_i$.
All scalar coefficients
$A_{00},~B_1, \cdots,~D_{333}$ in the right
hand sides of the above equations
are so-called Passarino-Veltman functions
(called as PV-function)~\cite{Denner:2005nn}.
The basic PV-functions are
well-known and have been implemented
into~{\tt LoopTools}~\cite{Hahn:1998yk}
as well as the library
{\tt Collier}~\cite{Denner:2016kdg}
for numerical evaluations.

In this calculation, analytic results
for one-loop form factors
are presented in terms of the basic
PV-functions with following
the short notations as:
\begin{eqnarray}
 \{A; B; C; D\}_{ijk\cdots}^{[f_1,
 f_2, \cdots](\cdots)}[q_i\cdot q_j] =
 \{A; B; C; D\}_{ijk\cdots}^{(\cdots)}
 [q_i\cdot q_j; m_{f_1}^2, m_{f_2}^2, \cdots].
\end{eqnarray}
We also expand scalar one-loop
three-point functions in terms of
the basic $f$-function as follows:
\begin{eqnarray}
\label{expandC0}
C_0(0,0,M^2,m^2,m^2,m^2)
&=&
- \dfrac{2}{M^2}
f\left(
\frac{4m^2}{M^2}
\right).
\end{eqnarray}
Where $f$-function
is defined as in \cite{Djouadi:2005gi}:
\begin{eqnarray}
\label{f-PV}
f(\tau) &=
\left\{ \begin{array}{rcl}
\arcsin^{2}\sqrt{\tau} & \mbox{for}
& \tau\leq 1, \\
& \\
-\frac{1}{4}
\left[\log\dfrac{1+\sqrt{1-\tau^{-1}}}
{1-\sqrt{{1-\tau^{-1}}}}-i\pi\right]^{2}
& \mbox{for} & \tau > 1.
\end{array}\right.
\end{eqnarray}
\section*{Appendix B: One-loop induced
of $\phi \rightarrow \gamma \gamma$ in 
the THDM}
\label{AppendixB}
In this appendix, one-loop induced of
$\phi (k) \rightarrow \gamma_\mu (k_1)
\gamma_\nu (k_2)$ in the THDM are shown.
The CP-even Higgses $\phi$ can be the
SM-like Higgs $h$ and another CP-even
Higgs $H$. General one-loop amplitude
for the decay channels
$\phi (k) \rightarrow
\gamma_\mu (k_1)
\gamma_\nu (k_2)$
can be expressed
as follows:
\begin{eqnarray}
\mathcal{A}_{\phi \rightarrow
\gamma \gamma}
&=&
F_{\phi \rightarrow
\gamma \gamma }
\cdot
\Big[
k_2^{\mu} k_1^{\nu}
-
\frac{k_{12}}{2}
\cdot
 g^{\mu\nu}
\Big]
\cdot
\varepsilon^*_{\mu}(k_1)
\varepsilon^*_{\nu}(k_2).
\end{eqnarray}
Where $k^2=k_{12} =(k_1+k_2)^2
= 2 k_1\cdot k_2=M_{\phi}^2$.
One-loop form factors are
given by
\begin{eqnarray}
 F_{\phi \rightarrow
 \gamma \gamma }
 &=&
 \dfrac{e^2}{2\pi^2}
\sum \limits_f
N^C_f\; Q_f^2
\cdot
g_{\phi ff}
\times
F_{\phi \rightarrow
\gamma \gamma }^{f}
\nonumber\\
&&
+
\dfrac{e^2}{16\pi^2}
\cdot
\dfrac{g}{M_W}
\cdot
\kappa_{\phi \, WW}
\times
F_{\phi \rightarrow
\gamma \gamma }^{W}
\nonumber\\
&&
+\frac{e^2}{4\pi^2}
\cdot
\left(
\dfrac{
g_{\phi \, H^\pm H^\mp}
}
{M_\phi^2}
\right)
\times
F_{\phi \rightarrow
\gamma \gamma }^{H^\pm}.
\end{eqnarray}
Here we already used
$g_{\phi \, WW}=\kappa_{\phi \, WW}\cdot
g_{h \, WW}^{\textrm{SM}}$
for $\phi=h, H$ and
$g_{h \, WW}^{\textrm{SM}}
= \dfrac{2M_W^2}{v}$.
In this equation,
each form factor is
given explicitly
as follows:
\begin{eqnarray}
F_{\phi \rightarrow
 \gamma \gamma }^{f}
&=&
\left(
\frac{m_f^2}{M_\phi^2}
\right)
\Big[
2 -
\big( M_\phi^2 - 4 m_f^2 \big)
C_0(0,0,M_\phi^2,m_f^2,m_f^2,m_f^2)
\Big],
\\
F_{\phi \rightarrow
\gamma \gamma }^{W}
&=&
\left(
\frac{2}{M_\phi^2}
\right)
\Big[
M_\phi^2
+
6 M_W^2
-6 M_W^2 (M_\phi^2-2 M_W^2)
C_0(0,0,M_\phi^2,M_W^2,M_W^2,M_W^2)
\Big],
\\
F_{\phi \rightarrow
\gamma \gamma }^{H^\pm}
&=&
-1
-
2 M_{H^\pm}^2
C_0 (0,0,M_\phi^2,
M_{H^\pm}^2,M_{H^\pm}^2,
M_{H^\pm}^2).
\end{eqnarray}
Expanding scalar one-loop three-point
functions in all above equations
as in Eq.~(\ref{expandC0}),
we the arrive at
\begin{eqnarray}
\label{onshellPHIgg}
F_{\phi \rightarrow
\gamma \gamma }
&=&
\dfrac{\alpha}{4 \pi}
\Bigg\{
\sum \limits_f
N^C_f
\,
Q_f^2
\cdot
g_{\phi ff}
\cdot
(
4 \tau_f
)
\cdot
\Big[
1
+
\Big(
1
-
\tau_f
\Big)
f ( \tau_f )
\Big]
\nonumber
\\
&&\hspace{0.5cm}
+
\left(
\dfrac{g_{\phi \, WW}}
{M_W^2}
\right)
\cdot
\Big[
2
+
3 \tau_W
+
3 \tau_W
\big( 2 - \tau_W \big)
f ( \tau_W )
\Big]
\nonumber\\
&&\hspace{0.5cm}
-
\left(
\dfrac{4 \;
g_{\phi H^{\pm}H^{\mp}}
}
{M_\phi^2}
\right)
\cdot
\Big[
1
-
\tau_{H^{\pm}}
f ( \tau_{H^{\pm}} )
\Big]
\Bigg\}.
\end{eqnarray}
\section*{Appendix C: Form factors
for $H \rightarrow h\gamma \gamma$
in the THDM}
\label{formfactorS}                
In this appendix, we shown in detail
analytic expressions for
one-loop form factors
for $H \rightarrow \gamma \gamma$
in the THDM.
\subsubsection*{
\underline{Form factors
$F_{ab}^{\textrm{Trig}}$ for
$ab = \{ 11, 23\}$}}
We first present analytic expressions
for one-loop form factors
$F_{23,(f,W, H^{\pm}) }
^{\textrm{Trig}}$
in the following paragraphs.
The form
factor $F_{23,f}^{\textrm{Trig}}$,
contributed from fermions in the loop,
is taken the form of
\begin{eqnarray}
F_{23,f}^{\textrm{Trig}}
&=&
\frac{1}{2 \pi^2}
\sum \limits_f
N^C_f
(e Q_f)^2
\cdot
g_{\phi ff}
\cdot
\left(
\frac{m_f^2}{q_{23}}
\right)
\Big[
2
+
\big( 4 m_f^2 - q_{23} \big)
C_0(0,0,q_{23},m_f^2,m_f^2,m_f^2)
\Big]
\nonumber
\\
&&
\\
&=&
\frac{1}{4 \pi^2}
\sum \limits_f
N^C_f
(e Q_f)^2
\cdot
g_{\phi ff}
\cdot
\lambda_f
\cdot
\Big[
1
+
\big(
1
-
\lambda_f
\big)
f ( \lambda_f )
\Big].
\end{eqnarray}
The contributions from $W$ boson
loop can be casted into the form of
\begin{eqnarray}
F_{23,W}^{\textrm{Trig}}
&=&
\left(
\dfrac{e^3}{8 \pi^2 M_W s_W}
\right)
\cdot
\left(
\dfrac{
\kappa_{\phi WW}
}
{
q_{23}
}
\right)
\times
\\
&&
\times
\Bigg[
M_{\phi}^2
+
6 M_W^2
+
2 M_W^2
\big( M_{\phi}^2
+
6 M_W^2
-
4 q_{23}
\big)
C_0(0,0,q_{23},M_W^2,M_W^2,M_W^2)
\Bigg]
\nonumber
\\
&&
\nonumber
\\
&=&
\left(
\dfrac{e^2}{16 \pi^2 M_W^2 }
\right)
\cdot
g_{\phi WW}
\cdot
\Bigg[
2 \rho_\phi
+
3 \lambda_W
+
\lambda_W
\big(
8
-
2 \rho_\phi
-
3 \lambda_W
\big)
f ( \lambda_W )
\Bigg].
\end{eqnarray}
We finally consider singly charged
Higgs propagating in the loop.
The form factors are given by
\begin{eqnarray}
F_{23,H^\pm}^{\textrm{Trig}}
&=& -
\left(
\dfrac{e^2}{4\pi^2}
\right)
\cdot
\left(
\dfrac{
g_{\phi \, H^\pm H^\mp}
}
{
q_{23}}
\right)
\cdot
\Big[
1
+
2 M_{H^\pm}^2
C_0 (0,0,q_{23},M_{H^\pm}^2,
M_{H^\pm}^2,M_{H^\pm}^2)
\Big]\\
&=&
-
\left(
\dfrac{e^2}{4\pi^2}
\right)
\cdot
\left(
\dfrac{
g_{\phi \, H^\pm H^\mp}
}
{
q_{23}
}
\right)
\cdot
\Big[
1
-
\lambda_{H^\pm}
\cdot
f ( \lambda_{H^\pm} )
\Big].
\end{eqnarray}
In all above equations, we have used
the variables like $\rho_\phi
= M_{\phi}^2 / q_{23}$
and $\lambda_i = 4 m_i^2/q_{23}$
with $m_i$ being
$m_f, M_W, M_{H^\pm}$ in this case.
We have also
expanded scalar one-loop
three-point function
in terms of $f$-function
as in Eq.~\ref{expandC0},
we then arrive at
the final result as
in Eq.~\ref{offPHIgg}.
\subsubsection*{
\underline{Form factors
$F_{ab,f}^{\textrm{Box}}$
for $ab = \{ 11, 23\}$
}}
We first mention one-loop form factors
$F_{ab,f}^{\textrm{Box}}$, collecting
from one-loop four point diagrams with
all fermion exchanging in the loop.
They are reading
\begin{eqnarray}
F_{23,f}^{\textrm{Box}}
&=&
C_{0}^{f,0}
+
C_{2}^{f,0}
-
2 D_{00}^{f,0}
+ 4 m_f^2
\Big[
\frac{1}{2}
D_{0}^{f,0}
+
2 \big(
D_{3}^{f,0}
+
D_{33}^{f,0}
\big)
+
\big(
2
+
\tau_{h,f}^{-1}
-
\tau_{H,f}^{-1}
\big)
D_{23}^{f,0}
\\
&&
+
\big(
2
-
\tau_{h,f}^{-1}
-
\tau_{H,f}^{-1}
\big)
D_{13}^{f,0}
+
\big(
\tau_{h,f}^{-1}
-
\zeta_f^{-1}
\big)
\big(
D_{12}^{f,0}
+
D_{22}^{f,0}
\big)
+
\tau_{h,f}^{-1}
D_{2}^{f,0}
\Big]
\nonumber
\\
&&
+
\Big\{
\big(
C_{0}^{f,1}
+
C_{1}^{f,1}
+
3 C_{2}^{f,1}
+
2 C_{12}^{f,1}
+
2 C_{22}^{f,1}
\big)
+
4 m_f^2
\Big[
\frac{1}{2}
D_{0}^{f,1}
+
2 \big(
D_{3}^{f,1}
+
D_{33}^{f,1}
\big)
\nonumber
\\
&&
+
\big(
\tau_{h,f}^{-1}
+
\zeta_f^{-1}
\big)
\big(
D_{2}^{f,1}
+
D_{22}^{f,1}
\big)
+
\big(
2
+
\tau_{h,f}^{-1}
+
\zeta_f^{-1}
\big)
D_{23}^{f,1}
\nonumber
\\
&&
+
\tau_{h,f}^{-1}
\big(
D_{1}^{f,1}
+
2 D_{12}^{f,1}
+
2 D_{13}^{f,1}
\big)
\Big]
\Big\}
+
\Big\{
q_{12} \leftrightarrow q_{13}
\Big\}-\text{term},
\nonumber\\
&&
\nonumber
\\
F_{11,f}^{\textrm{Box}}
&=&
4 D_{00}^{f,0}
-
2 C_{0}^{f,0}
+
4 m_f^2
\Big[
2
D_{3}^{f,0}
+
\big( 2 +
2 \tau_{H,f}^{-1}
\big)
D_{33}^{f,0}
\\
&&
+
\big(
\tau_{H,f}^{-1}
-
\tau_{h,f}^{-1}
+
\zeta_f^{-1}
-
\eta_f^{-1}
\big)
\big(
D_{13}^{f,0}
+
D_{12}^{f,0}
\big)
\nonumber
\\
&&
+
\big(
2
+
\tau_{H,f}^{-1}
+
\tau_{h,f}^{-1}
-
\zeta_f^{-1}
-
\eta_f^{-1}
\big)
D_{2}^{f,0}
+
\big(
2
+
\tau_{H,f}^{-1}
+
\tau_{h,f}^{-1}
+
\zeta_f^{-1}
-
\eta_f^{-1}
\big)
D_{22}^{f,0}
\nonumber
\\
&&
+
\big(
4
+
3 \tau_{H,f}^{-1}
+
\tau_{h,f}^{-1}
+
\zeta_f^{-1}
-
\eta_f^{-1}
\big)
D_{23}^{f,0}
\Big]
\nonumber
\\
&&
+ \Big\{
4 D_{00}^{f,1}
+
4 m_f^2
\Big[
2 D_{0}^{f,1}
+
\big( 4 + 2 \tau_{h,f}^{-1} \big)
D_{1}^{f,1}
+
\big( 4 + 2 \tau_{H,f}^{-1} \big)
D_{3}^{f,1}
+
\big( 2 + 2 \tau_{h,f}^{-1} \big)
D_{11}^{f,1}
\nonumber
\\
&&\hspace{+2.75cm}
+
\big( 4 + 2 \tau_{h,f}^{-1}
+ 2 \tau_{H,f}^{-1} \big)
D_{13}^{f,1}
+
\big(
4
+
3 \tau_{h,f}^{-1}
+
\tau_{H,f}^{-1}
+
\zeta_f^{-1}
-
\eta_f^{-1}
\big)
D_{12}^{f,1}
\nonumber
\\
&&
+
\big(
4
+
\tau_{h,f}^{-1}
+
\tau_{H,f}^{-1}
+
\zeta_f^{-1}
-
\eta_f^{-1}
\big)
D_{2}^{f,1}
+
\big(
2
+
\tau_{h,f}^{-1}
+
\tau_{H,f}^{-1}
+
\zeta_f^{-1}
-
\eta_f^{-1}
\big)
D_{22}^{f,1}
\nonumber
\\
&&
+
\big(
4
+
\tau_{h,f}^{-1}
+
3 \tau_{H,f}^{-1}
+
\zeta_f^{-1}
-
\eta_f^{-1}
\big)
D_{23}^{f,1}
+
\big(2 +
2\tau_{H,f}^{-1} \big)
D_{33}^{f,1}
\Big]
\Big\}
\nonumber
\\
&&
+
\Big\{
q_{12}
\leftrightarrow q_{13}
\Big\}-\text{term}.
\nonumber
\end{eqnarray}
In the above expressions,
we have denoted the abbreviation
notations as
\begin{eqnarray}
C_{ij\ldots}^{f,0}
& \equiv &
C_{ij\ldots}
(M_h^2, 0, q_{13}, m_f^2, m_f^2, m_f^2),
\\
C_{ij\ldots}^{f,1}
& \equiv &
C_{ij\ldots}
(0, 0, q_{23}, m_f^2, m_f^2, m_f^2),
\\
D_{ij\ldots}^{f,0}
& \equiv &
D_{ij\ldots}
(0, M_h^2, 0, M_H^2, q_{12},
q_{13}, m_f^2, m_f^2, m_f^2, m_f^2),
\\
D_{ij\ldots}^{f,1}
& \equiv &
D_{ij\ldots}
(M_h^2, 0, 0, M_H^2,
q_{12}, q_{23}, m_f^2,
m_f^2, m_f^2, m_f^2),
\end{eqnarray}
and used the following kinematic
invariant variables like
\begin{eqnarray}
\tau_{\phi,i}
=
4 m_i^2 / M_\phi^2, \quad
\zeta_i
=
4 m_i^2 / q_{12},
\quad
\eta_i =
4 m_i^2 / q_{13}
\quad
\lambda_i =
4 m_i^2 / q_{23}.
\end{eqnarray}
\subsubsection*{
\underline{Form factors
$F_{ab,H^\pm}^{\textrm{Box}}$
for $ab = \{ 11, 23\}$}}
Charged Higgs ($H^{\pm}$)
exchanging in the loop
is next considered. One-loop form factor
are expressed in terms of PV-functions
as follows:
\begin{eqnarray}
F_{23,H^\pm}^{\textrm{Box}}
&=&
- 2 g_{H h H^\pm H^\mp}
\cdot
C_{12}^{H^\pm,0}
\\
&&
+
2
g_{h H^\pm H^\mp}
\cdot
g_{H H^\pm H^\mp}
\cdot
\Big[
\;
D_{3}^{H^\pm,0}
+
D_{13}^{H^\pm,0}
+
D_{23}^{H^\pm,0}
+
D_{33}^{H^\pm,0}
\nonumber
\\
&&
+
\Big(
D_{3}^{H^\pm,1}
+
D_{23}^{H^\pm,1}
+
D_{33}^{H^\pm,1}
\Big)
+
\Big(
q_{12}
\leftrightarrow q_{13}
\Big)-\text{term}
\Big],
\nonumber
\end{eqnarray}
and
\begin{eqnarray}
\dfrac{
F_{11,H^\pm}^{\textrm{Box}}
}
{
2g_{h H^\pm H^\mp}
\cdot
g_{H H^\pm H^\mp}
}
&=&
D_{2}^{H^\pm,0}
+
D_{22}^{H^\pm,0}
+
2 D_{23}^{H^\pm,0}
+
D_{3}^{H^\pm,0}
+
D_{33}^{H^\pm,0}
\\
&&
+
\Big[
2
\big(
D_{1}^{H^\pm,1}
+
D_{12}^{H^\pm,1}
+
D_{13}^{H^\pm,1}
+
D_{2}^{H^\pm,1}
+
D_{23}^{H^\pm,1}
+
D_{3}^{H^\pm,1}
\big)
\nonumber\\
&&
+
D_{0}^{H^\pm,1}
+
D_{11}^{H^\pm,1}
+
D_{22}^{H^\pm,1}
+
D_{33}^{H^\pm,1}
\Big]
+
\Big[
q_{12}
\leftrightarrow
q_{13}
\Big]-\text{term}.
\nonumber
\end{eqnarray}
Where we have used
some abbreviation
notations as
\begin{eqnarray}
C_{ij\ldots}^{H^\pm,0}
& \equiv &
C_{ij\ldots}
(0, q_{23}, 0, M_{H^\pm}^2,
M_{H^\pm}^2, M_{H^\pm}^2),
\\
D_{ij\ldots}^{H^\pm,0}
& \equiv &
D_{ij\ldots}
(0, M_h^2, 0, M_H^2, q_{12},
q_{13}, M_{H^\pm}^2,
M_{H^\pm}^2, M_{H^\pm}^2,
M_{H^\pm}^2),
\\
D_{ij\ldots}^{H^\pm,1}
& \equiv &
D_{ij\ldots}
(M_h^2, 0, 0, M_H^2, q_{12},
q_{23}, M_{H^\pm}^2,
M_{H^\pm}^2, M_{H^\pm}^2,
M_{H^\pm}^2).
\end{eqnarray}

\subsubsection*{
\underline{Form factors
$F_{ab,W}^{\textrm{Box}}$
for $ab = \{ 11, 23\}$ } }
One-loop form factor from
the box Feynman diagrams with
$W$ boson propagating in the loop
are collected in the following
subsection. In detail, expressions
are reading as follows:
\begin{eqnarray}
F_{23,W}^{\textrm{Box}}
&=&
-
\dfrac{8 \,
}{M_W^2 g^2\; s_{2(\beta-\alpha)}}
\cdot
g_{hHG^{\pm}G^{\mp}}
\cdot
C_{12}^{W,5}
+
\frac{1}
{4 M_W^2}
\Big[
- 4 \big( C_{2}^{W,4}
+
2 D_{00}^{W,0}
\big)
\\
&&
+
\big(
2 C_{0}^{W,1}
-
2 C_{1}^{W,1}
+
C_{2}^{W,1}
+
3 C_{12}^{W,1}
+
3 C_{22}^{W,1}
\big)
\nonumber
\\
&&
+
\big(
2 C_{0}^{W,2}
-
2 C_{1}^{W,2}
+
C_{2}^{W,2}
+
3 C_{12}^{W,2}
+
3 C_{22}^{W,2}
\big)
\nonumber
\\
&&
+
2 \big(
C_{2}^{W,3}
-
C_{22}^{W,3}
\big)
-
2 \big(
4 C_{1}^{W,0}
+
7 C_{2}^{W,0}
+
3 C_{12}^{W,0}
+
3 C_{22}^{W,0}
\big)
\Big]
\nonumber
\\
&&
+
4 \Big[
\big(
2
+
2 \tau_{h,W}^{-1}
-
\zeta_W^{-1}
\big)
D_{0}^{W,0}
+
\tau_{h,W}^{-1}  \big(
D_{1}^{W,0}
+
3 D_{2}^{W,0}
\big)
\nonumber
\\
&&
+
\big(
-
\tau_{h,W}^{-1}
+
\zeta_W^{-1}
\big)
D_{11}^{W,0}
+
2 \zeta_W^{-1}
D_{12}^{W,0}
+
\big(
\tau_{h,W}^{-1}
+
\zeta_W^{-1}
\big)
D_{22}^{W,0}
\Big]
\nonumber
\\
&&
+
4 \Big[
3
+
\tau_{H,W}^{-1}
\big(
1
+
4 \tau_{h,W}^{-1}
\big)
-
\zeta_W^{-1}
-
2 \eta_W^{-1}
\Big]
D_{13}^{W,0}
\nonumber
\\
&&
+
4 \Big[
2
+
\big(
1
+
\tau_{H,W}^{-1}
\big)
\big(
1
+
4 \tau_{h,W}^{-1}
\big)
-
2 \zeta_W^{-1}
-
3 \eta_W^{-1}
\Big]
D_{23}^{W,0}
\nonumber
\\
&&
+
4 \Big[
3
+
2 \tau_{h,W}^{-1}
\big(
1
+
2 \tau_{H,W}^{-1}
\big)
+
\tau_{H,W}^{-1}
-
3 \zeta_W^{-1}
-
\eta_W^{-1}
\Big]
D_{3}^{W,0}
\nonumber
\\
&&
+
4 \Big[
3
+
2 \tau_{h,W}^{-1}
\big(
1
+
2 \tau_{H,W}^{-1}
\big)
-
2 \zeta_W^{-1}
-
2 \eta_W^{-1}
\Big]
D_{33}^{W,0}
\nonumber
\\
&&
+
\Big\{
8 D_{0}^{W,1}
-
4 \big(
\tau_{h,W}^{-1}
-
\zeta_W^{-1}
\big)
\big(
D_{2}^{W,1}
-
D_{12}^{W,1}
-
D_{22}^{W,1}
\big)
\nonumber\\
&&
+
\big(
5 \tau_{h,W}^{-1}
-
6 \zeta_W^{-1}
-
5 \eta_W^{-1}
\big)
D_{13}^{W,1}
\nonumber
\\
&&
- \big(
5  \tau_{h,W}^{-1}
+
2 \zeta_W^{-1}
+
3 \eta_W^{-1}
\big)
\big(
D_{123}^{W,1}
+
D_{133}^{W,1}
\big)
-
\big(
3 \tau_{h,W}^{-1}
+
3 \tau_{H,W}^{-1}
+
4 \zeta_W^{-1}
\big)
D_{223}^{W,1}
\nonumber
\\
&&
- \big(
3 \tau_{h,W}^{-1}
+
8 \tau_{H,W}^{-1}
+
6 \zeta_W^{-1}
+
3 \eta_W^{-1}
\big)
D_{233}^{W,1}
-
\big(
5 \tau_{H,W}^{-1}
+
2 \zeta_W^{-1}
+
3 \eta_W^{-1}
\big)
D_{333}^{W,1}
\nonumber
\\
&&
+
\Big[
12
+
8 \tau_{h,W}^{-1}
\big(
1
+
2 \tau_{H,W}^{-1}
\big)
-
4 \tau_{H,W}^{-1}
-
19 \zeta_W^{-1}
-
12 \eta_W^{-1}
\Big]
D_{23}^{W,1}
\nonumber
\\
&&
+
\Big[
12
+
\tau_{h,W}^{-1}
\big(
16 \tau_{H,W}^{-1}
-
5
\big)
+
3 \tau_{H,W}^{-1}
-
9 \zeta_W^{-1}
-
8 \eta_W^{-1}
\Big]
D_{3}^{W,1}
\nonumber
\\
&&
+
\Big[
12
+
\tau_{h,W}^{-1}
\big(
16 \tau_{H,W}^{-1}
+
5
\big)
-
6 \tau_{H,W}^{-1}
-
13 \zeta_W^{-1}
-
13 \eta_W^{-1}
\Big]
D_{33}^{W,1}
\nonumber
\\
&&
+
(2 M_W^2)^{-1}
\big(
-
9 D_{00}^{W,1}
-
3 D_{002}^{W,1}
-
5 D_{003}^{W,1}
\big)
\Big\}
+
\Big\{
q_{12}
\leftrightarrow
q_{13}
\Big\}-\text{term}.
\nonumber
\end{eqnarray}
Where the coupling
$g_{hHG^{\pm}G^{\mp}}$
is given by Eq.\ref{hHGG}.
Another form factor is given
\begin{eqnarray}
&&\hspace{-1cm}
F_{11,
W^\pm}^{\textrm{Box}}
=
\frac{1}{
4 M_W^2
}
\Big\{
16 D_{00}^{W,0}
-
8 C_{0}^{W,0}
+
3 C_{22}^{W,1}
+
3 C_{22}^{W,2}
+
8 C_{0}^{W,4}
+
3 C_{11}^{W,6}
+
3 C_{11}^{W,7}
\\
&&
- 2
\Big[
C_{0}^{W,3}
+
C_{11}^{W,3}
+
C_{22}^{W,3}
-
2 \big(
C_{1}^{W,3}
-
C_{12}^{W,3}
+
C_{2}^{W,3}
\big)
\Big]
\Big\}
\nonumber
\\
&&
+ 4 \Big[
\big(
2 \tau_{h,W}^{-1}
+
3 \tau_{H,W}^{-1}
-
\zeta_W^{-1}
-
2 \eta_W^{-1}
\big)
D_{0}^{W,0}
\nonumber
\\
&&
+
\big(
- \tau_{h,W}^{-1}
+
\tau_{H,W}^{-1}
+
\zeta_W^{-1}
-
\eta_W^{-1}
\big)
\big(
D_{1}^{W,0}
+
D_{12}^{W,0}
+
D_{13}^{W,0}
\big)
\Big]
\nonumber
\\
&&
+ 4 \Big[
3
+
\tau_{h,W}^{-1}
\big(
1
+
4  \tau_{H,W}^{-1}
\big)
+
3 \tau_{H,W}^{-1}
+
\zeta_W^{-1}
-
3 \eta_W^{-1}
+
\lambda_W^{-1}
\Big]
D_{2}^{W,0}
\nonumber
\\
&&
+ 4 \Big[
3
+
\tau_{h,W}^{-1}
\big(
3
+
4 \tau_{H,W}^{-1}
\big)
+
3 \tau_{H,W}^{-1}
-
2 \zeta_W^{-1}
-
4 \eta_W^{-1}
\Big]
D_{22}^{W,0}
\nonumber
\\
&&
+ 4 \Big[
6
+
\tau_{h,W}^{-1}
\big(
5
+
8 \tau_{H,W}^{-1}
\big)
+
7 \tau_{H,W}^{-1}
-
5 \zeta_W^{-1}
-
7 \eta_W^{-1}
\Big]
D_{23}^{W,0}
\nonumber
\\
&&
+ 4 \Big[
3
+
2 \tau_{h,W}^{-1}
\big(
1
+
2 \tau_{H,W}^{-1}
\big)
+
6  \tau_{H,W}^{-1}
-
2 \zeta_W^{-1}
-
4 \eta_W^{-1}
\Big]
D_{3}^{W,0}
\nonumber
\\
&&
+ 4 \Big[
3
+
2 \tau_{h,W}^{-1}
\big(
1
+
2 \tau_{H,W}^{-1}
\big)
+
4 \tau_{H,W}^{-1}
-
3 \zeta_W^{-1}
-
3 \eta_W^{-1}
\Big]
D_{33}^{W,0}
\nonumber
\\
&&
-
\Big\{
\big(
5 \tau_{h,W}^{-1}
+
2 \zeta_W^{-1}
+
3 \eta_W^{-1}
\big)
D_{111}^{W,1}
+
\big(
5 \tau_{H,W}^{-1}
+
2 \zeta_W^{-1}
+
3 \eta_W^{-1}
\big)
D_{333}^{W,1}
\nonumber
\\
&&
+ \big(
13 \tau_{h,W}^{-1}
+
3 \tau_{H,W}^{-1}
+
8 \zeta_W^{-1}
+
6 \eta_W^{-1}
\big)
D_{112}^{W,1}
+
\big(
10 \tau_{h,W}^{-1}
+
5 \tau_{H,W}^{-1}
+
6 \zeta_W^{-1}
+
9 \eta_W^{-1}
\big)
D_{113}^{W,1}
\nonumber
\\
&&
+ \big(
11 \tau_{h,W}^{-1}
+
6 \tau_{H,W}^{-1}
+
10 \zeta_W^{-1}
+
3 \eta_W^{-1}
\big)
D_{122}^{W,1}
+
4 \big(
4 \tau_{h,W}^{-1}
+
4 \tau_{H,W}^{-1}
+
4 \zeta_W^{-1}
+
3 \eta_W^{-1}
\big)
D_{123}^{W,1}
\nonumber
\\
&&
+ \big(
5 \tau_{h,W}^{-1}
+
10 \tau_{H,W}^{-1}
+
6 \zeta_W^{-1}
+
9 \eta_W^{-1}
\big)
D_{133}^{W,1}
+
\big(
3 \tau_{h,W}^{-1}
+
3 \tau_{H,W}^{-1}
+
4 \zeta_W^{-1}
\big)
D_{222}^{W,1}
\nonumber
\\
&&
+ \big(
6 \tau_{h,W}^{-1}
+
11 \tau_{H,W}^{-1}
+
10 \zeta_W^{-1}
+
3 \eta_W^{-1}
\big)
D_{223}^{W,1}
+
\big(
3 \tau_{h,W}^{-1}
+
13 \tau_{H,W}^{-1}
+
8 \zeta_W^{-1}
+
6 \eta_W^{-1}
\big)
D_{233}^{W,1}
\nonumber
\\
&&
- \Big[
12
+
\tau_{h,W}^{-1}
\big(
-
3
+
16 \tau_{H,W}^{-1}
\big)
-
3  \tau_{H,W}^{-1}
-
7 \zeta_W^{-1}
-
8 \eta_W^{-1}
\Big]
D_{0}^{W,1}
\nonumber
\\
&&
- \Big[
24
+
\tau_{h,W}^{-1}
\big(
-
7
+
32 \tau_{H,W}^{-1}
\big)
-
2  \tau_{H,W}^{-1}
-
20  \zeta_W^{-1}
-
19  \eta_W^{-1}
\Big]
D_{1}^{W,1}
\nonumber
\\
&&
- \Big[
12
+
\tau_{h,W}^{-1}
\big(
-
1
+
16 \tau_{H,W}^{-1}
\big)
+
5 \tau_{H,W}^{-1}
-
19 \zeta_W^{-1}
-
18 \eta_W^{-1}
\Big]
D_{11}^{W,1}
\nonumber
\\
&&
- \Big[
24
+
2 \tau_{h,W}^{-1}
\big(
1
+
16 \tau_{H,W}^{-1}
\big)
+
8 \tau_{H,W}^{-1}
-
42 \zeta_W^{-1}
-
34 \eta_W^{-1}
\Big]
D_{12}^{W,1}
\nonumber
\\
&&
- \Big[
24
+
4 \tau_{h,W}^{-1}
\big(
1
+
8 \tau_{H,W}^{-1}
\big)
+
4 \tau_{H,W}^{-1}
-
38 \zeta_W^{-1}
-
36 \eta_W^{-1}
\Big]
D_{13}^{W,1}
\nonumber
\\
&&
- \Big[
24
+
\tau_{h,W}^{-1}
\big(
-
5
+
32 \tau_{H,W}^{-1}
\big)
-
5 \tau_{H,W}^{-1}
-
22 \zeta_W^{-1}
-
16 \eta_W^{-1}
\Big]
D_{2}^{W,1}
\nonumber
\\
&&
- \Big[
12
+
\tau_{h,W}^{-1}
\big(
3
+
16 \tau_{H,W}^{-1}
\big)
+
3  \tau_{H,W}^{-1}
-
23 \zeta_W^{-1}
-
16 \eta_W^{-1}
\Big]
D_{22}^{W,1}
\nonumber
\\
&&
- \Big[
24
+
8 \tau_{h,W}^{-1}
\big(
1
+
4 \tau_{H,W}^{-1}
\big)
+
2 \tau_{H,W}^{-1}
-
42 \zeta_W^{-1}
-
34 \eta_W^{-1}
\Big]
D_{23}^{W,1}
\nonumber
\\
&&
- \Big[
24
+
2 \tau_{h,W}^{-1}
\big(
-
1
+
16 \tau_{H,W}^{-1}
\big)
-
7 \tau_{H,W}^{-1}
-
20 \zeta_W^{-1}
-
19 \eta_W^{-1}
\Big]
D_{3}^{W,1}
\nonumber
\\
&&
- \Big[
12
+
\tau_{h,W}^{-1}
\big(
5
+
16 \tau_{H,W}^{-1}
\big)
-
\tau_{H,W}^{-1}
-
19 \zeta_W^{-1}
-
18 \eta_W^{-1}
\Big]
D_{33}^{W,1}
\nonumber
\\
&&
+ \frac{1}{M_W^2}
\big(
3 D_{00}^{W,1}
+
5 D_{001}^{W,1}
+
5 D_{002}^{W,1}
+
5 D_{003}^{W,1}
\big)
\Big\}
-
\Big\{
q_{12} \leftrightarrow q_{13}
\Big\}-\text{term}
.
\nonumber
\end{eqnarray}
The abbreviation notations
are expressed as follows
\begin{eqnarray}
C_{ij\ldots}^{W,0}
& \equiv &
C_{ij\ldots}
(0,0,q_{23}, M_W^2, M_W^2, M_W^2),
\\
C_{ij\ldots}^{W,1}
& \equiv &
C_{ij\ldots}
(0,q_{12},M_H^2, M_W^2, M_W^2, M_W^2),
\\
C_{ij\ldots}^{W,2}
& \equiv &
C_{ij\ldots}
(0,q_{13},M_H^2, M_W^2, M_W^2, M_W^2),
\\
C_{ij\ldots}^{W,3}
& \equiv &
C_{ij\ldots}
(M_h^2,q_{23},M_H^2, M_W^2, M_W^2, M_W^2),
\\
C_{ij\ldots}^{W,4}
& \equiv &
C_{ij\ldots}
(M_h^2,0,q_{13}, M_W^2, M_W^2, M_W^2),
\\
C_{ij\ldots}^{W,5}
& \equiv &
C_{ij\ldots}
(0,q_{23},0, M_W^2, M_W^2, M_W^2),
\\
C_{ij\ldots}^{W,6}
& \equiv &
C_{ij\ldots}
(M_h^2,q_{12},0, M_W^2, M_W^2, M_W^2),
\\
C_{ij\ldots}^{W,7}
& \equiv &
C_{ij\ldots}
(M_h^2,q_{13},0, M_W^2, M_W^2, M_W^2),
\\
D_{ij\ldots}^{W,0}
& \equiv &
D_{ij\ldots}
(0,M_h^2,0,M_H^2,q_{12},q_{13},
M_W^2, M_W^2, M_W^2, M_W^2),
\\
D_{ij\ldots}^{W,1}
& \equiv &
D_{ij\ldots}
(M_h^2,0,0,M_H^2,q_{12},q_{23},
M_W^2, M_W^2, M_W^2, M_W^2).
\end{eqnarray}
\subsubsection*{
\underline{Form factors
$F_{ab,W H^\pm}^{\textrm{Box}}$
for $ab = \{ 11, 23\}$}}
We finally arrive at one-loop box diagrams
with $W$ boson and charged Higgs $H^\pm$
propagating in the loop diagrams. The
form factors can be expressed as follows:
\begin{eqnarray}
\label{WHform}
F_{ab,W H^\pm}^{\textrm{Box}}
&=&
F_{ab,W H^\pm}^{\textrm{Box},C}
+
\Big[
F_{ab,W H^\pm}^{\textrm{Box},D}
(0,M_h^2,0,M_H^2,q_{12},q_{13})
+
F_{ab,W H^\pm}^{\textrm{Box},D}|_{
(M_h^2\leftrightarrow M_H^2)}
\Big]
\nonumber\\
&&
+
\Big[
F_{ab,W H^\pm}^{\textrm{Box},D}
(0,0,M_H^2,M_h^2,q_{23},q_{12})
+
F_{ab,W H^\pm}^{\textrm{Box},D}
(M_h^2,0,0,M_H^2,q_{12},q_{23})
\Big]
\nonumber
\\
&&
+
\Big[
q_{12} \leftrightarrow q_{13}
\Big]-\text{term}
\end{eqnarray}
for $ab = \{ 11, 23\}$.

Where each form factor in this
equation is written explicitly
as
\begin{eqnarray}
&&
\hspace{-1cm}
\dfrac{
F_{23,W H^\pm}^{\textrm{Box},C}
}
{
M_W^2
}
=
2 \big(
4 C_{1}^{W,0}
+ 3 C_{2}^{W,0}
- C_{12}^{W,0}
- C_{22}^{W,0}
\big)
+
8 \big(
C_{2}^{H^\pm,0}
+ C_{12}^{H^\pm,0}
+ C_{22}^{H^\pm,0}
\big)
\\
&&
+
2 \big(
2 C_{2}^{W H^\pm,0}
+
C_{22}^{W H^\pm,1}
-
C_{2}^{W H^\pm,1}
\big)
+
\Big[
\big(
C_{1}^{W H^\pm,2}
+4 C_{2}^{W H^\pm,2}
- C_{11}^{W H^\pm,2}
- C_{12}^{W H^\pm,2}
\big)
\nonumber
\\
&&
+
2 \big(
C_{2}^{W H^\pm,3}
- C_{12}^{W H^\pm,3}
- C_{22}^{W H^\pm,3}
\big)
\Big]
+
\Big[
q_{12} \leftrightarrow q_{13}
\Big]-\text{term}.
\nonumber
\end{eqnarray}
Second form factor in
Eq.~\ref{WHform}
is given
by
\begin{eqnarray}
&& \hspace{-1cm}
\dfrac{
F_{11,W
H^\pm}^{\textrm{Box}, C}
}
{
M_W^2
}
=
\big(
C_{0}^{W H^\pm,0}
-
C_{11}^{W H^\pm,0}
-
C_{22}^{W H^\pm,0}
\big)
+ 2 \big(
4 C_{1}^{W H^\pm,0}
+
4 C_{2}^{W H^\pm,0}
-
C_{12}^{W H^\pm,0}
\big)
\nonumber
\\
&&
+
2
\big(
C_{0}^{W H^\pm,1}
+
C_{11}^{W H^\pm,1}
+
C_{22}^{W H^\pm,1}
\big)
- 4 \big(
C_{1}^{W H^\pm,1}
+ C_{2}^{W H^\pm,1}
- C_{12}^{W H^\pm,1}
\big)
\nonumber
\\
&&
+
\big(
C_{0}^{W H^\pm,2}
-
C_{11}^{W H^\pm,2}
-
C_{22}^{W H^\pm,2}
\big)
-
2 C_{12}^{W H^\pm,2}
+
2 \big(
C_{2}^{W H^\pm,3}
-
C_{22}^{W H^\pm,3}
\big)
\nonumber
\\
&&
+
\big(
C_{0}^{W H^\pm,4}
-
C_{11}^{W H^\pm,4}
-
C_{22}^{W H^\pm,4}
\big)
-
2 C_{12}^{W H^\pm,4}
+
2 \big(
C_{2}^{W H^\pm,6}
-
C_{22}^{W H^\pm,6}
\big)
\nonumber
\\
&&
+
\big(
C_{0}^{W H^\pm,5}
-
C_{11}^{W H^\pm,5}
-
C_{22}^{W H^\pm,5}
\big)
-
2 C_{12}^{W H^\pm,5}
+
2 \big(
C_{2}^{W H^\pm,8}
-
C_{22}^{W H^\pm,8}
\big)
\nonumber
\\
&&
+
2 \big(
5 C_{2}^{W H^\pm,7}
-
C_{22}^{W H^\pm,7}
\big)
+
8 C_{0}^{W,0}.
\end{eqnarray}
Moreover, one has
\begin{eqnarray}
&&
\hspace{-1cm}
\dfrac{
F_{23,W H^\pm}^{\textrm{Box},D}
(0,M_h^2,0,M_H^2,q_{12},q_{13})
}
{
8 M_W^4
}
= \\
&&
\hspace{-0.3cm}
=
\Bigg\{
\big(
2  \tau_{h,W}^{-1}
-
2  \zeta_W^{-1}
-
\lambda_W^{-1}
\big)
D_{113}^{W H^\pm,0}
-
\big(
\tau_{h,W}^{-1}
+
3  \tau_{H,W}^{-1}
+
3  \zeta_W^{-1}
+
\eta_W^{-1}
\big)
D_{233}^{W H^\pm,0}
\nonumber
\\
&&
+
\big(
\tau_{h,W}^{-1}
-
\tau_{H,W}^{-1}
-
4  \zeta_W^{-1}
-
\lambda_W^{-1}
\big)
D_{123}^{W H^\pm,0}
-
\big(
2  \tau_{H,W}^{-1}
+
\zeta_W^{-1}
+
\eta_W^{-1}
\big)
D_{333}^{W H^\pm,0}
\nonumber
\\
&&
+
\big(
\tau_{h,W}^{-1}
-
3  \tau_{H,W}^{-1}
-
2  \zeta_W^{-1}
\big)
D_{133}^{W H^\pm,0}
-
\big(
\tau_{h,W}^{-1}
+
\tau_{H,W}^{-1}
+
2  \zeta_W^{-1}
\big)
D_{223}^{W H^\pm,0}
\nonumber
\\
&&
+
\Big[
2 \tau_{{H^\pm},W}^{-1}
\big(
1
-
\,4  \tau_{{H^\pm},W}^{-1}
+
\,4  \tau_{H,W}^{-1}
\big)
+
2 \tau_{h,W}^{-1}
\big(
1
-
\,4  \tau_{H,W}^{-1}
+
\,4  \tau_{{H^\pm},W}^{-1}
\big)
+
\zeta_W^{-1}
-
\tau_{H,W}^{-1}
\Big]
D_{3}^{W H^\pm,0}
\nonumber
\\
&&
+
\Big[
2 \big(
\tau_{H,W}^{-1}
-
\tau_{{H^\pm},W}^{-1}
\big)
\big(
4  \tau_{{H^\pm},W}^{-1}
-
1
\big)
+
\tau_{h,W}^{-1}
\big(
1
-
8  \tau_{H,W}^{-1}
+
8  \tau_{{H^\pm},W}^{-1}
\big)
+
3 \eta_W^{-1}
\Big]
D_{13}^{W H^\pm,0}
\nonumber
\\
&&
+
\Big[
2 \big(
\tau_{H,W}^{-1}
-
\tau_{{H^\pm},W}^{-1}
\big)
\big(
4 \tau_{{H^\pm},W}^{-1}
-
1
\big)
-
\tau_{h,W}^{-1}
\big(
8 \tau_{H,W}^{-1}
-
8 \tau_{{H^\pm},W}^{-1}
+
5
\big)
+
\big(
\zeta_W^{-1}
+
4 \eta_W^{-1}
\big)
\Big]
D_{23}^{W H^\pm,0}
\nonumber
\\
&&
+
\Big[
2 \tau_{{H^\pm},W}^{-1}
\big(
1
-
4  \tau_{{H^\pm},W}^{-1}
+
4  \tau_{H,W}^{-1}
\big)
-
\tau_{h,W}^{-1}
\big(
8  \tau_{H,W}^{-1}
-
8 \tau_{{H^\pm},W}^{-1}
+
1
\big)
-
\zeta_W^{-1}
-
\lambda_W^{-1}
\Big]
D_{33}^{W H^\pm,0}
\Bigg\}
\nonumber
\\
&&
+ \frac{1}{2}
\big(
D_{00}^{W H^\pm,0}
-
D_{001}^{W H^\pm,0}
-
D_{002}^{W H^\pm,0}
-
2 D_{003}^{W H^\pm,0}
\big).
\nonumber
\end{eqnarray}
In additional, we have next
form factor in Eq.~\ref{WHform}
as
\begin{eqnarray}
&&
\hspace{-1cm}
\dfrac{
F_{11,W H^\pm}^{\textrm{Box}, D}
(0,M_h^2,0,M_H^2,q_{12},q_{13})
}
{
(-8 M_W^4)
}
=
\Bigg\{
\big(
2  \tau_{H,W}^{-1}
+
\zeta_W^{-1}
+
\eta_W^{-1}
\big)
D_{333}^{W H^\pm,0}
\\
&&
-
\big(
\tau_{h,W}^{-1}
-
\tau_{H,W}^{-1}
-
\zeta_W^{-1}
+
\eta_W^{-1}
\big)
\big(
D_{12}^{W H^\pm,0}
+
D_{13}^{W H^\pm,0}
+
D_{122}^{W H^\pm,0}
\big)
\nonumber
\\
&&
-
\big(
2  \tau_{h,W}^{-1}
-
2 \zeta_W^{-1}
-
\lambda_W^{-1}
\big)
\big(
D_{133}^{W H^\pm,0}
+
2 D_{123}^{W H^\pm,0}
\big)
+
\big(
\tau_{h,W}^{-1}
+
\tau_{H,W}^{-1}
+
2 \zeta_W^{-1}
\big)
D_{222}^{W H^\pm,0}
\nonumber
\\
&&
+
\big(
2 \tau_{h,W}^{-1}
+
4 \tau_{H,W}^{-1}
+
5 \zeta_W^{-1}
+
\eta_W^{-1}
\big)
D_{223}^{W H^\pm,0}
+
\big(
\tau_{h,W}^{-1}
+
5 \tau_{H,W}^{-1}
+
4 \zeta_W^{-1}
+
2 \eta_W^{-1}
\big)
D_{233}^{W H^\pm,0}
\nonumber
\\
&&
-
\Big[
2 \tau_{{H^\pm},W}^{-1}
\big(
1
-
4 \tau_{{H^\pm},W}^{-1}
\big)
\nonumber\\
&&
\hspace{+3.0cm}
+
2 \tau_{h,W}^{-1}
\big(
1
+
4 \tau_{{H^\pm},W}^{-1}
-
4 \tau_{H,W}^{-1}
\big)
+
\tau_{H,W}^{-1}
\big(
8 \tau_{{H^\pm},W}^{-1}
+
1
\big)
-
\zeta_W^{-1}
\Big]
D_{2}^{W H^\pm,0}
\nonumber
\\
&&
+
\Big[
2 \tau_{{H^\pm},W}^{-1}
\big(
4  \tau_{{H^\pm},W}^{-1}
-
4 \tau_{H,W}^{-1}
-
1
\big)
+
\tau_{h,W}^{-1}
\big(
8 \tau_{H,W}^{-1}
-
8 \tau_{{H^\pm},W}^{-1}
+
1
\big)
+
\zeta_W^{-1}
+
4 \lambda_W^{-1}
\Big]
D_{22}^{W H^\pm,0}
\nonumber
\\
&&
- \Big[
4 \tau_{{H^\pm},W}^{-1}
\big(
4  \tau_{H,W}^{-1}
-
4 \tau_{{H^\pm},W}^{-1}
+
1
\big)
\nonumber
\\
&&
\hspace{+3.0cm}
+
\big(
5 \zeta_W^{-1}
+
5 \eta_W^{-1}
-
7 \tau_{H,W}^{-1}
\big)
-
\tau_{h,W}^{-1}
\big(
16 \tau_{H,W}^{-1}
-
16 \tau_{{H^\pm},W}^{-1}
+
7
\big)
\Big]
D_{23}^{W H^\pm,0}
\nonumber
\\
&&
-
\Big[
2 \tau_{{H^\pm},W}^{-1}
\big(
1
-
4 \tau_{{H^\pm},W}^{-1}
\big)
+
\nonumber\\
&&
\hspace{3cm}
+
2 \tau_{h,W}^{-1}
\big(
1
+
4 \tau_{{H^\pm},W}^{-1}
-
4  \tau_{H,W}^{-1}
\big)
+
\tau_{H,W}^{-1}
\big(
8 \tau_{{H^\pm},W}^{-1}
+
1
\big)
-
\zeta_W^{-1}
\Big]
D_{3}^{W H^\pm,0}
\nonumber
\\
&&
+
\Big[
2 \tau_{{H^\pm},W}^{-1}
\big(
4 \tau_{{H^\pm},W}^{-1}
-
1
\big)
+
2 \tau_{H,W}^{-1}
\big(
1
-
4  \tau_{{H^\pm},W}^{-1}
\big)
\nonumber
\\
&&
\hspace{+3.0cm}
+
\tau_{h,W}^{-1}
\big(
8 \tau_{H,W}^{-1}
-
8 \tau_{{H^\pm},W}^{-1}
+
1
\big)
-
\zeta_W^{-1}
+
\lambda_W^{-1}
\Big]
D_{33}^{W H^\pm,0}
\Bigg\}
\nonumber
\\
&&
+
\big(
D_{00}^{W H^\pm,0}
+
2 D_{002}^{W H^\pm,0}
+
2 D_{003}^{W H^\pm,0}
\big).
\nonumber
\end{eqnarray}
We also have
the following
form factor
\begin{eqnarray}
&&
\hspace{-1cm}
\dfrac{
F_{23,W H^\pm}^{\textrm{Box},D}
(0,M_H^2,0,M_h^2,q_{12},q_{13})
}
{
(4 M_W^4)
}
=
\Bigg\{
2 \big(
\tau_{h,W}^{-1}
+
\tau_{H,W}^{-1}
+
2 \zeta_W^{-1}
\big)
D_{122}^{W H^\pm,1}
\\
&&
-
4 \big(
\tau_{h,W}^{-1}
+
\zeta_W^{-1}
\big)
D_{22}^{W H^\pm,1}
+
2 \big(
3 \tau_{h,W}^{-1}
+
\tau_{H,W}^{-1}
-
\lambda_W^{-1}
\big)
D_{123}^{W H^\pm,1}
\nonumber
\\
&&
-
2 \big(
2 \tau_{H,W}^{-1}
-
2 \zeta_W^{-1}
-
\lambda_W^{-1}
\big)
D_{112}^{W H^\pm,1}
+
\big(
1
+
4 \tau_{h,W}^{-1}
-
4 \tau_{{H^\pm},W}^{-1}
-
8 \zeta_W^{-1}
\big)
D_{2}^{W H^\pm,1}
\nonumber
\\
&&
+ 2
\Big[
2 \tau_{{H^\pm},W}^{-1}
\big(
4 \tau_{{H^\pm},W}^{-1}
-
4 \tau_{H,W}^{-1}
-
1
\big)
+
\big(
2 \tau_{H,W}^{-1}
-
3 \zeta_W^{-1}
-
2 \eta_W^{-1}
\big)
\nonumber
\\
&&
+
\tau_{h,W}^{-1}
\big(
8 \tau_{H,W}^{-1}
-
8 \tau_{{H^\pm},W}^{-1}
+
1
\big)
\Big]
D_{12}^{W H^\pm,1}
\Bigg\}
+
\big(
D_{001}^{W H^\pm,1}
-
D_{002}^{W H^\pm,1}
-
2 M_h^2
D_{23}^{W H^\pm,1}
\big).
\nonumber
\end{eqnarray}
Another factor is presented
as
\begin{eqnarray}
&&
\hspace{-1cm}
\dfrac{
F_{11,W H^\pm}^{\textrm{Box}, D}
(0,M_H^2,0,M_h^2,q_{12},q_{13})
}
{
(-8 M_W^4)
}
=
\\
\nonumber \\
&&
\hspace{-0.3cm}
=
\Bigg\{
\big(
3  \tau_{h,W}^{-1}
+
\tau_{H,W}^{-1}
-
\lambda_W^{-1}
\big)
D_{333}^{W H^\pm,1}
+
\big(
\tau_{h,W}^{-1}
-
5  \tau_{H,W}^{-1}
+
5 \zeta_W^{-1}
-
\eta_W^{-1}
\big)
\big(
D_{12}^{W H^\pm,1}
+
D_{13}^{W H^\pm,1}
\big)
\nonumber
\\
&&
-
2 \big(
2 \tau_{H,W}^{-1}
-
2 \zeta_W^{-1}
-
\lambda_W^{-1}
\big)
D_{123}^{W H^\pm,1}
+
\big(
\tau_{h,W}^{-1}
-
\tau_{H,W}^{-1}
+
\zeta_W^{-1}
-
\eta_W^{-1}
\big)
\big(
D_{122}^{W H^\pm,1}
+
D_{133}^{W H^\pm,1}
\big)
\nonumber
\\
&&
+
\big(
\tau_{h,W}^{-1}
+
\tau_{H,W}^{-1}
+
2 \zeta_W^{-1}
\big)
D_{222}^{W H^\pm,1}
+
\big(
4 \tau_{h,W}^{-1}
+
2 \tau_{H,W}^{-1}
+
5 \zeta_W^{-1}
+
\eta_W^{-1}
\big)
D_{223}^{W H^\pm,1}
\nonumber
\\
&&
+
\big(
5  \tau_{h,W}^{-1}
+
\tau_{H,W}^{-1}
+
4 \zeta_W^{-1}
+
2 \eta_W^{-1}
\big)
D_{233}^{W H^\pm,1}
\nonumber
\\
&&
-
\Big[
2 \big(
\tau_{H,W}^{-1}
-
\tau_{{H^\pm},W}^{-1}
\big)
\big(
1
+
4 \tau_{{H^\pm},W}^{-1}
\big)
-
\tau_{h,W}^{-1}
\big(
8 \tau_{H,W}^{-1}
-
8 \tau_{{H^\pm},W}^{-1}
-
1
\big)
-
5 \zeta_W^{-1}
+
1
\Big]
D_{2}^{W H^\pm,1}
\nonumber
\\
&&
+
\Big[
2 \tau_{{H^\pm},W}^{-1}
\big(
4 \tau_{{H^\pm},W}^{-1}
-
1
\big)
+
8 \tau_{h,W}^{-1}
\big(
\tau_{H,W}^{-1}
-
\tau_{{H^\pm},W}^{-1}
\big)
\nonumber\\
&&
\hspace{5cm}
+
\big(
9 \zeta_W^{-1}
+
4 \lambda_W^{-1}
\big)
+
\tau_{H,W}^{-1}
\big(
1
-
8 \tau_{{H^\pm},W}^{-1}
\big)
\Big]
D_{22}^{W H^\pm,1}
\nonumber
\\
&&
+
\Big[
4 \tau_{{H^\pm},W}^{-1}
\big(
4  \tau_{{H^\pm},W}^{-1}
-
1
\big)
+
2 \tau_{H,W}^{-1}
\big(
1
-
8 \tau_{{H^\pm},W}^{-1}
\big)
\nonumber
\\
&&\hspace{5.0cm}
+
2 \tau_{h,W}^{-1}
\big(
8 \tau_{H,W}^{-1}
-
8 \tau_{{H^\pm},W}^{-1}
+
3
\big)
+
12 \zeta_W^{-1}
+
5 \lambda_W^{-1}
\Big]
D_{23}^{W H^\pm,1}
\nonumber
\\
&&
-
\Big[
2 \big(
\tau_{H,W}^{-1}
-
\tau_{{H^\pm},W}^{-1}
\big)
\big(
4 \tau_{{H^\pm},W}^{-1}
+
1
\big)
+
\tau_{h,W}^{-1}
\big(
1
-
8 \tau_{H,W}^{-1}
+
8 \tau_{{H^\pm},W}^{-1}
\big)
-
5 \zeta_W^{-1}
+
1
\Big]
D_{3}^{W H^\pm,1}
\nonumber
\\
&&
+
\Big[
2 \tau_{{H^\pm},W}^{-1}
\big(
4  \tau_{{H^\pm},W}^{-1}
-
1
\big)
+
\tau_{H,W}^{-1}
\big(
1
-
8 \tau_{{H^\pm},W}^{-1}
\big)
\nonumber
\\
&&\hspace{5.0cm}
+
2 \tau_{h,W}^{-1}
\big(
4  \tau_{H,W}^{-1}
-
4  \tau_{{H^\pm},W}^{-1}
+
3
\big)
+
3 \zeta_W^{-1}
+
\lambda_W^{-1}
\Big]
D_{33}^{W H^\pm,1}
\Bigg\}
\nonumber
\\
&&
+\frac{1}{M_W^2}
\big(
3 D_{00}^{W H^\pm,1}
+
2 D_{002}^{W H^\pm,1}
+
2 D_{003}^{W H^\pm,1}
\big).
\nonumber
\end{eqnarray}
We also have the following
factor:
\begin{eqnarray}
&&
\hspace{-1cm}
\dfrac{
F_{23,W
H^\pm}^{\textrm{Box},D}
(0,0,M_H^2,M_h^2,
q_{23},q_{12})
}
{
(4 M_W^4)
}
=
\\
\nonumber \\
&=& 4 \Big(
2\zeta_W^{-1}
+
\lambda_W^{-1}
-
2  \tau_{h,W}^{-1}
\Big)
D_{112}^{W H^\pm,2}
-
4 \big(
3  \tau_{h,W}^{-1}
-
3  \tau_{H,W}^{-1}
-
\zeta_W^{-1}
+
\eta_W^{-1}
\big)
D_{122}^{W H^\pm,2}
\nonumber
\\
&&
+
8 \Big(
\tau_{H,W}^{-1}
-
\tau_{h,W}^{-1}
\Big)
D_{222}^{W H^\pm,2}
-
4 \Big(
3 \tau_{h,W}^{-1}
+
\tau_{H,W}^{-1}
-
\lambda_W^{-1}
\Big)
\Big[
D_{23}^{W H^\pm,2}
+
D_{123}^{W H^\pm,2}
+
D_{223}^{W H^\pm,2}
\Big]
\nonumber
\\
&&
- \Big[
16 \tau_{{H^\pm},W}^{-2}
-
4  \tau_{H,W}^{-1}
\big(
4  \tau_{{H^\pm},W}^{-1}
+
1
\big)
+
4  \tau_{h,W}^{-1}
\big(
4  \tau_{H,W}^{-1}
-
4  \tau_{{H^\pm},W}^{-1}
+
3
\big)
-
8 \zeta_W^{-1}
-
1
\Big]
D_{12}^{W H^\pm,2}
\nonumber
\\
&&
-
\Big[
16  \tau_{{H^\pm},W}^{-2}
+
4  \tau_{h,W}^{-1}
\big(
4  \tau_{H,W}^{-1}
-
4  \tau_{{H^\pm},W}^{-1}
+
3
\big)
-
4  \tau_{H,W}^{-1}
\big(
4  \tau_{{H^\pm},W}^{-1}
+
3
\big)
+
4  \lambda_W^{-1}
-
1
\Big]
D_{22}^{W H^\pm,2}
\nonumber
\\
&&
-
\Big[
16  \tau_{{H^\pm},W}^{-2}
+
4  \tau_{h,W}^{-1}
\big(
4  \tau_{H,W}^{-1}
-
4  \tau_{{H^\pm},W}^{-1}
+
1
\big)
-
4  \tau_{H,W}^{-1}
\big(
4  \tau_{{H^\pm},W}^{-1}
+
1
\big)
+
4  \lambda_W^{-1}
-
1
\Big]
D_{2}^{W H^\pm,2}
\nonumber
\\
&&
+ \frac{2}{M_W^2}
\Big[
D_{00}^{W H^\pm,2}
+
D_{001}^{W H^\pm,2}
+
2 D_{002}^{W H^\pm,2}
\Big].
\nonumber
\end{eqnarray}
Furthermore, one has
\begin{eqnarray}
&&
\hspace{-1cm}
\dfrac{
F_{11,W
H^\pm}^{\textrm{Box}, D}
(0,0,M_H^2,M_h^2,q_{23},q_{12})
}
{
(4 M_W^4 )
}
=
-\frac{8}{M_W^2}
D_{003}^{W H^\pm,2}
+
4 \big(
\tau_{H,W}^{-1}
-
\tau_{h,W}^{-1}
+
\zeta_W^{-1}
-
\eta_W^{-1}
\big)
D_{133}^{W H^\pm,2}
\\
&&
\hspace{3.3cm}
-
4 \big(
3 \tau_{h,W}^{-1}
+
\tau_{H,W}^{-1}
-
\lambda_W^{-1}
\big)
D_{333}^{W H^\pm,2}
+
8 \big(
\tau_{H,W}^{-1}
-
\tau_{h,W}^{-1}
\big)
D_{233}^{W H^\pm,2}
\nonumber
\\
&&
\hspace{3.3cm}
-
\Big[
4 \lambda_W^{-1}
+
\big(
4 \tau_{H,W}^{-1}
-
4 \tau_{{H^\pm},W}^{-1}
+
1
\big)
\big(
4 \tau_{h,W}^{-1}
-
4 \tau_{{H^\pm},W}^{-1}
-
1
\big)
\Big]
D_{33}^{W H^\pm,2}.
\nonumber
\end{eqnarray}
Next one is shown
\begin{eqnarray}
&&
\hspace{-1cm}
\dfrac{
F_{23,W
H^\pm}^{\textrm{Box},D}
(M_h^2,0,0,M_H^2,q_{12},q_{23})
}
{
(4 M_W^4)
}
=
\\
&&=
4 \big(
\zeta_W^{-1}
-
\tau_{h,W}^{-1}
\big)
\big[
D_{12}^{W H^\pm,3}
+
D_{22}^{W H^\pm,3}
-
D_{2}^{W H^\pm,3}
\big]
-
\big(
13 \tau_{h,W}^{-1}
-
2 \zeta_W^{-1}
-
\eta_W^{-1}
\big)
D_{13}^{W H^\pm,3}
\nonumber
\\
&&
- \big(
3 \tau_{h,W}^{-1}
+
2 \zeta_W^{-1}
+
\eta_W^{-1}
\big)
\big[
D_{123}^{W H^\pm,3}
+
D_{133}^{W H^\pm,3}
\big]
-
\big(
\tau_{h,W}^{-1}
+
\tau_{H,W}^{-1}
+
4 \zeta_W^{-1}
\big)
D_{223}^{W H^\pm,3}
\nonumber
\\
&&
-
\big(
\tau_{h,W}^{-1}
+
4 \tau_{H,W}^{-1}
+
6 \zeta_W^{-1}
+
\eta_W^{-1}
\big)
D_{233}^{W H^\pm,3}
-
\big(
3 \tau_{H,W}^{-1}
+
2 \zeta_W^{-1}
+
\eta_W^{-1}
\big)
D_{333}^{W H^\pm,3}
\nonumber
\\
&&
-\dfrac{1}{4}
\Big[
4 \tau_{{H^\pm},W}^{-1}
\big(
16 \tau_{{H^\pm},W}^{-1}
-
5
\big)
-
16 \tau_{H,W}^{-1}
\big(
4 \tau_{{H^\pm},W}^{-1}
+
1
\big)
\nonumber
\\
&&
\hspace{+4.0cm}
+
32 \tau_{h,W}^{-1}
\big(
2 \tau_{H,W}^{-1}
-
2 \tau_{{H^\pm},W}^{-1}
+
1
\big)
+
4 \zeta_W^{-1}
+
16 \lambda_W^{-1}
+
1
\Big]
D_{23}^{W H^\pm,3}
\nonumber
\\
&&
+
\dfrac{1}{4}
\Big[
4  \tau_{{H^\pm},W}^{-1}
\big(
16 \tau_{H,W}^{-1}
-
16 \tau_{{H^\pm},W}^{-1}
+
3
\big)
\nonumber
\\
&&
\hspace{+4.0cm}
-
4 \tau_{h,W}^{-1}
\big(
16 \tau_{H,W}^{-1}
-
16 \tau_{{H^\pm},W}^{-1}
-
5
\big)
+
4  \zeta_W^{-1}
-
12 \tau_{H,W}^{-1}
+
1
\Big]
D_{3}^{W H^\pm,3}
\nonumber
\\
&&
+ \dfrac{1}{4}
\Big[
4 \tau_{{H^\pm},W}^{-1}
\big(
16 \tau_{H,W}^{-1}
-
16 \tau_{{H^\pm},W}^{-1}
+
5
\big)
\nonumber
\\
&&\hspace{+2.7cm}
-
4  \tau_{h,W}^{-1}
\big(
16 \tau_{H,W}^{-1}
-
16 \tau_{{H^\pm},W}^{-1}
+
5
\big)
+
4 \zeta_W^{-1}
+
4 \eta_W^{-1}
-
8 \tau_{H,W}^{-1}
-
1
\Big]
D_{33}^{W H^\pm,3}
\nonumber
\\
&&
+ \frac{1}{2 M_W^2}
\big[
5 D_{00}^{W H^\pm,3}
- D_{002}^{W H^\pm,3}
-3 D_{003}^{W H^\pm,3}
\big].
\nonumber
\end{eqnarray}
Further factor is
expressed as follows:
\begin{eqnarray}
&&
\hspace{-1cm}
\dfrac{
F_{11,W H^\pm}^{\textrm{Box}, D}
(M_h^2,0,0,M_H^2,q_{12},q_{23})
}
{
(-4 M_W^4)
}
= \\
&&
\hspace{-0.3cm}
= \big(
3 \tau_{h,W}^{-1}
+
2  \zeta_W^{-1}
+
\eta_W^{-1}
\big)
D_{111}^{W H^\pm,3}
+
\big(
3 \tau_{H,W}^{-1}
+
2  \zeta_W^{-1}
+
\eta_W^{-1}
\big)
D_{333}^{W H^\pm,3}
\nonumber
\\
&&
+
\big(
7  \tau_{h,W}^{-1}
+
\tau_{H,W}^{-1}
+
8 \zeta_W^{-1}
+
2 \eta_W^{-1}
\big)
D_{112}^{W H^\pm,3}
+
3 \big(
2 \tau_{h,W}^{-1}
+
\tau_{H,W}^{-1}
+
2 \zeta_W^{-1}
+
\eta_W^{-1}
\big)
D_{113}^{W H^\pm,3}
\nonumber
\\
&&
+
\big(
5  \tau_{h,W}^{-1}
+
2  \tau_{H,W}^{-1}
+
10  \zeta_W^{-1}
+
\eta_W^{-1}
\big)
D_{122}^{W H^\pm,3}
+
4 \big(
2  \tau_{h,W}^{-1}
+
2 \tau_{H,W}^{-1}
+
4 \zeta_W^{-1}
+
\eta_W^{-1}
\big)
D_{123}^{W H^\pm,3}
\nonumber
\\
&&
+
3 \big(
\tau_{h,W}^{-1}
+
2  \tau_{H,W}^{-1}
+
2  \zeta_W^{-1}
+
\eta_W^{-1}
\big)
D_{133}^{W H^\pm,3}
+
\big(
2 \tau_{h,W}^{-1}
+
5 \tau_{H,W}^{-1}
+
10 \zeta_W^{-1}
+
\eta_W^{-1}
\big)
D_{223}^{W H^\pm,3}
\nonumber
\\
&&
+
\big(
\tau_{h,W}^{-1}
+
\tau_{H,W}^{-1}
+
4 \zeta_W^{-1}
\big)
D_{222}^{W H^\pm,3}
+
\big(
\tau_{h,W}^{-1}
+
7 \tau_{H,W}^{-1}
+
8 \zeta_W^{-1}
+
2 \eta_W^{-1}
\big)
D_{233}^{W H^\pm,3}
\nonumber
\\
&&
- \dfrac{1}{4}
\Big[
4 \tau_{{H^\pm},W}^{-1}
\big(
16 \tau_{H,W}^{-1}
-
16 \tau_{{H^\pm},W}^{-1}
+
5
\big)
\nonumber
\\
&&
\hspace{+3cm}
+
4 \tau_{h,W}^{-1}
\big(
16 \tau_{{H^\pm},W}^{-1}
-
16 \tau_{H,W}^{-1}
+
3
\big)
+
12 \tau_{H,W}^{-1}
-
4 \zeta_W^{-1}
-
1
\Big]
D_{0}^{W H^\pm,3}
\nonumber
\\
&&
- \dfrac{1}{4}
\Big[
2 \big(
4  \tau_{H,W}^{-1}
-
4  \tau_{{H^\pm},W}^{-1}
+
1
\big)
\big(
16 \tau_{{H^\pm},W}^{-1}
+
1
\big)
\nonumber
\\
&&\hspace{6cm}
-
4 \tau_{h,W}^{-1}
\big(
32 \tau_{H,W}^{-1}
-
32 \tau_{{H^\pm},W}^{-1}
+
1
\big)
-
4 \eta_W^{-1}
\Big]
D_{1}^{W H^\pm,3}
\nonumber
\\
&&
- \dfrac{1}{4}
\Big[
4 \tau_{{H^\pm},W}^{-1}
\big(
16 \tau_{H,W}^{-1}
-
16 \tau_{{H^\pm},W}^{-1}
+
5
\big)
\nonumber
\\
&&
\hspace{+2cm}
-
4 \tau_{h,W}^{-1}
\big(
16 \tau_{H,W}^{-1}
-
16 \tau_{{H^\pm},W}^{-1}
+
15
\big)
+
12 \zeta_W^{-1}
+
8 \eta_W^{-1}
-
20 \tau_{H,W}^{-1}
-
1
\Big]
D_{11}^{W H^\pm,3}
\nonumber
\\
&&
+ \dfrac{1}{2}
\Big[
4  \tau_{{H^\pm},W}^{-1}
\big(
16  \tau_{{H^\pm},W}^{-1}
-
16  \tau_{H,W}^{-1}
-
5
\big)
\nonumber
\\
&&\hspace{+2.5cm}
+
4  \tau_{h,W}^{-1}
\big(
16  \tau_{H,W}^{-1}
-
16 \tau_{{H^\pm},W}^{-1}
+
13
\big)
+
32 \tau_{H,W}^{-1}
-
4  \zeta_W^{-1}
-
20  \eta_W^{-1}
+
1
\Big]
D_{12}^{W H^\pm,3}
\nonumber
\\
&&
+ \dfrac{1}{2}
\Big[
4  \tau_{{H^\pm},W}^{-1}
\big(
16 \tau_{{H^\pm},W}^{-1}
-
16  \tau_{H,W}^{-1}
-
5
\big)
\nonumber
\\
&&
\hspace{+2.5cm}
+
8  \tau_{h,W}^{-1}
\big(
8 \tau_{H,W}^{-1}
-
8 \tau_{{H^\pm},W}^{-1}
+
5
\big)
+
40  \tau_{H,W}^{-1}
-
12  \zeta_W^{-1}
-
8  \eta_W^{-1}
+
1
\Big]
D_{13}^{W H^\pm,3}
\nonumber
\\
&&
- \dfrac{1}{4}
\Big[
8 \tau_{{H^\pm},W}^{-1}
\big(
3
-
16 \tau_{{H^\pm},W}^{-1}
\big)
+
4  \tau_{H,W}^{-1}
\big(
32 \tau_{{H^\pm},W}^{-1}
+
1
\big)
\nonumber
\\
&&
\hspace{6cm}
-
4  \tau_{h,W}^{-1}
\big(
32  \tau_{H,W}^{-1}
-
32  \tau_{{H^\pm},W}^{-1}
-
1
\big)
-
8 \zeta_W^{-1}
+
2
\Big]
D_{2}^{W H^\pm,3}
\nonumber
\\
&&
+ \dfrac{1}{4}
\Big[
4  \tau_{{H^\pm},W}^{-1}
\big(
16 \tau_{{H^\pm},W}^{-1}
-
16 \tau_{H,W}^{-1}
-
5
\big)
\nonumber
\\
&&
\hspace{+2.5cm}
+
4  \tau_{h,W}^{-1}
\big(
16 \tau_{H,W}^{-1}
-
16  \tau_{{H^\pm},W}^{-1}
+
11
\big)
+
44  \tau_{H,W}^{-1}
+
4 \zeta_W^{-1}
-
32 \eta_W^{-1}
+
1
\Big]
D_{22}^{W H^\pm,3}
\nonumber
\\
&&
+ \dfrac{1}{2}
\Big[
4  \tau_{{H^\pm},W}^{-1}
\big(
16  \tau_{{H^\pm},W}^{-1}
-
16 \tau_{H,W}^{-1}
-
5
\big)
\nonumber
\\
&&
\hspace{+2.5cm}
+
4  \tau_{h,W}^{-1}
\big(
16  \tau_{H,W}^{-1}
-
16  \tau_{{H^\pm},W}^{-1}
+
3
\big)
+
32 \tau_{H,W}^{-1}
+
16  \zeta_W^{-1}
+
20  \lambda_W^{-1}
+
1
\Big]
D_{23}^{W H^\pm,3}
\nonumber
\\
&&
- \dfrac{1}{4}
\Big[
8 \tau_{{H^\pm},W}^{-1}
\big(
16  \tau_{H,W}^{-1}
-
16 \tau_{{H^\pm},W}^{-1}
+
3
\big)
\nonumber
\\
&&
\hspace{+2.5cm}
-
8 \tau_{h,W}^{-1}
\big(
16 \tau_{H,W}^{-1}
-
16 \tau_{{H^\pm},W}^{-1}
-
1
\big)
-
4 \tau_{H,W}^{-1}
-
4  \eta_W^{-1}
+
2
\Big]
D_{3}^{W H^\pm,3}
\nonumber
\\
&&
- \dfrac{1}{4}
\Big[
4  \tau_{{H^\pm},W}^{-1}
\big(
16  \tau_{H,W}^{-1}
-
16  \tau_{{H^\pm},W}^{-1}
+
5
\big)
\nonumber
\\
&&
\hspace{+2.5cm}
-
4  \tau_{h,W}^{-1}
\big(
16 \tau_{H,W}^{-1}
-
16 \tau_{{H^\pm},W}^{-1}
+
5
\big)
+
12  \zeta_W^{-1}
+
8  \eta_W^{-1}
-
60 \tau_{H,W}^{-1}
-
1
\Big]
D_{33}^{W H^\pm,3}
\nonumber
\\
&&
+\frac{1}{M_W^2}
\big[
5 D_{00}^{W H^\pm,3}
+
3 D_{001}^{W H^\pm,3}
+
3 D_{002}^{W H^\pm,3}
+
3 D_{003}^{W H^\pm,3}
\big].
\nonumber
\end{eqnarray}
In all above expressions,
we have denoted
\begin{eqnarray}
C_{ij\ldots}^{W,0}
& \equiv &
C_{ij\ldots}
(0,0,q_{23}, M_W^2, M_W^2, M_W^2),
\\
C_{ij\ldots}^{H^\pm,0}
& \equiv &
C_{ij\ldots}
(0,0,q_{23}, M_{H^\pm}^2, M_{H^\pm}^2, M_{H^\pm}^2),
\\
C_{ij\ldots}^{W H^\pm,0}
& \equiv &
C_{ij\ldots}
(M_h^2,0,q_{13}, M_{H^\pm}^2, M_W^2, M_W^2),
\\
C_{ij\ldots}^{W H^\pm,1}
& \equiv &
C_{ij\ldots}
(M_h^2,q_{23},M_H^2, M_{H^\pm}^2, M_W^2, M_W^2),
\\
C_{ij\ldots}^{W H^\pm,2}
& \equiv &
C_{ij\ldots}
(M_H^2,0,q_{12}, M_{H^\pm}^2, M_W^2, M_W^2),
\\
C_{ij\ldots}^{W H^\pm,3}
& \equiv &
C_{ij\ldots}
(0,q_{12},M_H^2, M_{H^\pm}^2, M_{H^\pm}^2, M_W^2),
\\
C_{ij\ldots}^{W H^\pm,4}
& \equiv &
C_{ij\ldots}
(M_h^2,0,q_{12}, M_{H^\pm}^2, M_W^2, M_W^2),
\\
C_{ij\ldots}^{W H^\pm,5}
& \equiv &
C_{ij\ldots}
(M_H^2,0,q_{13}, M_{H^\pm}^2, M_W^2, M_W^2),
\\
C_{ij\ldots}^{W H^\pm,6}
& \equiv &
C_{ij\ldots}
(0,q_{12},M_h^2, M_{H^\pm}^2, M_{H^\pm}^2, M_W^2),
\\
C_{ij\ldots}^{W H^\pm,7}
& \equiv &
C_{ij\ldots}
(0,q_{13},M_h^2, M_{H^\pm}^2, M_{H^\pm}^2, M_W^2),
\\
C_{ij\ldots}^{W H^\pm,8}
& \equiv &
C_{ij\ldots}
(0,q_{13},M_H^2, M_{H^\pm}^2, M_{H^\pm}^2, M_W^2),
\\
D_{ij\ldots}^{W H^\pm,0}
& \equiv &
D_{ij\ldots}
(0,M_h^2,0,M_H^2,q_{12},q_{13},
M_{H^\pm}^2, M_{H^\pm}^2, M_W^2, M_W^2),
\\
D_{ij\ldots}^{W H^\pm,1}
& \equiv &
D_{ij\ldots}
(0,M_H^2,0,M_h^2,q_{12},q_{13},
M_{H^\pm}^2, M_{H^\pm}^2, M_W^2, M_W^2),
\\
D_{ij\ldots}^{W H^\pm,2}
& \equiv &
D_{ij\ldots}
(0,0,M_H^2,M_h^2,q_{23},q_{12},
M_{H^\pm}^2, M_{H^\pm}^2, M_{H^\pm}^2, M_W^2),
\\
D_{ij\ldots}^{W H^\pm,3}
& \equiv &
D_{ij\ldots}
(M_h^2,0,0,M_H^2,q_{12},q_{23},
M_{H^\pm}^2, M_W^2, M_W^2, M_W^2).
\end{eqnarray}
\section*{Appendix D: Effective
Lagrangian in the THDM}
Deriving all couplings in
Tables~\ref{THDM-coupling1},
\ref{THDM-coupling2}
for the THDM are presented
in this appendix. From the
kinematic terms, one has
\begin{eqnarray}
\mathcal{L}_{K}
&=&
(D_\mu\Phi_1)^\dagger(D^\mu\Phi_1)
+(D_\mu\Phi_2)^\dagger(D^\mu\Phi_2)
\\
&\supset&
\frac{2M_W^2}{v}
s_{\beta-\alpha}
\;
hW_{\mu}^{\pm}W^{\mp,\mu}
+
\frac{2M_W^2}{v}c_{\alpha-\beta}
\;
HW_{\mu}^{\pm}W^{\mp,\mu}
+
\frac{M_Z^2}{v}s_{\beta-\alpha}
\;
hZ_{\mu}Z^{\mu}
\notag\\
&&
+
\frac{M_Z^2}{v}c_{\alpha-\beta}
\;
HZ_{\mu}Z^{\mu}
+
i\frac{M_Zc_{2W}}{v}
Z^{\mu}(H^{\mp}
\partial_{\mu}{H^\pm}
-
H^{\pm} \partial_{\mu}
H^{\mp})
\nonumber\\
&&
+
i\frac{M_Z
s_{2W}}{v}
A^{\mu}
(
H^{\mp}\partial_{\mu} H^\pm
-
H^{\pm}\partial_{\mu}H^{\mp}
)
+
\frac{4M_W^2s_W^2}{v^2}
H^{\pm}H^{\mp}A_\mu{A^{\mu}}
\nonumber\\
&&
-
i\frac{
M_W s_{\beta-\alpha}
}{v}
(
HW^{\mp,\mu}\partial_{\mu}H^{\pm}
-HW^{\pm,\mu}\partial_{\mu}H^{\mp}
+H^{\mp}W^{\pm,\mu}\partial_{\mu}H
-H^{\pm}W^{\mp,\mu}\partial_{\mu}H
)
\nonumber\\
&&
-
i
\frac{M_Wc_{\beta-\alpha}}{v}
(-hW^{\mp,\mu}\partial_{\mu}H^{\pm}
+hW^{\pm,\mu}\partial_{\mu}H^{\mp}
-H^{\mp}W^{\pm,\mu}\partial_{\mu}h
+H^{\pm}W^{\mp,\mu}\partial_{\mu}h)
\nonumber\\
&&
+\frac{2M_W^2s_W
c_{\beta-\alpha}}{v^2}
hH^{\mp}W_{\mu}^{\pm}A^{\mu}
-
\frac{2M_W^2s_Ws_{\beta-\alpha}}{v^2}
HH^{\mp}W_{\mu}^{\pm}A^{\mu}
+ \cdots
\end{eqnarray}
From scalar potential, we have
\begin{eqnarray}
-\mathcal{V}(\Phi_1,\Phi_2)
&\supset&
-\frac{\lambda_1}{2}(\Phi_1^\dagger\Phi_1)^2
-\frac{\lambda_2}{2}(\Phi_2^\dagger\Phi_2)^2
-\lambda_3(\Phi_1^\dagger\Phi_1)
(\Phi_2^\dagger\Phi_2)
\\
&&
-\lambda_4(\Phi_1^\dagger\Phi_2)
(\Phi_2^\dagger\Phi_1)
-\frac{\lambda_5}{2}[(\Phi_1^\dagger\Phi_2)^2
+(\Phi_2^\dagger\Phi_1)^2]
\notag\\
&\supset&
-\lambda_{hHH}hHH
-\lambda_{Hhh}Hhh
-\lambda_{hH^{\pm}H^{\mp}}hH^{\pm}H^{\mp}
\notag\\
&&
-\lambda_{hH^{\pm}H^{\mp}}HH^{\pm}H^{\mp}
-\lambda_{HhH^{\pm}H^{\mp}}
HhH^{\pm}H^{\mp}+ \cdots
\end{eqnarray}
where all coefficient couplings are
presented explicitly in terms of
bare parameters as well as physical
parameters as follows:
\begin{eqnarray}
-\lambda_{hHH}
&=&
\frac{3\lambda_1v}{2}
s_{\alpha} c_{\alpha}^2 c_{\beta}
-\frac{3\lambda_2 v}{2}
s_{\beta}s_{\alpha}^2c_{\alpha}
-\frac{
\lambda_{345}
}{2}v
[c_{\beta}(2s_{\alpha}c_{\alpha}^2 -
s_{\alpha}^3) + s_{\beta}
(c_{\alpha}^3-2s_{\alpha}^2c_{\alpha})
]
\nonumber
\\
&& \\
&=&
\frac{s_{\beta-\alpha}}{2v}
\Big[ (2M^2-2 M_H^2 - M_h^2)
s_{\beta-\alpha}^2
+
(3 M^2 - 2 M_H^2 - M_h^2)
\cot{(2\beta)}\;
s_{2(\beta-\alpha)}
\nonumber
\\
&&
\hspace{1cm}
-(4M^2 - 2 M_H^2 - M_h^2)
c_{\beta-\alpha}^2
\Big]
\\
&=&
\frac{s_{\alpha-\beta}
[s_{2\alpha}(3M^2-M_h^2-2M_H^2) +
M^2s_{2\beta}]}{v\;s_{2\beta}}
,
\\
-\lambda_{Hhh} &=&
-\frac{3\lambda_1v}{2}
c_{\beta}c_{\alpha}s_{\alpha}^2
-\frac{3\lambda_2v}{2}s_{\beta}
c_{\alpha}^2s_{\alpha}
-\frac{
\lambda_{345}
}{2}v
[s_{\beta}(s_{\alpha}^3
-2c_{\alpha}^2s_{\alpha})
-c_{\beta}(2c_{\alpha}s_{\alpha}^2
-c_{\alpha}^3)]
\nonumber\\
&&
\\
&=&
\frac{c_{\alpha-\beta}
[s_{2\alpha}(3M^2-M_H^2-2m_h^2)
-M^2s_{2\beta}]}{v\; s_{2\beta}}
,
\\
-\lambda_{HH^{\pm}H^{\mp}}
&=& -\lambda_1vc_{\beta}c_{\alpha}s_{\beta}^2
-\lambda_2vs_{\beta}s_{\alpha}c_{\beta}^2
-\lambda_3v(s_{\beta}s_{\alpha}s_{\beta}^2
+c_{\beta}c_{\alpha}c_{\beta}^2)
+
\frac{\lambda_{45}}{2}
v s_{(2\beta)}\;
s_{\beta+\alpha}
\\
&=& -\frac{1}{v}
\Big[2(M^2-M_H^2)
\cot{(2\beta)}\;
s_{\beta-\alpha}
+(2M_{H^\pm}^2 + M_H^2-2M^2)
c_{\beta-\alpha}
\Big]
\\
&=&
\frac{s_{\alpha+\beta}
(4M^2-3M_H^2-2M_{H^\pm}^2)
+(2M_{H^\pm}^2 - M_H^2)s_{\alpha-3\beta}}
{2vs_{(2\beta)}},
\\
-\lambda_{hH^{\pm}H^{\mp}}
&=& \lambda_1v
c_{\beta}s_{\alpha}s_{\beta}^2
-\lambda_2 v s_{\beta}c_{\alpha}c_{\beta}^2
-\lambda_3v(s_{\beta}c_{\alpha}s_{\beta}^2
-c_{\beta}s_{\alpha}c_{\beta}^2)
+
\frac{\lambda_{45}
}{2}
v \; s_{(2\beta)}
\; c_{(\beta+\alpha)}
\\
&=&
\frac{1}{v}
\Big[ 2(M^2 - M_h^2)
\cot{2\beta}c_{\beta-\alpha}
+(2M^2-2M_{H^\pm}^2- M_h^2)
s_{\beta-\alpha}
\Big]
\\
&=&
\frac{c_{\alpha+\beta}
(4M^2-3M_h^2 - 2M_{H^\pm}^2)
+(2M_{H^\pm}^2-M_h^2)
c_{(\alpha-3\beta)}}
{2vs_{2\beta}},
\end{eqnarray}
and
\begin{eqnarray}
-\lambda_{HhH^{\pm}H^{\mp}}
&=&
\lambda_1s_{\beta}^2s_{\alpha}c_{\alpha}
-\lambda_2c_{\beta}^2s_{\alpha}c_{\alpha}
+\lambda_3s_{\alpha}c_{\alpha}c_{2\beta}
+(\lambda_4+\lambda_5)s_{\beta}c_{\beta}c_{2\alpha}
\\
\label{gHhSS}
&=&
\frac{s_{2\alpha}(3c_{2\alpha}
+ c_{2(\alpha-2\beta)}
- 4c_{2\beta})}{4v^2s_{2\beta}^2} M_H^2
-\frac{s_{2\alpha}(3c_{2\alpha}
+ c_{2(\alpha-2\beta)}+4c_{2\beta})}
{4v^2s_{2\beta}^2} M_h^2
\nonumber\\
&&
+\frac{s_{2(\alpha-\beta)}}{v^2}
M_{H^\pm}^2 + \frac{(s_{2(\alpha-3\beta)}
+2s_{2(\alpha-\beta)}+5s_{2(\alpha+\beta)})}
{4v^2s_{2\beta}^2}M^2,
\\
-\lambda_{HhG^{\pm}G^{\mp}}
&=&
\lambda_1c_{\beta}^2s_{\alpha}c_{\alpha}
-\lambda_2s_{\beta}^2s_{\alpha}c_{\alpha}
-\lambda_3s_{\alpha}c_{\alpha}c_{2\beta}
-(\lambda_4+\lambda_5)s_{\beta}
c_{\beta}c_{2\alpha} \nonumber\\
\label{hHGG}
&=&\frac{1}{2v^2s_{2\beta}}
s_{2(\alpha-\beta)}
[(M_h^2- M_H^2)s_{2\alpha}
+ 2(M^2-M_{H^\pm}^2)s_{2\beta}].
\end{eqnarray}

The relationship between the
bare parameters in the Higgs
potential and the physical
parameters is
\begin{eqnarray}
\lambda_1 &=&
\frac{1}{v^2 c_{\beta}^2}
(c_{\alpha}^2\; M_H^2+s_{\alpha}^2\;
M_h^2 - s_{\beta}^2 M^2),
\\
\lambda_2 &=&
\frac{1}{v^2s_{\beta}^2}
(s_{\alpha}^2 M_H^2 + c_{\alpha}^2 M_h^2
-c_{\beta}^2M^2),
\\
\lambda_3 &=&
\frac{s_{2\alpha}}{v^2s_{2\beta}}(M_H^2 - M_h^2)
-\frac{1}{v^2}(M^2-2M_{H^\pm}^2),
\\
\lambda_4 &=&
\frac{1}{v^2}(M_{A^0}^2-2 M_{H^\pm}^2 + M^2),
\\
\lambda_5
&=&
\frac{1}{v^2}(M^2-M_{A^0}^2).
\end{eqnarray}
Where the $M^2$ parameter is given
by $M^2 =
m_{12}^2/(s_{\beta}c_{\beta})$.


\begin{thebibliography}{200}
\bibitem{Liss:2013hbb}
A.~Liss \textit{et al.} [ATLAS],
[arXiv:1307.7292 [hep-ex]].
\bibitem{CMS:2013xfa}
[CMS],
[arXiv:1307.7135 [hep-ex]].
\bibitem{Baer:2013cma}
H.~Baer, T.~Barklow, K.~Fujii, Y.~Gao, A.~Hoang, S.~Kanemura, J.~List, H.~E.~Logan, A.~Nomerotski and M.~Perelstein, \textit{et al.}
[arXiv:1306.6352 [hep-ph]].


\bibitem{ATLAS:2020qcv}
G.~Aad \textit{et al.} [ATLAS],
Phys. Lett. B \textbf{809} (2020), 135754
doi:10.1016/j.physletb.2020.135754
[arXiv:2005.05382 [hep-ex]].


\bibitem{ATLAS:2023yqk}
G.~Aad \textit{et al.} [ATLAS and CMS],
Phys. Rev. Lett. \textbf{132} (2024), 021803
doi:10.1103/PhysRevLett.132.021803
[arXiv:2309.03501 [hep-ex]].

\bibitem{CMS:2014fzn}
V.~Khachatryan \textit{et al.} [CMS],
Eur. Phys. J. C \textbf{75} (2015) no.5, 212
doi:10.1140/epjc/s10052-015-3351-7
[arXiv:1412.8662 [hep-ex]].
\bibitem{ATLAS:2015egz}
G.~Aad \textit{et al.} [ATLAS],
Eur. Phys. J. C \textbf{76} (2016) no.1, 6
doi:10.1140/epjc/s10052-015-3769-y
[arXiv:1507.04548 [hep-ex]].
\bibitem{CMS:2021kom}
A.~M.~Sirunyan \textit{et al.} [CMS],
JHEP \textbf{07} (2021), 027
doi:10.1007/JHEP07(2021)027
[arXiv:2103.06956 [hep-ex]].
\bibitem{D0:2008swt}
V.~M.~Abazov \textit{et al.} [D0],
Phys. Lett. B \textbf{671} (2009), 349-355
doi:10.1016/j.physletb.2008.12.009
[arXiv:0806.0611 [hep-ex]].
\bibitem{CMS:2013rmy}
S.~Chatrchyan \textit{et al.} [CMS],
Phys. Lett. B \textbf{726} (2013), 587-609
doi:10.1016/j.physletb.2013.09.057
[arXiv:1307.5515 [hep-ex]].
\bibitem{ATLAS:2017zdf}
M.~Aaboud \textit{et al.} [ATLAS],
JHEP \textbf{10} (2017), 112
doi:10.1007/JHEP10(2017)112
[arXiv:1708.00212 [hep-ex]].
\bibitem{ATLAS:2020qcv}
G.~Aad \textit{et al.} [ATLAS],
Phys. Lett. B \textbf{809} (2020), 135754
doi:10.1016/j.physletb.2020.135754
[arXiv:2005.05382 [hep-ex]].

\bibitem{CMS:2015tzs}
V.~Khachatryan \textit{et al.} [CMS],
Phys. Lett. B \textbf{753} (2016), 341-362
doi:10.1016/j.physletb.2015.12.039
[arXiv:1507.03031 [hep-ex]].
\bibitem{CMS:2017dyb}
A.~M.~Sirunyan \textit{et al.} [CMS],
JHEP \textbf{09} (2018), 148
doi:10.1007/JHEP09(2018)148
[arXiv:1712.03143 [hep-ex]].
\bibitem{CMS:2018myz}
A.~M.~Sirunyan \textit{et al.} [CMS],
JHEP \textbf{11} (2018), 152
doi:10.1007/JHEP11(2018)152
[arXiv:1806.05996 [hep-ex]].
\bibitem{ATLAS:2021wwb}
G.~Aad \textit{et al.} [ATLAS],
Phys. Lett. B \textbf{819} (2021), 136412
doi:10.1016/j.physletb.2021.136412
[arXiv:2103.10322 [hep-ex]].

\bibitem{Cen:2018okf}
J.~Y.~Cen, J.~H.~Chen, X.~G.~He, G.~Li, J.~Y.~Su and W.~Wang,
JHEP \textbf{01} (2019), 148
doi:10.1007/JHEP01(2019)148
[arXiv:1811.00910 [hep-ph]].

\bibitem{CMS:2020imj}
A.~M.~Sirunyan \textit{et al.} [CMS],
JHEP \textbf{07} (2020), 126
doi:10.1007/JHEP07(2020)126
[arXiv:2001.07763 [hep-ex]].

\bibitem{CMS:2021wlt}
A.~M.~Sirunyan \textit{et al.} [CMS],
Eur. Phys. J. C \textbf{81} (2021) no.8, 723
doi:10.1140/epjc/s10052-021-09472-3
[arXiv:2104.04762 [hep-ex]].

\bibitem{CMS:2022jqc}
A.~Tumasyan \textit{et al.} [CMS],
JHEP \textbf{09} (2023), 032
doi:10.1007/JHEP09(2023)032
[arXiv:2207.01046 [hep-ex]].

\bibitem{ATLAS:2022zuc}
G.~Aad \textit{et al.} [ATLAS],
Eur. Phys. J. C \textbf{83} (2023) no.7, 633
doi:10.1140/epjc/s10052-023-11437-7
[arXiv:2207.03925 [hep-ex]].


\bibitem{ATLAS:2023bzb}
G.~Aad \textit{et al.} [ATLAS],
JHEP \textbf{09} (2023), 004
doi:10.1007/JHEP09(2023)004
[arXiv:2302.11739 [hep-ex]].


\bibitem{CMS:2019kca}
A.~M.~Sirunyan \textit{et al.} [CMS],
JHEP \textbf{03} (2020), 065
doi:10.1007/JHEP03(2020)065
[arXiv:1910.11634 [hep-ex]].

\bibitem{CMS:2019qcx}
A.~M.~Sirunyan \textit{et al.} [CMS],
Eur. Phys. J. C \textbf{79} (2019) no.7, 564
doi:10.1140/epjc/s10052-019-7058-z
[arXiv:1903.00941 [hep-ex]].

\bibitem{CMS:2019ogx}
A.~M.~Sirunyan \textit{et al.} [CMS],
JHEP \textbf{03} (2020), 055
doi:10.1007/JHEP03(2020)055
[arXiv:1911.03781 [hep-ex]].

\bibitem{ATLAS:2019tpq}
G.~Aad \textit{et al.} [ATLAS],
Phys. Rev. D \textbf{102} (2020) no.3, 032004
doi:10.1103/PhysRevD.102.032004
[arXiv:1907.02749 [hep-ex]].

\bibitem{ATLAS:2020zms}
G.~Aad \textit{et al.} [ATLAS],
Phys. Rev. Lett. \textbf{125} (2020) no.5, 051801
doi:10.1103/PhysRevLett.125.051801
[arXiv:2002.12223 [hep-ex]].

\bibitem{ATLAS:2020gxx}
G.~Aad \textit{et al.} [ATLAS],
Eur. Phys. J. C \textbf{81} (2021) no.5, 396
doi:10.1140/epjc/s10052-021-09117-5
[arXiv:2011.05639 [hep-ex]].

\bibitem{ATLAS:2020tlo}
G.~Aad \textit{et al.} [ATLAS],
Eur. Phys. J. C \textbf{81} (2021) no.4, 332
doi:10.1140/epjc/s10052-021-09013-y
[arXiv:2009.14791 [hep-ex]].


\bibitem{ATLAS:2022rws}
G.~Aad \textit{et al.} [ATLAS],
JHEP \textbf{07} (2023), 203
doi:10.1007/JHEP07(2023)203
[arXiv:2211.01136 [hep-ex]].

\bibitem{ATLAS:2023zkt}
G.~Aad \textit{et al.} [ATLAS],
JHEP \textbf{02} (2024), 197
doi:10.1007/JHEP02(2024)197
[arXiv:2311.04033 [hep-ex]].

\bibitem{Krause:2018wmo}
M.~Krause, M.~M\"uhlleitner and M.~Spira,
Comput. Phys. Commun. \textbf{246} (2020), 106852
doi:10.1016/j.cpc.2019.08.003
[arXiv:1810.00768 [hep-ph]].

\bibitem{Athron:2021kve}
P.~Athron, A.~B\"uchner, D.~Harries, W.~Kotlarski, D.~St\"ockinger and A.~Voigt,
Comput. Phys. Commun. \textbf{283} (2023), 108584
doi:10.1016/j.cpc.2022.108584
[arXiv:2106.05038 [hep-ph]].

\bibitem{Denner:2019fcr}
A.~Denner, S.~Dittmaier and A.~M\"uck,
Comput. Phys. Commun. \textbf{254} (2020), 107336
doi:10.1016/j.cpc.2020.107336
[arXiv:1912.02010 [hep-ph]].


\bibitem{Kanemura:2017gbi}
S.~Kanemura, M.~Kikuchi, K.~Sakurai and K.~Yagyu,
Comput. Phys. Commun. \textbf{233} (2018), 134-144
doi:10.1016/j.cpc.2018.06.012
[arXiv:1710.04603 [hep-ph]].


\bibitem{Kanemura:2019slf}
S.~Kanemura, M.~Kikuchi, K.~Mawatari, K.~Sakurai and K.~Yagyu,
Comput. Phys. Commun. \textbf{257} (2020), 107512
doi:10.1016/j.cpc.2020.107512
[arXiv:1910.12769 [hep-ph]].


\bibitem{Kanemura:2022ldq}
S.~Kanemura, M.~Kikuchi and K.~Yagyu,
Nucl. Phys. B \textbf{983} (2022), 115906
doi:10.1016/j.nuclphysb.2022.115906
[arXiv:2203.08337 [hep-ph]].

\bibitem{Aiko:2023xui}
M.~Aiko, S.~Kanemura, M.~Kikuchi, K.~Sakurai and K.~Yagyu,
[arXiv:2311.15892 [hep-ph]].

\bibitem{Phan:2021xwc}
K.~H.~Phan, L.~Hue and D.~T.~Tran,
PTEP \textbf{2021} (2021) no.10, 103B07
doi:10.1093/ptep/ptab121
[arXiv:2106.14466 [hep-ph]].

\bibitem{VanOn:2021myp}
V.~Van On, D.~T.~Tran, C.~L.~Nguyen and K.~H.~Phan,
Eur. Phys. J. C \textbf{82} (2022) no.3, 277
doi:10.1140/epjc/s10052-022-10225-z
[arXiv:2111.07708 [hep-ph]].
\bibitem{Kachanovich:2020xyg}
A.~Kachanovich, U.~Nierste and I.~Ni\v{s}and\v{z}i\'c,
Phys. Rev. D \textbf{101} (2020) no.7, 073003
doi:10.1103/PhysRevD.101.073003
[arXiv:2001.06516 [hep-ph]].

\bibitem{Hue:2023tdz}
L.~T.~Hue, D.~T.~Tran, T.~H.~Nguyen and K.~H.~Phan,
PTEP \textbf{2023} (2023) no.8, 083B06
doi:10.1093/ptep/ptad106
[arXiv:2305.04002 [hep-ph]].
\bibitem{Chiang:2012qz}
C.~W.~Chiang and K.~Yagyu,
Phys. Rev. D \textbf{87} (2013) no.3, 033003
doi:10.1103/PhysRevD.87.033003
[arXiv:1207.1065 [hep-ph]].
\bibitem{Benbrik:2022bol}
R.~Benbrik, M.~Boukidi, M.~Ouchemhou, L.~Rahili and O.~Tibssirte,
Nucl. Phys. B \textbf{990} (2023), 116154
doi:10.1016/j.nuclphysb.2023.116154
[arXiv:2211.12546 [hep-ph]].


\bibitem{Akeroyd:1999xf}
A.~G.~Akeroyd, A.~Arhrib and C.~Dove,
Phys. Rev. D \textbf{61} (2000), 071702
doi:10.1103/PhysRevD.61.071702
[arXiv:hep-ph/9910287 [hep-ph]].

\bibitem{Akeroyd:2001aka}
A.~G.~Akeroyd, A.~Arhrib and M.~Capdequi Peyranere,
Phys. Rev. D \textbf{64} (2001), 075007
[erratum: Phys. Rev. D \textbf{65} (2002), 099903]
doi:10.1103/PhysRevD.65.099903
[arXiv:hep-ph/0104243 [hep-ph]].

\bibitem{Yin:2002sq}
J.~Yin, W.~G.~Ma, R.~Y.~Zhang and H.~S.~Hou,
Phys. Rev. D \textbf{66} (2002), 095008
doi:10.1103/PhysRevD.66.095008

\bibitem{Arhrib:2002ti}
A.~Arhrib,
Phys. Rev. D \textbf{67} (2003), 015003
doi:10.1103/PhysRevD.67.015003
[arXiv:hep-ph/0207330 [hep-ph]].

\bibitem{Farris:2003pn}
T.~Farris, J.~F.~Gunion, H.~E.~Logan and S.~f.~Su,
Phys. Rev. D \textbf{68} (2003), 075006
doi:10.1103/PhysRevD.68.075006
[arXiv:hep-ph/0302266 [hep-ph]].

\bibitem{Sasaki:2017fvk}
K.~Sasaki and T.~Uematsu,
Phys. Lett. B \textbf{781} (2018), 290-294
doi:10.1016/j.physletb.2018.04.005
[arXiv:1712.00197 [hep-ph]].

\bibitem{Abouabid:2020eik}
H.~Abouabid, A.~Arhrib, R.~Benbrik, J.~El Falaki, B.~Gong, W.~Xie and Q.~S.~Yan,
JHEP \textbf{05} (2021), 100
doi:10.1007/JHEP05(2021)100
[arXiv:2009.03250 [hep-ph]].

\bibitem{Bernreuther:2018ynm}
W.~Bernreuther, L.~Chen and Z.~G.~Si,
JHEP \textbf{07} (2018), 159
doi:10.1007/JHEP07(2018)159
[arXiv:1805.06658 [hep-ph]].


\bibitem{Accomando:2020vbo}
E.~Accomando, M.~Chapman, A.~Maury and S.~Moretti,
Phys. Lett. B \textbf{818} (2021), 136342
doi:10.1016/j.physletb.2021.136342
[arXiv:2002.07038 [hep-ph]].


\bibitem{Aiko:2022gmz}
M.~Aiko, S.~Kanemura and K.~Sakurai,
Nucl. Phys. B \textbf{986} (2023), 116047
doi:10.1016/j.nuclphysb.2022.116047
[arXiv:2207.01032 [hep-ph]].

\bibitem{Akeroyd:2023kek}
A.~G.~Akeroyd, S.~Alanazi and S.~Moretti,
J. Phys. G \textbf{50} (2023) no.9, 095001
doi:10.1088/1361-6471/ace3e1
[arXiv:2301.00728 [hep-ph]].

\bibitem{Esmail:2023axd}
W.~Esmail, A.~Hammad and S.~Moretti,
JHEP \textbf{11} (2023), 020
doi:10.1007/JHEP11(2023)020
[arXiv:2305.13781 [hep-ph]].

\bibitem{Biekotter:2023eil}
T.~Biek\"otter, S.~Heinemeyer, J.~M.~No, K.~Radchenko, M.~O.~O.~Romacho and G.~Weiglein,
JHEP \textbf{01} (2024), 107
doi:10.1007/JHEP01(2024)107
[arXiv:2309.17431 [hep-ph]].

\bibitem{Phan:2024zus}
K.~H.~Phan, D.~T.~Tran and T.~H.~Nguyen,
[arXiv:2404.02417 [hep-ph]].

\bibitem{Brignole:1992zv}
A.~Brignole and F.~Zwirner,
Phys. Lett. B \textbf{299} (1993), 72-82
doi:10.1016/0370-2693(93)90885-L
[arXiv:hep-ph/9210266 [hep-ph]].

\bibitem{He:2016sqr}
S.~P.~He and S.~h.~Zhu,
Phys. Lett. B \textbf{764} (2017), 31-37
[erratum: Phys. Lett. B \textbf{797} (2019), 134782]
doi:10.1016/j.physletb.2016.11.007
[arXiv:1607.04497 [hep-ph]].

\bibitem{Falaki:2023tyd}
J.~E.~Falaki,
Phys. Lett. B \textbf{840} (2023), 137879
doi:10.1016/j.physletb.2023.137879
[arXiv:2301.13773 [hep-ph]].



\bibitem{Moretti:2004wa}
M.~Moretti, S.~Moretti, F.~Piccinini, R.~Pittau and A.~D.~Polosa,
JHEP \textbf{02} (2005), 024
doi:10.1088/1126-6708/2005/02/024
[arXiv:hep-ph/0410334 [hep-ph]].

\bibitem{Binoth:2006ym}
T.~Binoth, S.~Karg, N.~Kauer and R.~Ruckl,
Phys. Rev. D \textbf{74} (2006), 113008
doi:10.1103/PhysRevD.74.113008
[arXiv:hep-ph/0608057 [hep-ph]].

\bibitem{Lopez-Val:2009xtx}
D.~Lopez-Val and J.~Sola,
Phys. Rev. D \textbf{81} (2010), 033003
doi:10.1103/PhysRevD.81.033003
[arXiv:0908.2898 [hep-ph]].

\bibitem{Ahmed:2021crg}
I.~Ahmed, U.~Nawaz, T.~Khurshid and S.~F.~Qazi,
Adv. High Energy Phys. \textbf{2022} (2022), 9735729
doi:10.1155/2022/9735729
[arXiv:2110.03920 [hep-ph]].
\bibitem{Branco:2011iw}
G.~C.~Branco, P.~M.~Ferreira, L.~Lavoura, M.~N.~Rebelo, M.~Sher and J.~P.~Silva,
Phys. Rept. \textbf{516} (2012), 1-102
doi:10.1016/j.physrep.2012.02.002
[arXiv:1106.0034 [hep-ph]].
\bibitem{Nie:1998yn}
S.~Nie and M.~Sher,
Phys. Lett. B \textbf{449} (1999), 89-92
doi:10.1016/S0370-2693(99)00019-2
[arXiv:hep-ph/9811234 [hep-ph]].
\bibitem{Kanemura:1999xf}
S.~Kanemura, T.~Kasai and Y.~Okada,
Phys. Lett. B \textbf{471} (1999), 182-190
doi:10.1016/S0370-2693(99)01351-9
[arXiv:hep-ph/9903289 [hep-ph]].
\bibitem{Akeroyd:2000wc}
A.~G.~Akeroyd, A.~Arhrib and E.~M.~Naimi,
Phys. Lett. B \textbf{490} (2000), 119-124
doi:10.1016/S0370-2693(00)00962-X
[arXiv:hep-ph/0006035 [hep-ph]].
\bibitem{Ginzburg:2005dt}
I.~F.~Ginzburg and I.~P.~Ivanov,
Phys. Rev. D \textbf{72} (2005), 115010
doi:10.1103/PhysRevD.72.115010
[arXiv:hep-ph/0508020 [hep-ph]].
\bibitem{Kanemura:2011sj}
S.~Kanemura, Y.~Okada, H.~Taniguchi and K.~Tsumura,
Phys. Lett. B \textbf{704} (2011), 303-307
doi:10.1016/j.physletb.2011.09.035
[arXiv:1108.3297 [hep-ph]].
\bibitem{Kanemura:2015ska}
S.~Kanemura and K.~Yagyu,
Phys. Lett. B \textbf{751} (2015), 289-296
doi:10.1016/j.physletb.2015.10.047
[arXiv:1509.06060 [hep-ph]].
\bibitem{Bian:2016awe}
L.~Bian and N.~Chen,
JHEP \textbf{09} (2016), 069
doi:10.1007/JHEP09(2016)069
[arXiv:1607.02703 [hep-ph]].
\bibitem{Xie:2018yiv}
W.~Xie, R.~Benbrik, A.~Habjia, S.~Taj, B.~Gong and Q.~S.~Yan,
Phys. Rev. D \textbf{103} (2021) no.9, 095030
doi:10.1103/PhysRevD.103.095030
[arXiv:1812.02597 [hep-ph]].
\bibitem{Chun:2012jw}
E.~J.~Chun, H.~M.~Lee and P.~Sharma,
JHEP \textbf{11} (2012), 106
doi:10.1007/JHEP11(2012)106
[arXiv:1209.1303 [hep-ph]].
\bibitem{Chen:2013dh}
C.~S.~Chen, C.~Q.~Geng, D.~Huang and L.~H.~Tsai,
Phys. Lett. B \textbf{723} (2013), 156-160
doi:10.1016/j.physletb.2013.05.007
[arXiv:1302.0502 [hep-ph]].
\bibitem{Arhrib:2011uy}
A.~Arhrib, R.~Benbrik, M.~Chabab, G.~Moultaka, M.~C.~Peyranere, L.~Rahili and J.~Ramadan,
Phys. Rev. D \textbf{84} (2011), 095005
doi:10.1103/PhysRevD.84.095005
[arXiv:1105.1925 [hep-ph]].
\bibitem{Arhrib:2011vc}
A.~Arhrib, R.~Benbrik, M.~Chabab, G.~Moultaka and L.~Rahili,
JHEP \textbf{04} (2012), 136
doi:10.1007/JHEP04(2012)136
[arXiv:1112.5453 [hep-ph]].
\bibitem{Akeroyd:2012ms}
A.~G.~Akeroyd and S.~Moretti,
Phys. Rev. D \textbf{86} (2012), 035015
doi:10.1103/PhysRevD.86.035015
[arXiv:1206.0535 [hep-ph]].
\bibitem{Akeroyd:2011zza}
A.~G.~Akeroyd and H.~Sugiyama,
Phys. Rev. D \textbf{84} (2011), 035010
doi:10.1103/PhysRevD.84.035010
[arXiv:1105.2209 [hep-ph]].
\bibitem{Akeroyd:2011ir}
A.~G.~Akeroyd and S.~Moretti,
Phys. Rev. D \textbf{84} (2011), 035028
doi:10.1103/PhysRevD.84.035028
[arXiv:1106.3427 [hep-ph]].
\bibitem{Aoki:2011pz}
M.~Aoki, S.~Kanemura and K.~Yagyu,
Phys. Rev. D \textbf{85} (2012), 055007
doi:10.1103/PhysRevD.85.055007
[arXiv:1110.4625 [hep-ph]].
\bibitem{Kanemura:2012rs}
S.~Kanemura and K.~Yagyu,
Phys. Rev. D \textbf{85} (2012), 115009
doi:10.1103/PhysRevD.85.115009
[arXiv:1201.6287 [hep-ph]].
\bibitem{Chabab:2014ara}
M.~Chabab, M.~C.~Peyranere and L.~Rahili,
Phys. Rev. D \textbf{90} (2014) no.3, 035026
doi:10.1103/PhysRevD.90.035026
[arXiv:1407.1797 [hep-ph]].
\bibitem{Han:2015hba}
Z.~L.~Han, R.~Ding and Y.~Liao,
Phys. Rev. D \textbf{91} (2015), 093006
doi:10.1103/PhysRevD.91.093006
[arXiv:1502.05242 [hep-ph]].
\bibitem{Chabab:2015nel}
M.~Chabab, M.~C.~Peyran\`ere and L.~Rahili,
Phys. Rev. D \textbf{93} (2016) no.11, 115021
doi:10.1103/PhysRevD.93.115021
[arXiv:1512.07280 [hep-ph]].
\bibitem{Haba:2016zbu}
N.~Haba, H.~Ishida, N.~Okada and Y.~Yamaguchi,
Eur. Phys. J. C \textbf{76} (2016) no.6, 333
doi:10.1140/epjc/s10052-016-4180-z
[arXiv:1601.05217 [hep-ph]].
\bibitem{Ghosh:2017pxl}
D.~K.~Ghosh, N.~Ghosh, I.~Saha and A.~Shaw,
Phys. Rev. D \textbf{97} (2018) no.11, 115022
doi:10.1103/PhysRevD.97.115022
[arXiv:1711.06062 [hep-ph]].
\bibitem{Ashanujjaman:2021txz}
S.~Ashanujjaman and K.~Ghosh,
JHEP \textbf{03} (2022), 195
doi:10.1007/JHEP03(2022)195
[arXiv:2108.10952 [hep-ph]].
\bibitem{Zhou:2022mlz}
R.~Zhou, L.~Bian and Y.~Du,
JHEP \textbf{08} (2022), 205
doi:10.1007/JHEP08(2022)205
[arXiv:2203.01561 [hep-ph]].
\bibitem{Aoki:2012jj}
M.~Aoki, S.~Kanemura, M.~Kikuchi and K.~Yagyu,
Phys. Rev. D \textbf{87} (2013) no.1, 015012
doi:10.1103/PhysRevD.87.015012
[arXiv:1211.6029 [hep-ph]].
\bibitem{Hahn:2000kx}
T.~Hahn,
Comput. Phys. Commun. \textbf{140} (2001), 418-431
doi:10.1016/S0010-4655(01)00290-9
[arXiv:hep-ph/0012260 [hep-ph]].
\bibitem{Hahn:1998yk}
T.~Hahn and M.~Perez-Victoria,
Comput. Phys. Commun. \textbf{118} (1999), 153-165.
\bibitem{Denner:2016kdg}
A.~Denner, S.~Dittmaier and L.~Hofer,
Comput. Phys. Commun. \textbf{212} (2017), 220-238
doi:10.1016/j.cpc.2016.10.013
[arXiv:1604.06792 [hep-ph]].
\bibitem{Mertig:1990an}
R.~Mertig, M.~Bohm and A.~Denner,
Comput. Phys. Commun. \textbf{64} (1991), 345-359
doi:10.1016/0010-4655(91)90130-D
\bibitem{Haller:2018nnx}
J.~Haller, A.~Hoecker,
R.~Kogler, K.~M\"onig,
T.~Peiffer and J.~Stelzer,
Eur. Phys. J. C \textbf{78} (2018) no.8, 675
doi:10.1140/epjc/s10052-018-6131-3
[arXiv:1803.01853 [hep-ph]].
\bibitem{Denner:2005nn}
A.~Denner and S.~Dittmaier,
Nucl. Phys. B \textbf{734} (2006), 62-115
doi:10.1016/j.nuclphysb.2005.11.007
[arXiv:hep-ph/0509141 [hep-ph]].
\bibitem{Djouadi:2005gi}
A.~Djouadi,
Phys. Rept. \textbf{457} (2008), 1-216
doi:10.1016/j.physrep.2007.10.004
[arXiv:hep-ph/0503172 [hep-ph]].

\bibitem{Phan:2024vfy}
K.~H.~Phan, D.~T.~Tran and T.~H.~Nguyen,
[arXiv:2409.00662 [hep-ph]].
\end{thebibliography}
\end{document}